\documentstyle[12pt,epsfig]{elsart}
\def\Journal#1#2#3#4{{#1} {\bf #2}, #3 (#4)}

\def\NCA{\em Nuovo Cimento}

\def\NIMA{{\em Nucl. Instrum. Methods} A}
\def\NIMB{{\em Nucl. Instrum. Methods} B}
\def\NPB{{\em Nucl. Phys.} B}
\def\NPBP{{\em Nucl. Phys.} B (Proc. Suppl.)}
\def\NPA{{\em Nucl. Phys.} A}
\def\PLB{{\em Phys. Lett.}  B}
\def\PRL{\em Phys. Rev. Lett.}
\def\PRD{{\em Phys. Rev.} D}
\def\PRC{{\em Phys. Rev.} C}
\def\PRA{{\em Phys. Rev.} A}
\def\PR{{\em Phys. Rev. }}
\def\PRP{{\em Phys. Rep. }}
\def\EPC{{\em Europ. Phys. J.} C}
\def\ZPC{{\em Z. Phys.} C}
\def\ZPA{{\em Z. Phys.} A}
\def\ZP{{\em Z. Phys. }}
\def\PS{{\em Physics Scripta }}
\def\PAN{\em Phys. Atom. Nucl.}
\def\ARAA{{\em Ann. Rev. Astr. Astroph.}}
\def\ARNP{{\em Ann. Rev. Nucl. Part. Phys.}}

\def\PNPP{\em Prog. Nucl. Part. Phys.}
\def\PTP{\em Prog. Theo. Phys.}
\def\PTPS{\em Prog. Theo. Phys. Suppl.}
\def\APJ{\em ApJ}
\def\AA{\em Astron. Astroph.}
\def\AAS{\em Astron. Astroph. Suppl. Ser.}
\def\RPP{\em Rep. Prog. Phys.}
\def\RMP{\em Rev. Mod. Phys.}
\def\ASP{\em Astroparticle Physics}
\def\EPL{\em Europhys. Lett.}
\def\IJMP{\em Int. Journal of Modern Physics}
\def\JPG{\em Journal of Physics G}
\def\NAT{\em Nature}
\def\SCI{\em Science}
\def\JETP{\em JETP Lett.}
\def\ANYAS{\em Ann. NY. Acad. Sci}
\def\BASI{\em Bull. Astr. Soc. India}
\def\BAPS{\em Bull. Am. Phys. Soc.}
\def\SJNP{\em Sov. Journal of Nucl. Phys.}
\def\ZETZ{\em Zh. Eksp Teor. Fiz.}

\def\ra{\rightarrow}
\def\be{\begin{equation}}
\def\ee{\end{equation}}
\def\bea{\begin{eqnarray}}
\def\eea{\end{eqnarray}}
\newcommand{\ECT}{Proc. {\it ECT Workshop on double beta decay 
and related topics}, Trento, eds. H.V. Klapdor-Kleingrothaus and 
S.Stoica, World Scientific, 1995 }
\newcommand{\gsim}{\mbox{$\stackrel{>}{\sim}$ }}
\newcommand{\lsim}{\mbox{$\stackrel{<}{\sim}$ }}
\newcommand{\expe}{experiment }
\newcommand{\expek}{experiment, }
\newcommand{\tran}{transition }
\newcommand{\trans}{transitions }

\newcommand{\exps}{experiments }
\newcommand{\expsp}{experiments. }
\newcommand{\gut}{grand unified theory }
\newcommand{\bb}{double beta decay }
\newcommand{\eos}{equation of state }
\newcommand{\obb}{0\mbox{$\nu\beta\beta$ decay} }
\newcommand{\zbb}{2\mbox{$\nu\beta\beta$ decay} }
\newcommand{\nbb}{neutrinoless double beta decay }
\newcommand{\majo}{Majorana }

\newcommand{\mas}{Majorana neutrinos }
\newcommand{\lsnd}{LSND }
\newcommand{\gno}{GNO }
\newcommand{\ica}{ICARUS }
\newcommand{\SNO}{Sudbury Neutrino Observatory }

\newcommand{\ugm}{upward going muons }

\newcommand{\osz}{oscillation }
\newcommand{\oszsp}{oscillations. }
\newcommand{\oszs}{oscillations }

\newcommand{\com}{component }
\newcommand{\hdm}{hot dark matter }
\newcommand{\lbls}{long baseline experiments }
\newcommand{\vev}{vacuum expectation value }

\newcommand{\ssm}{see-saw-mechanism }
\newcommand{\sn}{supernova }
\newcommand{\sna}{SN 1987A }
\newcommand{\sne}{supernovae }
\newcommand{\crs}{cosmic rays }
\newcommand{\agn}{AGN }
\newcommand{\enu}{\mbox{$E_{\nu}$} }
\newcommand{\tnu}{\mbox{$T_{\nu}$} }
\newcommand{\tg}{\mbox{$T_{\gamma}$} }
\newcommand{\msun}{M_{\odot} }
\newcommand{\pmts}{photomultipliers }
\newcommand{\adn}{almost degenerated neutrinos }
\newcommand{\delm}{\mbox{$\Delta m^2$} }
\newcommand{\nul}{\mbox{$\nu_L$} }
\newcommand{\nulb}{\mbox{$\bar{\nu}_L$} }

\newcommand{\nui}{\mbox{$\nu_i$} }
\newcommand{\nua}{\mbox{$\nu_{\alpha}$} }

\newcommand{\nulc}{\mbox{$(\nu_L)^C$} }

\newcommand{\nr}{\mbox{$N_R$} }
\newcommand{\me}{\mbox{$m_{\nu_e}$} }
\newcommand{\mbe}{\mbox{$m_{\bar{\nu}_e}$} }
\newcommand{\mmu}{\mbox{$m_{\nu_\mu}$} }
\newcommand{\mtau}{\mbox{$m_{\nu_\tau}$} }
\newcommand{\bnel}{\mbox{$\bar{\nu}_e$} }
\newcommand{\bnmu}{\mbox{$\bar{\nu}_\mu$} }
\newcommand{\bntau}{\mbox{$\bar{\nu}_\tau$} }
\newcommand{\nel}{\mbox{$\nu_e$} }
\newcommand{\nmu}{\mbox{$\nu_\mu$} }
\newcommand{\ntau}{\mbox{$\nu_\tau$} }
\newcommand{\sint}{\mbox{$sin^2 2\theta$} }
\newcommand{\sk}{Super-Kamiokande }
\newcommand{\munu}{\mbox{$\mu_{\nu}$} }
\newcommand{\mub}{\mbox{$\mu_B$} }
\newcommand{\ton}{\mbox{$T_{1/2}^{0\nu}$} }
\newcommand{\tzn}{\mbox{$T_{1/2}^{2\nu}$} }
\newcommand{\snp}{solar neutrino problem }
\newcommand{\mamo}{magnetic moment }
\newcommand{\lh}{left-handed }
\newcommand{\rh}{right-handed }
\newcommand{\neu}{neutrino }
\newcommand{\neus}{neutrinos }
\newcommand{\neusph}{neutrinosphere }
\newcommand{\neusp}{neutrinos. }

\newcommand{\ema}{\mbox{$\langle m_{\nu_e} \rangle$ }}

\newcommand{\gess}{\mbox{$^{76}Ge$ }}
\newcommand{\rnzhz}{\mbox{$^{222}Rn$ }}
\newcommand{\moeh}{\mbox{$^{100}Mo$ }}
\newcommand{\ndhf}{\mbox{$^{150}Nd$ }}
\newcommand{\caav}{\mbox{$^{48}Ca$ }}
\newcommand{\gdhs}{\mbox{$^{160}Gd$ }}
\newcommand{\hohds}{\mbox{$^{163}Ho$ }}
\newcommand{\rehsa}{\mbox{$^{187}Re$ }}
\newcommand{\tehaz}{\mbox{$^{128}Te$ }}
\newcommand{\tehd}{\mbox{$^{130}Te$ }}
\newcommand{\cdhsz}{\mbox{$^{116}Cd$ }}
\newcommand{\xehsd}{\mbox{$^{136}Xe$ }}
\newcommand{\xehed}{\mbox{$^{131}Xe$ }}
\newcommand{\cshed}{\mbox{$^{131}Cs$ }}
\newcommand{\cshsd}{\mbox{$^{137}Cs$ }}
\newcommand{\clsd}{\mbox{$^{37}Cl$ }}
\newcommand{\arsd}{\mbox{$^{36}Ar$ }}
\newcommand{\arad}{\mbox{$^{38}Ar$ }}
\newcommand{\arsid}{\mbox{$^{37}Ar$ }}
\newcommand{\arv}{\mbox{$^{40}Ar$ }}
\newcommand{\kav}{\mbox{$^{40}K$ }}
\newcommand{\gaes}{\mbox{$^{71}Ga$ }}
\newcommand{\cref}{\mbox{$^{51}Cr$ }}
\newcommand{\seza}{\mbox{$^{82}Se$ }}
\newcommand{\zsn}{\mbox{$^{96}Zr$ }}
\newcommand{\csz}{\mbox{$^{12}C$ }}
\newcommand{\sfz}{\mbox{$^{15}O$ }}
\newcommand{\sfs}{\mbox{$^{16}O$ }}

\newcommand{\sszw}{\mbox{$^{12}N$ }}
\newcommand{\ssz}{\mbox{$^{16}N$ }}
\newcommand{\gees}{\mbox{$^{71}Ge$ }}
\newcommand{\ihsz}{\mbox{$^{127}I$ }}
\newcommand{\yhss}{\mbox{$^{176}Yb$ }}
\newcommand{\hev}{\mbox{$^4He$ }}
\newcommand{\hed}{\mbox{$^3He$ }}
\newcommand{\bes}{\mbox{$^7Be$ }}
\newcommand{\lis}{\mbox{$^7Li$ }}
\newcommand{\ba}{\mbox{$^8B$ }}
\newcommand{\cms}{\mbox{$cm^{-2}s^{-1}$ }}

\begin{document}
\newpage
\begin{frontmatter}
\title{On the physics of massive neutrinos}
\author{K. Zuber}
\address{Lehrstuhl f\"ur Experimentelle Physik IV, Universit\"at Dortmund,
Otto-Hahn Str.4, \protect\newline
44221 Dortmund, Germany}
\begin{abstract}
Massive neutrinos open up the possibility for a variety of new physical
phenomena. Among them are oscillations and double beta decay. Furthermore
they influence several fields from particle physics to cosmology. In this
article the concept of massive neutrinos is given and the present state of
experimental research is extensively reviewed. This includes astrophysical
studies of
solar, supernova and very high energy neutrinos. Future perspectives are
also outlined. 
\end{abstract}
\end{frontmatter}

\section{Introduction}
The birth of the \neu due to W. Pauli in 1930 was a rather desperate
attempt to explain the
continuous $\beta$-spectrum \cite{pau77}:\\
{\it '' ... I have considered ... a way out for saving the law of
conservation of energy. Namely, the 
possibility that there could exist in the nuclei electrically neutral
particles, that I
will call neutrons
(which are today called neutrinos) which have spin 1/2 and follow the
exclusion principle. The
continous $\beta$-spectrum would then be understandable assuming that in
$\beta$-decay
together with the electron, in all cases, also a neutron is emitted
in such a way that the sum
of energy of neutron and of electron remains constant... I admit that my
solution appears to
you not very probable... But only who dares wins, and the gravity of the
situation in regard to
continuous $\beta$-spectrum...''}\\
The experimental discovery of the \neu by Cowan and Reines \cite{rei56}
in 1956 and the observation that there
exist
different types of \neus by Danby et al. \cite{dan62} were important milestones. The
last important step about \neus stems from
the
LEP-\exps 
measuring the $Z^o$-width  which results in $2.993 \pm 0.011$ flavours
for \neu masses below 45 GeV
\cite{cer97}.\\
From all particles of the standard model, \neus are the most unknown.
Because they
are treated as massless particles, the physical phenomena associated with
them are rather
limited. On the other hand in case of massive \neus , which are predicted 
by several Grand Unified
Theories, several new effects can occur. This article reviews the effects 
of massive \neus as well as the present knowledge
and experimental status of \neu mass searches.
 
\section{Theoretical models of neutrinos}
The presently very successful standard model of particle physics contains
fermions as
\lh chiral projections in doublets and \rh charged fermions as singlets under
SU(3)$_C\otimes$SU(2)$_L\otimes$U(1)$_Y$ transformations. 
Neutrinos only show up in the doublets which does not allow any Yukawa
coupling and therefore no mass with the minimal particle content of the
standard model. Moreover, because \neus are the only uncharged fundamental
fermions, they
might be their
own antiparticles.\\
In the following chapter, a theoretical description of 
\neus is given as well as possible extensions of the standard model to
generate \neu masses. A second requirement will be to explain 
why \neu masses are so much smaller than the
corresponding charged fermion masses. The most promising way is given by the
see-saw-mechanism.
\subsection{Weyl-, Majorana- and Dirac-\neus}
The \neu states observed in weak interactions are \neus with
helicity -1 and antineutrinos with helicity +1. For massless \neus
and the absence of right-handed currents there is no chance to distinguish
between Dirac- and
\majo \neus. Because V-A theory is maximal parity violating
the other two states (\neus with helicity +1 and antineutrinos with
helicity -1),
if they exist, are unobservable.
If \neus are massless a 2-component spinor
(Weyl-spinor) is sufficient for description, first discussed for the general
case 
of massless spin 1/2 particles by Weyl \cite{wey29}, which are the
helicity -1(+1) projections for particles (antiparticles) out of a
4-\com spinor $\Psi$. They are given by
\be
\Psi_{L,R} = \frac{1}{2}(1 \pm \gamma_5) \Psi 
\ee
The eigenvalues of $\gamma_5$ (chirality) agree with those of helicity
in the massless case. Here the Dirac equation
decouples into two seperate equations for $\psi_{L,R}$
respectively. An alternative 2-component description was developed by Majorana 
\cite{maj37}
to describe a particle identical to its antiparticle.
If \neus acquire a mass, in general both helicity states for \neus and
antineutrinos can exist, making a 4-\com description necessary.
Here a 4-\com Dirac-spinor can be treated as a sum
of two 2-\com Weyl-spinors or as composed out of two degenerated \majo \neusp
However it is still an open question whether \neus 
are Dirac or \majo particles. 
The \majo condition, for a particle to be its own
antiparticle, can be written as
\be
C^{-1} \gamma_{\mu} C = - \gamma_{\mu}^T
\ee
with $C$ as the charge conjugation operator.
The real charge conjugated state $(\psi_{L,R})^C$ is not
obtained by a C operation but by CP, because
pure charge conjugation results in the wrong helicity state.
In the case of a Dirac-neutrino, the fields $\psi_R$ and
$\psi_L^C$ are sterile with
respect to weak interactions and therefore they are sometimes called
$N_R$ and $N_L^C$.
The most general mass term in the Lagrangian including both
Dirac- and \majo fields is given by
\begin{eqnarray*}
{\cal L} &=& - \frac{1}{2} (m^D (\bar{\psi_L} \psi_R +
\bar{\psi^C_L}\psi^C_R) + m_L^M \bar{\psi_L} \psi^C_R +
m_R^M
\bar{\psi^C_L} \psi_R) + h.c. \\
&=& \bar{\Psi}_L {\cal M} \Psi^C_R +  \bar{\Psi}^C_R {\cal M} \Psi_L
\end{eqnarray*}
with 
\be
\label{eq:massma}\Psi_R = {\psi_R \choose \psi^C_R} , \Psi_L = {\psi_L \choose 
\psi^C_L} \quad
\mbox{and} \quad
{\cal M} = 
\left( \begin{array}{cc}
m_L^M & m^D \\
m^D & m_R^M 
\end{array} \right)
\ee
In the general case of $n_a$ active \neus and $n_s$ sterile \neus 
${\cal M}$
is a ($n_a + n_s) \times (n_a + n_s)$ matrix
(see \cite{bil87}).
Assuming only one \neu generation, diagonalisation of ${\cal M}$ results in the 
eigenvalues
\be
m_{1,2} = \frac{1}{2} [(m_L^M + m_R^M) \pm \sqrt{(m_L^M -
m_R^M)^2 + 4 (m^D)^2}]
\ee
Four different cases can be considered:
\begin{itemize}
\item{$m_L^M=m_R^M=0 \ra m_{1,2}=m^D$: The result is a pure
Dirac-neutrino
which can be seen as two degenerated Majorana fields.}
\item{$m^D \gg m_L^M,m_R^M \ra m_{1,2}\approx m^D$: Neutrinos
are Pseudo-Dirac-Neutrinos.}
\item{$m^D=0 \ra m_{1,2} = m_L^M,m_R^M$: Neutrinos are pure
\majo particles.}
\item{$m^M_R \gg m^D,m_L^M=0$: This leads to the \ssm .}
\end{itemize}
The \ssm \cite{gel78,yan80} results in two eigenvalues 
\bea
m_1 = \frac{(m^D)^2}{m^M_R} \\
m_2 = m^M_R (1 + \frac{(m^D)^2}{(m^M_R)^2}) \approx m^M_R
\eea
Because \neu masses should be embedded in GUT-theories, the
latter offers two scales for $m^D$ and $m^M_R$. All fermions
out of a multiplet get their Dirac-mass via the coupling to
the same Higgs vacuum expectation value. Therefore the \neu
Dirac mass is expected to be of the order of the charged
lepton and quark masses. The heavy \majo mass can take values up to
the GUT-scale, which is in the simplest models about
$10^{16}$ GeV. Assuming three families and a unique $m^M_R$
the classical quadratic see-saw
\be
\me : \mmu : \mtau = m_u^2 : m_c^2 : m_t^2
\ee
emerges. This is only a rough estimate because several effects
influence this relation. Instead of the quark-masses, the 
charged lepton masses could be used. In scenarios where
$m^M_R$ is proportional to $m^D$ for the different families, a linear
see-saw relation results. Depending on the GUT-model, the mass scale
of $m^M_R$ need not be related to the GUT-scale but might
be in connection with some intermediate symmetry breaking
scale (Table 1). Last not least the relation holds at the GUT scale, to
get a prediction at the electroweak scale, the evolution has
to be calculated with the help of the renormalisation group
equations. Especially the third term can experience a
significant change depending on the used GUT model like \cite{blu92}
\be
\me : \mmu : \mtau = 0.05m_u^2 : 0.09m_c^2 : 0.38m_t^2 \quad \mbox{SUSY-GUT}
\ee
\be
\me : \mmu : \mtau = 0.05m_u^2 : 0.07m_c^2 : 0.18m_t^2 \quad \mbox{SO(10)-GUT}
\ee
A further \ssm resulting in \adn is discussed in chapter 
\ref{ch:23}.
\begin{center}
\begin{table}
{\small
\label{tab:lan88}
\caption{Predictions for \neu masses according to different see-saw models. As can be seen, the
quantity \ema, measured in \nbb , corresponds in most cases to \me (after \protect \cite{lan88}).}
\begin{tabular}{ccccc}
\hline
model & \me & \ema & \mmu & \mtau \\
\hline
Dirac & 1-10 MeV & 0 & 100 MeV-1 GeV & 1-100 GeV\\
pure Majorana (Higgs triplet) & arbitrary & \me & arbitrary & arbitrary\\
GUT seesaw ($M \approx 10^{16}$GeV) & $10^{-14}$ eV & \me & $10^{-9}$ eV &
$10^{-6}$ eV\\
Intermed. seesaw ($M \approx 10^9$GeV) & $10^{-7}$ eV & \me & $10^{-2}$
eV & 10 eV\\
SU(2)$\otimes$SU2$\otimes$U(1) (M $\approx$ 1 TeV) & $10^{-1}$ eV & \me &
10 keV & 1 MeV\\
light seesaw ($M \ll$ 1 GeV) & 1-10 MeV & $\ll$ \me & & \\
charged Higgs & $<$ 1eV& $\ll$ \me& & \\
\hline
\end{tabular}}
\end{table}
\end{center}

\subsection{Massive neutrinos in the standard model}
In the present standard model with minimal particle content, \neus 
remain massless.
The simplest extension to create \neu masses is the inclusion
of SU(2) singlet states denoted by $N_R$. Because of hypercharge
zero they remain singlets
of the entire gauge group and have no new interaction with the gauge
bosons. New Yukawa-couplings of the form
\be
{\cal{L}} = h_{\nu} (\nulb \bar{e}_L) {\Phi^0 \choose \Phi^-} \nr + h.c.
\ee
result in a Dirac mass term of $m^D = h_{\nu} v_2$ where $v_2 \approx 246$ 
GeV is the \vev of
the
neutral \com of the standard model Higgs-doublet. In
order to produce an eV-\neu , the Yukawa-coupling $h_{\nu}$ has to be
smaller than $10^{-10}$.
Some fine-tuning is required for this, on the other hand the generation of 
the mass pattern
is still unknown and such a small $h_{\nu}$ might be possible. An
immediate
consequence of a mass term is, that similar to the quark sector, a mixing 
between the 
mass eigenstates \nui and flavour eigenstates \nua can occur
\be
\nua = \sum U_{\alpha i} \nui
\ee
allowing several new phenomena, e.g. \neu \oszs , which will be discussed
later. Nevertheless the global lepton number L remains a conserved quantity.\\
Without introducing additional fermion singlets, it is only possible to generate 
\majo mass
terms,
because only \nul and its charge conjugate \nulc exist. These terms necessarily
violate L and therefore also B-L by two units.
The only fermionic bilinears carrying a B-L net
quantum
number are
\be
\bar{\Psi}_L (\Psi)^C_R \quad , \quad \bar{(\Psi)}^C_L \Psi_R
\ee
The necessary extensions of the Higgs-sector to produce gauge invariant Yukawa 
couplings
therefore offer
three possibilities: a) a triplet b) a single charged singlet and c) a
double charged singlet.\\
Case a: The additional Higgs triplet $\Delta = 
(\Delta^0,\Delta^-,\Delta^{- -})$ 
carries hypercharge -2 and the neutral component develops a \vev of $v_3$. It
is this \vev which
enters the Yukawa-coupling for the mass generation of \neus. There is no 
prediction for the
masses or $v_3$, but
it can be much smaller than $v_2$ and therefore explain the lightness of
\neusp
This additional \vev would also modify the relation between the gauge boson
masses to \cite{moh91}
\be
\frac{m_W^2}{m_Z^2 cos^2 \theta_W} = \frac{1+2 \frac{v_3^2}{v_2^2}}{1+4 
\frac{v_3^2}{v_2^2}}
\ee
which, by using experimental values, results in 
\be
\frac{v_3}{v_2} < 0.07
\ee
Case b: This corresponds to the Zee-model \cite{zee80}. By introducing a
single
charged higgs $h_-$ and 
additional higgs doublets, \majo masses can be
gene\-rated at the
one-loop
level by self-energy diagrams. If only one higgs couples to
leptons, a mass matrix
of the following form can be derived \cite{wol80}
\be
M = m_0 \left( \begin{array}{ccc}
0 & \sigma & cos \alpha \\
\sigma & 0 & sin \alpha \\
cos \alpha & sin \alpha & 0
\end{array} \right)
\ee
with 
\bea
tan \alpha = \frac{f_{\mu\tau}}{f_{e \tau}}( 1- 
\frac{m_\mu^2}{m_\tau^2})\\
\sigma = \frac{f_{e\mu}}{f_{e \tau}} \frac{m_\mu^2}{m_\tau^2}
cos \alpha \\
m_0 = A m_\tau^2 f_{e \tau}/cos \alpha
\eea
where $f$ are the Yukawa coupling constants and the electron 
mass is neglected.
This in general implies
two nearly degenerated \neus and one which is much lighter.\\
Case c: By including an additional double charged higgs $k_{++}$ with (B-L)
quantum number 2,
it is possible to generate masses on the 2-loop level which are
therefore small
\cite{bab88}. It can be shown
that for three flavours one eigenvalue is zero or at least much smaller
than the others.\\
All the solutions described above violate B-L by introducing B-L breaking 
terms in ${\cal{L}}$.
On the other hand, the vacuum could be non-invariant under B-L, for example
as a spontaneous
breaking of a global B-L symmetry. This is discussed
in more detail in connection with the associated Goldstone boson, called 
majoron, in chapter
\ref{ch:cha41}.

\subsection{Neutrino masses in grand unified theories}
\label{ch:23}
As already seen in the description of the see-saw-mechanism, by choosing a
large $m_R^M$ it is
possible to get small \neu masses. To find a scale for $m_R^M$, an
implementation of this mechanism into
grand unified theories seems reasonable.
The simplest grand unified theory is SU(5) even if
the minimal SU(5)-model is ruled out by proton-decay \expsp Because all
the fundamental fermions can be
arranged in one multiplet there is no room for a \rh \neu and consequently no
Dirac-masses.
Minimal SU(5) is also B-L conserving which is given by the multiplets 
and the gauge invariance of
the higgs field couplings. For this reason \majo mass terms also do not
exist. Therefore in the minimal
SU(5) \neus remain massless. By extending the higgs-sector it is possible
to create mass terms via
radiative corrections as in the Zee model. Nevertheless the proton decay
bound remains.\\
The next higher \gut relies on SO(10). All fundamental fermions can be 
arranged in a 16-multiplet,
where the 16th element can be associated to a \rh \neu . This allows 
the generation of Dirac
masses. In SO(10) B-L is not necessarily conserved opening the chance for
\majo mass terms as
well.
The breaking of SO(10) allows different schemes like
\be
SO(10) \ra SU(5) \ra SU(3) \otimes SU(2) \otimes U(1)
\ee
or into a left-right symmetric version after the Pati-Salam model 
\cite{pat74}
\be
SO(10) \ra SU(2)_L \otimes SU(2)_R \otimes SU(4)
\ee 
This generates a \rh weak interaction with \rh gauge bosons. These models
create \neu mass matrices
like \cite{moh95,moh81}
\be
\left( \begin{array}{cc}
f v_L & m^D \\  m^{DT} & f v_R 
\end{array} \right) \quad \mbox{and} \quad v_L = \frac{\lambda
(m_W^2)_L}{v_R}
\ee
where f is a 3$\times$3 matrix and $v_L,v_R$ are the \vev of the \lh and
\rh higgses respectively. Diagonalisation leads to
masses for
the light \neus
of the form
\be
m_{\nu} \approx \frac{f \lambda (m_W^2)_L}{v_R} - m^D f^{-1} (m^D)^T/v_R +
...
\ee
Two important things emerge from this. First of all, the first term
dominates over the second,
the latter is corresponding to the quadratic \ssm . Because no \neu masses
are involved in the first term and if f is diagonal, 
no scaling is included resulting in a model with almost degenerated \neus 
in leading order.
This is sometimes
called type II \ssm \cite{moh95}. In case the first term vanishes, we end up
with
the normal \ssm. For a more extensive discussion on \neu mass generation 
in GUTs see \cite{moh91}.

\section{Kinematical tests of neutrino masses}
\subsection{Beta decay}
\begin{figure}
\begin{center}
\epsfig{file=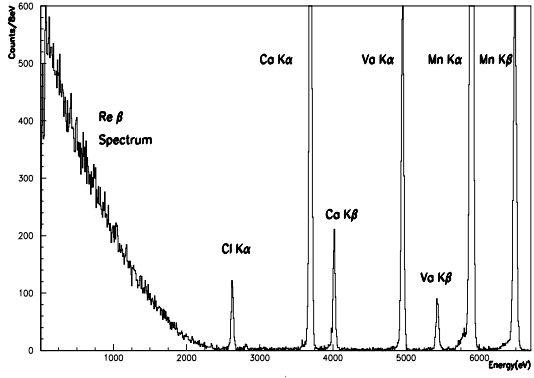,width=7cm,height=5cm}
\caption{\rehsa spectrum obtained with a cryogenic
bolometer. The big spikes correspond to calibration peaks (from \protect
\cite{meu98}).}
\label{pic:re187sp}
\end{center}
\end{figure}

The classical way to determine the mass of $\bar{\nel}$ is the
investigation of the
electron spectrum in beta decay.
A finite \neu mass will reduce the phase space and leads to a 
change of the shape
of the electron spectra, which for small masses can be investigated
best near the Q-value of the
transition.
In case several mass
eigenstates contribute, the total electron spectrum is given by a 
superposition
of the individual
contributions
\be
N(E) \propto F(E,Z) \cdot p \cdot E \cdot (Q-E) \cdot \sum^3_{i=1} 
\sqrt{(Q-E)^2 - m_i^2}
\mid U_{ei}^2
\mid 
\ee
where F(E,Z) is the Fermi-function, the $m_i$ are the mass eigenvalues 
and $U_{ei}^2$ are
the mixing matrix elements. The different involved $m_i$ produce kinks 
in the Kurie-plot 
where the size of the kinks is a measure
for the corresponding mixing
angle. This was discussed in connection with the now ruled out 17 keV
- \neu \cite{fra95,wie96}.\\
The search for an eV-\neu near the endpoint region is complicated 
due to several
effects \cite{hol92,ott95}. The number of electrons in an energy
interval $\Delta E$ near
the Q value scales with
\be
\label{eq:endp}n(Q-\Delta E) \propto (\frac{\Delta E}{Q})^3 
\ee
making a small Q-value advantageous, but even for tritium with the
relatively low endpoint energy of
about 18.6 keV only a fraction of $10^{-9}$ of all electrons lies in 
a region of 20 eV below the endpoint.
A further advantage of tritium is Z=1, making the distortions of the 
$\beta$ - spectrum due
to Coulomb - interactions small and allow a sufficiently accurate quantum
mechanical treatment.
Furthermore, the half-life is relatively short and the involved matrix
element is energy independent
(the decay is a superallowed transition between mirror nuclei). 
All these arguments make tritium the
favoured isotope for investigation.\\
For a precise measurement, the resolution function of the used
spectrometer has to be known quite accurately.
Additionally also the energy loss of electrons in the used
source, consisting of molecular tritium $T_2$, is important. Effects 
of molecular binding
have to be taken into account and only about 58 \% of the decays near 
the endpoint lead to 
the ground state of the \hed $T^+$-ion, making a detailed treatment of 
final states
necessary.
A compilation of the obtained limits within the last years is given in Table 2. 
As can be seen, most \exps end up with negative $m_{\nu}^2$ fit values,
which need
not to have a common explanation. 
\begin{figure}
\begin{center}
\epsfig{file=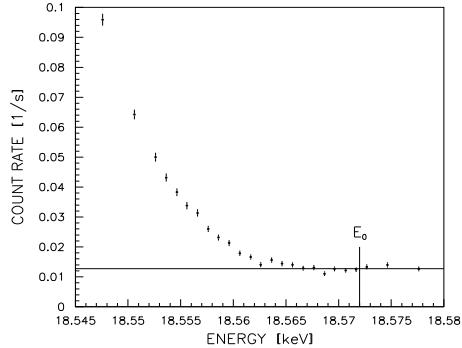,width=7cm,height=5cm}
\caption{Mainz 1998 raw data prior to publication (private communication by 
Ch. Weinheimer). The signal/background ratio is increased by a factor 
of 10 in comparison with the
1994 data. $E_0$ corresponds to the center of mass of the rotation-vibration 
excitations of the
molecular ground state of the daughter ion $^3HeT^+$.}
\label{pic:mainz}
\end{center}
\end{figure}
For a detailed discussion of the \exps see \cite{hol92,ott95}.
While until
1990 mostly magnetic spectrometers were used
for the measurements, the new \exps in Mainz and Troitzk use electrostatic retarding
spectrometers \cite{lob85,pic92}. Fig. \ref{pic:mainz} shows the
present electron spectrum near the
endpoint as obtained with the Mainz spectrometer. 
The negative $m_{\nu}^2$ values for a larger interval below the endpoint are understood for both
\expsp While in the Troitzk \expek using a gaseous $T_2$ source, the energy
loss of trapped
electrons in the spectrometer was underestimated, for 
the Mainz experiment, using a thin film of $T_2$, roughening transitions in the film seem to be
the reason. More recently, the Troitzk \expe observed excess counts in the
region of interest,
which can be attributed to a monoenergetic line short below the endpoint. This is currently under
study in the
Mainz \expe which after some upgrades might explore
a \bnel mass region down to 2 eV. \\
A complementary result would be the measurement of $\beta$-decay in \rehsa .
Because of its endpoint
energy of only 2.6 keV, according to
eq.(\ref{eq:endp}) it allows
a high statistics search near the endpoint. A cryogenic bolometer in form of a Re-foil together with
a NTD-germanium thermistor
readout has been successfully
constructed and a measurement of the $\beta$-spectrum above 100 eV was obtained
(Fig. \ref{pic:re187sp}) \cite{swi97}. Because this \expe measures the total released energy
reduced by the \neu rest mass, energy loss and final state effects are not important.
\\
CPT-invariance assures that \mbe = \me. A direct measurement of \me as proposed by
\cite{deR81} is the internal bremsstrahlungs - spectrum in EC-processes
\be
Z + e^- \ra (Z-1) + \nel + \gamma
\ee 
The most convenient isotope is \hohds and the limit obtained is \cite{spr87}
\be
\me <225 eV \quad (95 \% CL)
\ee
This is rather weak in comparison with beta decay. Astrophysical limits on \me will be
discussed in chapter \ref{cha7}.
\begin{table}
\begin{center}
\label{tab:tritiumlim}
\caption{Compilation of existing $\beta$-decay results of tritium and the corresponding 
\bnel mass limit.}
\begin{tabular}{ccc}
\hline
\expe & $m^2_{\bar{\nu}_e} (eV^2)$ & $m_{\bar{\nu}_e}$ (eV) \\
\hline
Tokyo (INS) & $-65 \pm 85 \pm 65 $ & $<13.1$\\
Los Alamos (LANL) & $-147 \pm 68 \pm 41$ & $<9.3$ \\
Z\"urich & $-24 \pm 48 \pm 61$ & $< 11.7$ \\
Livermore (LLNL) & $-130 \pm 20 \pm 15$ & $<7.0$ \\
Mainz & $ -22 \pm 17 \pm 14$ & $<5.6$ \\
Troitzk & $1.5 \pm 5.9 \pm 3.6$ & $<3.9$ \\
\hline
\end{tabular}
\end{center}
\end{table}

\subsection{Pion decay}
The easiest way to obtain limits on \mmu is given by the two-body decay of
the $\pi^+$.
For pion decay at rest the \neu mass is determined by
\be
\mmu^2 = m_{\pi^+}^2 +  m_{\mu^+}^2 - 2 m_{\pi^+} (p_{\mu^+}^2 +
m_{\mu^+}^2)^{(1/2)}
\ee
Therefore a precise measurement of the muon momentum $p_{\mu}$ and
knowledge of $m_{\mu}$
and
$m_{\pi}$ is required. 
These measurements were done at the PSI resulting in a limit of \cite{ass96}
\be
\mmu^2 = (-0.016 \pm 0.023) MeV^2 \quad \ra \quad \mmu < 170 keV (90
\%CL)
\ee
where the largest uncertainty comes from the pion mass. 
Investigations of pionic atoms
lead to two values of $m_\pi = 139.56782 \pm 0.00037$ MeV and $m_\pi =
139.56995 \pm 0.00035$ MeV respectively \cite{jec94}, but a recent independent
measurement supports
the higher value by measuring  $m_\pi = 139.57071 \pm 0.00053$ MeV \cite{len98}.

\subsection{Tau-decays}
\begin{figure}
\begin{center}  
\epsfig{file=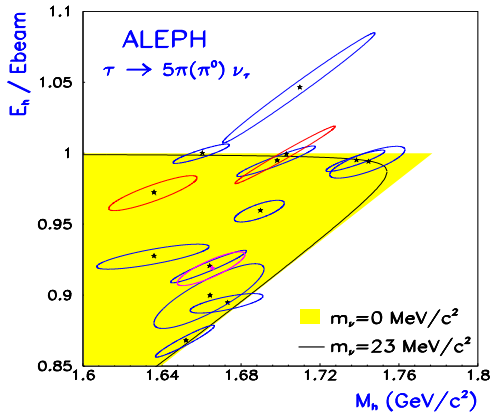,width=7cm,height=7cm}
\caption{2-dimensional plot of the hadronic energy
versus the invariant mass of the 5(6) $\pi$-system. The error ellipses are
positively correlated, because
both the hadronic mass and the hadronic energy are determined from the 
momenta of the particles
composing the hadronic system (from \protect \cite{bar98}).} 
\label{pic:ntau2d}
\end{center}
\end{figure}

Before discussing the mass of \ntau it should be mentioned that the
direct detection of \ntau via CC reactions still
has not been observed and all
evidences are indirect. The goal of E872 (DONUT) at Fermilab is to detect
exactly this reaction. With their presently accumulated data
of $4.55 \cdot 10^{17}$
protons on target, about 60 \ntau CC events should be observed.
The present knowledge of the mass of \ntau stems from measurements with
ARGUS (DORIS II) \cite{alb92}, CLEO(CESR)\cite{cin93}, OPAL \cite{ale96},
DELPHI \cite{pas97} and ALEPH
\cite{bar98} (LEP).
Practically all \exps use the $\tau$-decay into five charged pions
\be
\tau \ra \ntau + 5\pi^{\pm} (\pi^0)
\ee
with a branching ratio of BR = ($9.7 \pm 0.7) \cdot 10^{-4}$. To increase the
statistics CLEO, OPAL, DELPHI
and ALEPH
extended their search by including the 3 $\pi$ decay mode. But even with the 
disfavoured statistics,
the 5 prong-decay is much more sensitive, because the mass of the 
hadronic system peaks at about 1.6 
GeV, while the 3-prong system is dominated by the $a_1$ resonance at 
1.23 GeV. While ARGUS obtained their limit by investigating the invariant mass of the 
5 $\pi$-system, ALEPH, CLEO and OPAL 
performed a two-dimensional analysis by including the energy of 
the hadronic system
(Fig. \ref{pic:ntau2d}). A finite \neu mass leads to a distortion of 
the edge of the triangle. A
compilation of the resulting limits is given in Table 3 with the most 
stringent one given by ALEPH \cite{bar98}
\be
m_{\ntau} < 18.2 MeV (95 \% CL)
\ee
\begin{table}
\begin{center}
\label{tab:ntaulim} 
\caption{Comparision of \ntau mass limits (95 \% CL) as measured by various \expsp Numbers with $^{\ast}$ include also events from 3
$\pi$-decay.}
\begin{tabular}{ccc}
\hline
\expe & number of events & $m_{\ntau}$ limit (MeV)\\
\hline
ARGUS & 20 & 31\\
CLEO & 266 & 30\\
OPAL & 2514$^{\ast}$ + 5 & 27.6\\
DELPHI & 6534$^{\ast}$ & 27\\
ALEPH & 2939$^{\ast}$ + 41 & 18.2\\
\hline
\end{tabular}
\end{center}
\end{table}

Plans for a future charm-factory and B-factories might allow
to explore \mtau down to 1-5 MeV.\\
Independent bounds on a possible \ntau mass in the MeV region arise from
primordial
nucleosynthesis in the early universe. Basically, three effects influence
the detailed
predictions of the abundance of light elements \cite{gyu95}. An
unstable \ntau or its daughters would contribute to the energy density and
therefore influence the Hubble-expansion. Moreover, if they decay
radiatively
or into
$e^+e^-$ pairs, they would lower the baryon/photon ratio. A decay into
final states containing \nel or \bnel would influence the neutron fraction
and therefore the \hev abundance. Recent analysis seems to rule out Dirac
masses larger than 0.3 MeV and \majo masses larger than 0.95 MeV at 95 \%
CL for \ntau \cite{fie97}. An independent constraint from \bb , only valid for \majo
neutrinos, is
discussed in chapter \ref{ch:cha413}.

\section{Experimental tests of the neutrino character}
\subsection{Double beta decay}
\label{ch:cha41}
\begin{figure}
\begin{center}  
\epsfig{file=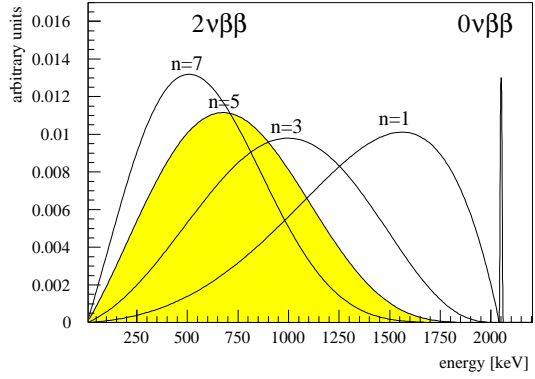,width=7cm,height=5cm}
\caption{Different spectral shapes of observable sum
energy spectra of emitted electrons in \bb . The n=1,3,7 forms correspond to
different majoron accompanied modes, n=5 (grey) is the \zbb and the \obb
results in a peak. The energy values are taken for \gess .}
\label{pic:bbshape}
\end{center}
\end{figure}

The most promising way to distinguish between Dirac and \mas is \nbb.
For extensive reviews see
\cite{doi83,doi85,mut88}.
Double beta decay was first discussed
by Goeppert-Mayer \cite{goe35} as a
process of second order Fermi theory given by 
\be
(Z,A) \ra (Z+2,A) + 2 e^- + 2 \bar{\nu_e} \quad (\zbb)
\ee
and subsequently in the form of
\be
(Z,A) \ra (Z+2,A) + 2 e^-  \quad (\obb)
\ee
first discussed by Furry \cite{fur39}. Clearly, the second process violates lepton
number conservation by 2
units and is only possible if 
\neus are massive \majo particles as discussed later. In principle V+A
currents could also
mediate neutrinoless \bb, but in gauge theories both are connected and a
positive signal would prove a finite \majo mass \cite{sch82,tak84}.
To observe double beta decay, single beta decay has to be
forbidden
energetically or at least strongly suppressed by large angular momentum
differences between the initial and
final state like in \caav. Because of nuclear pairing energies, all possible
double
beta emitters are gg-nuclei and the
transition is
dominated by $0^+ \ra 0^+$ ground-state transitions. The \zbb can be seen as two
subsequent Gamow-Teller
transitions allowing only
virtual $1^+$-states in the intermediate nucleus, because isospin 
selection rules forbid
or at least strongly suppress
any Fermi-transitions. The matrix elements for the \zbb can be written in
the form \cite{mut88}
\be
M^{2\nu}_{GT} = \sum_j \frac{\langle 0^+_f \| t\_ \sigma \|1^+_j \rangle 
\langle 1^+_j \| t\_
\sigma \| 0^+_i \rangle}{E_j + Q/2 + m_e - E_i}
\ee
and for the \obb as
\bea
M^{0\nu}_{GT} = \sum_{m,n} \langle 0^+_f \| t\_ _mt\_ _n
H(r) \sigma_m \cdot \sigma_n \|0^+_i \rangle \\
M^{0\nu}_F = \sum_{m,n} \langle 0^+_f \| t\_ _m t\_ _n H(r) \|0^+_i \rangle
(\frac{g_V}{g_A})^2
\eea
with $t\_$ as the isospin ladder operator converting a neutron into a proton, 
$\sigma$ as spin
operator,
$r
=\mid \vec{r}_m - \vec{r}_n \mid$ and
$H(r)$ the \neu potential.
In the neutrinoless case
because of the \neu potential also other intermediate states
than $1^+$ might be populated \cite{kla95}.\\ Typical energies for \bb are 
in the region of a few
MeV
distributed
among the four leptons which are therefore emitted as s-waves. The phase space 
depends on the available Q-value of the
decay as $\propto Q^5$ ($0\nu\beta\beta$ decay) and $\propto Q^{11}$ 
($2\nu\beta\beta$ decay),
numerical values can be found in \cite{boe92}.  
From the experimental point of view, the sum energy spectrum of the two
emitted electrons has a continuous
spectrum for the \zbb , while the \obb mode results in a peak at the 
position corresponding
to the Q-value of the involved
\tran (Fig.\ref{pic:bbshape}). The single electron spectrum for the two 
nucleon-mechanism is
given by
\cite{tre95}
\be
(T_1,T_2,cos \theta) = (T_1 + 1)^2(T_2 + 1)^2 \times
\delta(Q-T_1-T_2)(1-\beta_1\beta_2cos\theta) \quad (0\nu\beta\beta)
\ee
\be
(T_1,T_2,cos \theta) = (T_1 + 1)^2(T_2 + 1)^2 \times
(Q-T_1-T_2)^5(1-\beta_1\beta_2cos\theta) \quad (2\nu\beta\beta)
\ee
where $T_1, T_2$ are the kinetic energies in units of the electron mass,
$\beta_{1,2}$ is the velocity and $\theta$ the angle between the two electrons.
Some favourite isotopes are given in Table 4.
\paragraph{Experimental considerations}
A rough estimate of the expected half-lives for the \zbb mode results 
in the order of $10^{20}$
years. Therefore it is an extremely rare process making low-level counting 
techniques necessary.
To obtain reasonable
chances for detection, isotopes with large phase space factors (high Q-value) 
and large matrix
elements should be used. Also it
should be possible to use a significant amount of source material, which is 
improved by second
generation \bb \exps
using isotopical enriched materials. One of the main concerns is possible 
background. Background
sources are normally cosmic ray muons,
man-made activities like \cshsd , the natural decay chains of U and Th,
cosmogenic produced unstable isotopes within the detector components, \rnzhz 
and \kav.
The cosmic
ray muons can be shielded by going underground,
the natural decay chains of U and Th are reduced by intensive selection of 
only very clean materials
used in the different
detector components, which is also valid for \kav , and by using a 
lead shield. To avoid
cosmogenics, the exposure of all
detector components to cosmic
rays should be minimized. This is important for semiconductor devices. \rnzhz can 
be reduced by
working in an
air-free environment, which can be done by using pure nitrogen. For more details on
low-level counting techniques see \cite{heu95}.\\
The \exps focusing on electron detection can be either active or passive.  
Active detectors have the
advantage that source and detector are identical as in the case of \gess,
but only
measure the sum energy of both electrons. On the other hand passive
detectors allow more
information like measuring energy and tracks of both electrons seperately, 
but usually have
smaller source strength. 
Under the assumption of a flat background in the peak region, the sensitivity for
the 0$\nu\beta\beta$
half-life limit can be estimated from experimental
quantities to be 
\be
\ton \propto a \sqrt{\frac{M \cdot t}{B \cdot \Delta E}}
\ee
where $a$ is the isotopical abundance, $M$ the used mass, $t$ the 
measuring time, $\Delta E$ the
energy resolution at the
peak position and $B$ the background index normally given in counts/year/kg/keV.
Some experiments will be described in a little more detail.\\
{\it Semiconductor \exps}: In this type of \expe , first done by Fiorini et
al. \cite{fio67}, 
source and
detector are the
same, the isotope
under investigation is \gess .
The big advantage is the excellent energy resolution
(typically about 5 keV at 2 MeV). However, the technique only
allows the measurement of the sum energy of the two
electrons. A big step forward was done by using enriched germanium 
(natural abundance of \gess : 7.8 \%).
The 
Heidelberg-Moscow \expe \cite{gue97} in the Gran Sasso Laboratory is using 
11 kg of Ge
enriched to 86 \% in form of five HP-detectors.
A background as low as 0.2 counts/year/kg/keV at the peak position has 
been achieved. To improve
further on background
reduction, a pulse shape analysis system was developed to distinguish 
between single site
events (like double beta decay) and
multiple site events (like multiple Compton scattering) which seems to 
improve $B$ by another
factor of five.
The IGEX collaboration is using about 6 kg in form of enriched detectors \cite{avi97}.\\
Moreover, there is always the possibility to deposit a \bb emitter
near a
semiconductor detector
to study the decay, but
then only \trans to excited states can be observed by detecting the 
corresponding gamma rays.\\
{\it Scintillator \exps}: Some \bb isotopes can be used as part of 
scintillators. Experiments were done with
\caav 
in Form of CaF$_2$ \cite{you91} and \cdhsz in Form of CdWO$_4$
\cite{geo95}.\\ 
{\it Cryogenic detectors}: A technique which might become more important 
in the future can be bolometers running
at very low
temperature. Such detectors normally have a very good energy resolution. At
present only one such \expe is
running as a
10 mK bolometer using twenty 334g TeO$_2$ crystals to search for the \tehd
decay \cite{ale97}.\\
{\it Ionisation \exps}: These passive \exps are mostly built in form of 
TPCs where the emitter is either the
filling gas
or is included in thin foils. The advantage is that energy
measurements and tracking of the two electrons is possible. Moreover,
disadvantages are the energy resolution and the limited source strength 
by using thin foils. 
An \expe using a TPC with an active volume of 180 l filled
with Xe (enriched to
62.5 \% in \xehsd which corresponds to 3.3 kg) under a pressure of 5 atm 
is done in the Gotthard-tunnel
\cite{far97}. A TPC at UC Irvine was used to study \seza , \moeh and \ndhf .  
A combination of drift chambers, plastic scintillators and NaI-detectors 
is used in the ELEGANT V detector,
investigating samples of the order of 100 g enriched in \moeh and 
\cdhsz \cite{eji97}. Enriched foils of
\moeh, \seza , \cdhsz and \zsn are also used by the NEMO-2 \expe
\cite{arn95}.\\
{\it Geochemical \exps}: An alternative approach relies on the detection 
of the daughter nucleus.
The geochemical method is using very
old
ores, which have accumulated a significant amount of daughter nuclei. 
Clearly the advantage of such \exps is the
long
exposure time of up to a billion years. However several new uncertainties
are coming into consideration
like an accurate age determination, to exclude other processes producing 
the daughter, avoid a
high initial
concentration of the daughter and to have a significant source strength.
From all these considerations, only Se and
Te-ores are usable. \seza, \tehaz and \tehd decay to inert noble gases 
($^{82}Kr,^{128,130}Xe$) and the detection
is based on isotopical
anomalies due to \bb which are measured by mass spectrometry \cite{kir86}.\\
{\it Radiochemical \exps}: This method takes advantage of the radioactive 
decay of the
daughter nuclei, allowing a shorter ''measuring'' time than geochemical \expsp They
focus on the decay $^{232}Th \ra ^{232}U$ and
$^{238}U \ra ^{238}Pu$ with Q-values of 850 keV and 1.15 MeV respectively. For 
the detection of the $^{238}U \ra
^{238}$Pu decay, the emission of a 5.5 MeV $\alpha$-particle from the  
$^{238}$Pu decay is used as a signal. Of course geo- and radiochemical methods 
are not able to
distinguish between
the different \bb
modes and are finally limited in their sensitivity by \zbb .
\paragraph{\zbb}
The predicted half-life for the \zbb is given by
\be
(\tzn)^{-1} = G^{2\nu} (M_{GT}^{2\nu})^2
\ee
where $G^{2\nu}$ corresponds to the phase space and $M_{GT}^{2\nu}$ is 
the matrix element
describing the transition.
The main uncertainties in predicting accurate life-times are given by the 
errors on the matrix elements.
A reliable knowledge 
of the matrix elements is necessary, because it influences the 
extractable \neu mass limit in the \obb as well.
In the past, it was quite common to work in the closure approximation, the
replacement of the energies of the
virtual intermediate states by an average energy, allowing the summation over all
intermediate states because $\sum \mid 1^+
\rangle \langle 1^+ \mid =1$. Therefore only the wavefunctions of
the initial and final state have to be known.
But because interference between the different contributions has to
be taken into account, all amplitudes have to be 
weighted with the correct energy and closure fails as a good description.
The present determination of the matrix
elements are done with QRPA-calculations. 
For details see \cite{mut88,vog86,gro89,suh98}.
All calculations depend on the
strength of a particle-particle force which is a free parameter and 
has to be adjusted. A
complete list of
half-life calculations for A $\geq$ 70 can be found in \cite{sta90}.\\
The first evidence for \bb was observed in geochemical \exps using
selenium and tellurium-ores \cite{kir67,kir68}. Newer
measurements can be found in \cite{kir86,lin88,ber92}. Because
of phase space arguments,
the detection of the \tehd decay has to be attributed to \zbb . 
A radiochemical detection of \bb
using $^{238}U$ with a half-life of $2.0 \pm 0.6 \cdot 10^{21}$ y 
\cite{tur92} is consistent with
theoretical predictions for \zbb . In 1987 the first direct 
laboratory detection by using \seza was reported \cite{ell87}. 
Meanwhile \zbb has been observed in several 
isotopes which
are listed in Table 4. The highest statistics is obtained by
the Heidelberg-Moscow \expe which has
observed
more than 20000 events (for comparison the first observation in 1987 only consisted of 36
events). 
\begin{table}
\begin{center}
\label{tab:znuehwz}
\caption{Compilation of observed \zbb half-lives in several
isotopes. $^{\ast}$ corresponds to geochemical measurements.}
\begin{tabular}{ccc}
\hline
Isotope & Experiment & $T_{1/2} (10^{20}y)$\\
\hline
$^{48}$Ca & Calt.-KIAE& $0.43^{+0.24}_{-0.11} \pm 0.14$\\
\hline
$^{76}$Ge & MPIK-KIAE & 17.7 $\pm 0.1 ^{+1.3}_{-1.1}$\\
$^{76}$Ge & IGEX & 11 $\pm$ 1.5 \\
\hline
$^{82}$Se & NEMO 2 & $0.83 \pm 0.10 \pm 0.07$ \\
\hline
$^{100}$Mo & ELEGANT V& $0.115^{+0.03}_{-0.02}$ \\
$^{100}$Mo & NEMO 2& 0.095 $\pm$ 0.004 $\pm$ 0.009 \\
$^{100}$Mo & UCI& $0.0682^{+0.0038}_{-0.0053} \pm$ 0.0068 \\
\hline
$^{116}Cd$ & NEMO 2 & 0.375 $\pm$ 0.035 $\pm$ 0.021 \\
$^{116}Cd$ & ELEGANT V& $0.26^{+0.09}_{-0.05}$ \\
\hline
$^{128}$Te$^{\ast}$ & Wash. Uni-Tata & 77000 $\pm$ 4000 \\
\hline
$^{150}$Nd & ITEP/INR & $0.188^{+0.066}_{-0.039} \pm 0.019$\\
$^{150}$Nd & UCI & $0.0675^{+0.0037}_{-0.0042} \pm 0.0068$\\
\hline
\end{tabular}
\end{center}
\end{table}

\begin{figure}
\begin{center}
\epsfig{file=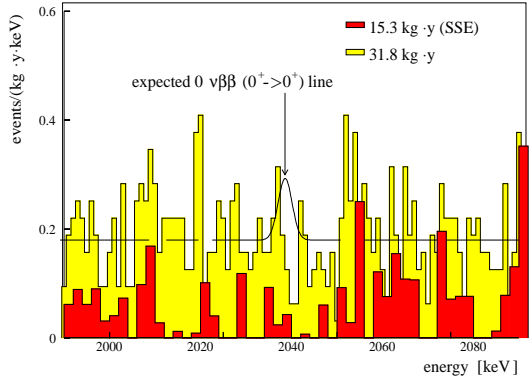,width=7cm,height=5cm}
\caption{Observed spectrum near the expected \obb peak
of the Heidelberg-Moscow collaboration. No signal can be seen. The two
different spectra correspond to measuring periods with (black) and 
without (grey) pulse
shape discrimination (with kind permission of H.V. Klapdor-Kleingrothaus).}
\label{pic:heimo}
\end{center}
\end{figure}

\paragraph{\obb}
\label{ch:cha413}
The half-life for the \obb is given by (assuming $m_{\nu} \lsim$ 1 MeV)
\be
\label{eq:t12}
(\ton)^{-1} = G^{0\nu} (M_{GT}-M_F)^2 \left( \frac{\ema}{m_e} \right) ^2
\ee
where the effective \majo \neu mass \ema is given by
\be
\label{eq:ema}\ema = \mid \sum_i U_{ei}^2 \eta_i m_i \mid
\ee
with the relative CP-phases $\eta_i = \pm 1$, $U_{ei}$ as the mixing
matrix elements and
$m_i$ as the
corresponding mass eigenvalues. The expression can be generalised if \rh currents are
included to
\begin{eqnarray*}
\label{eq:cmm}
(\ton)^{-1} & = & C_{mm} (\frac{\ema}{m_e})^2 + C_{\eta\eta} \langle 
\eta \rangle^2 + C_{\lambda
\lambda}
\langle
\lambda \rangle ^2 \\
& & + C_{m\eta}(\frac{\ema}{m_e})\langle \eta \rangle + 
C_{m\lambda}(\frac{\ema}{m_e})\langle
\lambda\rangle 
+ C_{\eta\lambda}\langle \eta \rangle \langle \lambda \rangle
\end{eqnarray*}
where the coefficients C contain the phase space factors and the matrix elements, 
\be
\langle \eta \rangle = \eta \sum_j U_{ej}V_{ej} \quad \langle \lambda 
\rangle = \lambda \sum_j
U_{ej}V_{ej}
\ee
with $V_{ej}$ as the mixing matrix elements between \rh \neusp Eq.(\ref{eq:cmm})
reduces to eq.(\ref{eq:t12}) in case
$\langle \eta \rangle,\langle
\lambda \rangle$ = 0. Also in \obb the matrix element calculations are 
done with QRPA-calculations
\cite{sta90,tom91,vog95,fae95}. The
general agreement
between the calculations done by different groups are within a factor 2-3.\\
From the experimental point, the evidence for \obb in the sum energy
spectrum of the electrons is a peak
at the position
corresponding to the
Q-value of the involved transition. The half-life limits obtained so far for several
different
isotopes are shown
in Table 5.
The best limit is coming from the Heidelberg-Moscow \expe resulting in a 
bound of \cite{bau97}
(Fig.\ref{pic:heimo}) 
\be
\label{eq:thalb}
\ton > 1.1 \cdot 10^{25} y \ra \ema < 0.47 eV \quad (90 \% CL)
\ee 
using the matrix elements of \cite{sta90}. Because in most see-saw models 
\ema corresponds
to \me \cite{lan88} (see Table 1), this bound is
much stronger than single beta decay but applies only if \neus are \majo particles. 
\begin{table}
\begin{center}
\label{tab:top10}
\caption{Compilation of \nbb half-life and mass limits of the most
investigated isotopes. The phase space
factors and Q-values are taken from \protect \cite{boe92}. $^\ddagger$ after \protect \cite{mut91},
$^\ast$ corresponds to geochemical measurement.}
\begin{tabular}{ccclc}
\hline
Decay & Q-value (keV)& (G$^{0\nu})^{-1} (y)$
& $T_{1/2}^{0\nu}$ (y) & 
$\langle m_{\nu}\rangle$ (eV)\\
\hline
$_{20}^{48}$Ca$\rightarrow _{22}^{48}$Ti& 
4271 $\pm$ 4 & 4.10E24&
$>9.5\cdot10^{21} (76\%)$& $<12.8 (76\%)^{\ddagger}$\\
$_{32}^{76}$Ge$\rightarrow _{34}^{76}$Se&
2039.6 $\pm$ 0.9 & 4.09E25 &
$>1.1\cdot10^{25} \hfill(90\%)$ & $<0.5 \hfill(90\%)$ \\
$_{34}^{82}$Se$\rightarrow _{36}^{82}$Kr&
2995 $\pm$ 6 & 9.27E24&
$>2.7\cdot10^{22} \hfill(68\%)$	&
$<5.0 \hfill(68\%)$\\
$_{~42}^{100}$Mo$\rightarrow _{~44}^{100}$Ru&
3034 $\pm$ 6 & 5.70E24 
& $>5.2\cdot10^{22} \hfill(68\%)$
&$<5.0 \hfill(68\%)$\\
$_{~48}^{116}$Cd$\rightarrow _{~50}^{116}$Sn&
2802 $\pm$ 4 & 5.28E24&
$>2.9\cdot10^{22} \hfill(90\%)$ 
& $<4.1 \hfill(90\%)$\\
$_{~52}^{128}$Te$\rightarrow _{~54}^{128}$Xe&
868 $\pm$ 4 & 1.43E26 &
$>7.7\cdot10^{24} \hfill(68\%)$	
& $< 1.1\hfill(68\%)^{\ast}$\\
$_{~52}^{130}$Te$\rightarrow _{~54}^{130}$Xe&
2533 $\pm$ 4 & 5.89E24 &
$>5.6\cdot10^{22} \hfill(90\%)$	
& $< 3.0 \hfill(90\%)$\\
$_{~54}^{136}$Xe$\rightarrow _{~56}^{136}$Ba&
2479 $\pm$ 8 & 5.52E24 
& $>4.4 \cdot10^{23} \hfill(90\%)$ 
& $<2.3 \hfill(90\%)$\\
$_{~60}^{150}$Nd$\rightarrow _{~62}^{150}$Sm&
3367.1 $\pm$ 2.2 & 1.25E24 &
$>2.1\cdot10^{21} \hfill(90\%)$
& $<4.1\hfill(90\%)$ \\
\hline
\end{tabular}
\end{center}
\end{table}

\begin{figure}
\begin{center}  
\begin{tabular}{cc}
\epsfig{file=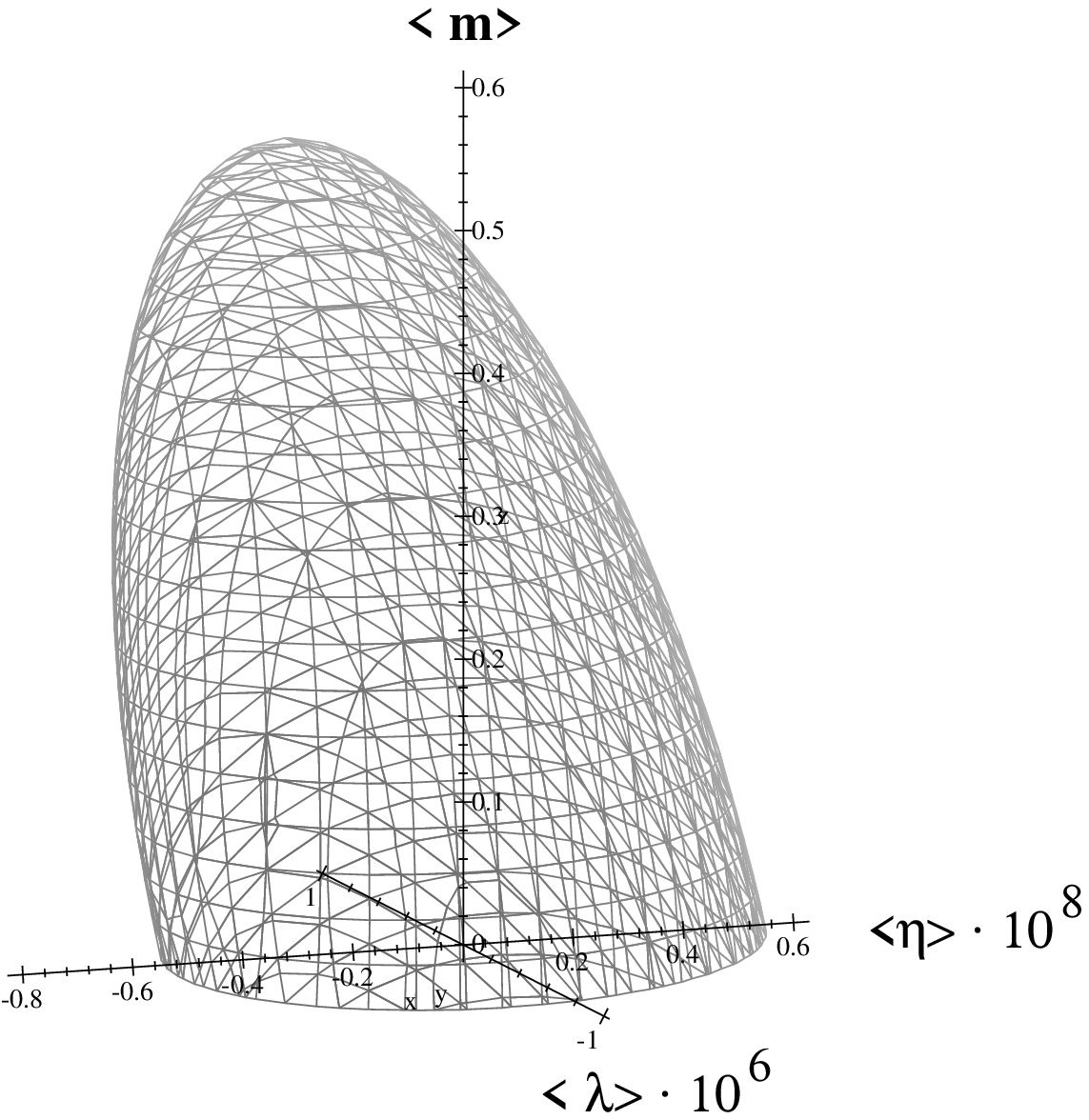,width=7cm,height=5cm} &
\epsfig{file=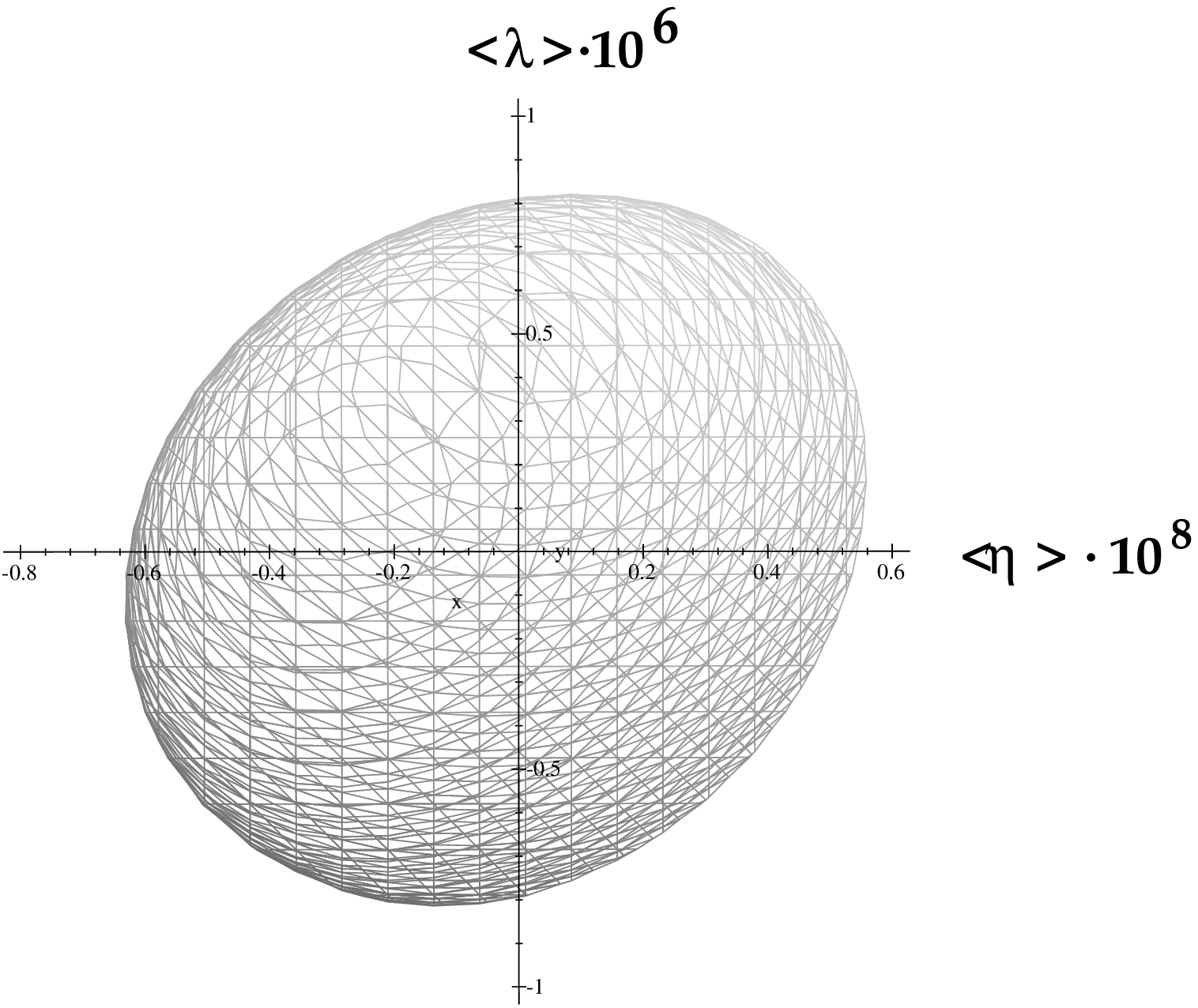,width=7cm,height=5cm}
\end{tabular}
\caption{Obtained neutrino mass \ema as a function of
the two right-handed current parameters $\langle\lambda\rangle, 
\langle \eta \rangle$. On
the right
the projection on the
$\langle\lambda\rangle,\langle\eta \rangle$ plane is shown.}
\label{pic:ellipse}
\end{center}
\end{figure}

\begin{figure}
\begin{center}
\epsfig{file=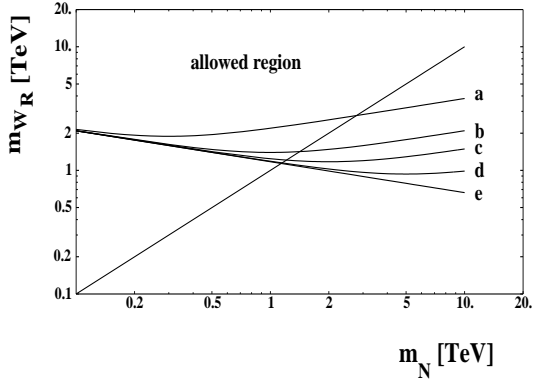,width=7cm,height=5cm}
\caption{Bound on the mass of a right-handed W as a function of the
heavy \neu mass $m_N$. under the
assumption of different masses of the double charged higgs $\Delta^{--}$. 
Shown are the regions
excluded by \bb (lower left) and  
from vacuum stability (lower right). The curves a-e correspond to masses 
of 0.3 TeV, 1 TeV, 2 TeV,
5 TeV and
infinite for the $\Delta^{--}$ (from  \protect \cite{hir96}).}
\label{pic:bbhiggs}
\end{center}
\end{figure}

Allowing also \rh currents to contribute, \ema is fixed by an ellipsoid which is shown
in Fig. \ref{pic:ellipse}. As can be seen, the largest mass allowed occurs
for $\langle \lambda\rangle, \langle
\eta \rangle
\neq 0$. In this case the half-life of eq. \ref{eq:thalb} corresponds to 
\bea
\ema < 0.56 eV\\
\langle \eta \rangle < 6.5 \cdot 10^{-9}\\
\langle \lambda \rangle < 8.2 \cdot 10^{-7}
\eea
The limit also allows a bound on a possible \rh $W_R$ which is shown 
in Fig. \ref{pic:bbhiggs}.
Together with vacuum stability arguments a mass for $W_R$ lower than 
about 1 TeV can be excluded.
The influence of
double charged Higgses,
which also can contribute to \nbb is shown as well \cite{hir96}. 
Limits on other interesting
quantities like
the R-parity violating SUSY parameter $\lambda'_{111}$ \cite{hir96a} and leptoquarks
\cite{hir96b} can be derived.
From the point of \rh currents the investigation of the \tran to the first 
excited state is important,
because the mass term here vanishes in first order. The phase space for this transition is
smaller, but the de-excitation photon might allow a good experimental signal. 
For a compilation of
existing
bounds on \trans to excited states see \cite{bar96a}. As long as no 
signal is seen, bounds from
ground
state
\trans are
much more stringent on \rh parameters.\\
Eq.(\ref{eq:ema}) has to be modified in case of heavy \neus ($m_{\nu}
\gsim $1 MeV). For such heavy \neus the mass can no longer be neglected in the
\neu propagator resulting in an A-dependent
contribution
\be
\ema =  \mid \sum_{i=1,light}^N U^2_{ei} m_i + \sum_{h=1,heavy}^M F (m_h,A) 
U^2_{eh} m_h  \mid
\ee
By comparing these limits for isotopes with different atomic mass, 
interesting limits
on the mixing angles and \ntau parameters for an MeV \ntau 
can be obtained \cite{hal83,zub97}.\\
A complete new class of decays emerges in connection with majoron emission in double beta
decay \cite{doi88}. 
The majoron $\chi$ is the Goldstone-boson of a spontaneous breaking of a
global lepton-number
symmetry. Depending on
its transformation properties under weak isospin, singlet \cite{chi80}, doublet \cite{san88}
and triplet
\cite{gel81} models exist. The
triplet and
pure doublet model are excluded by the measurements of the Z-width at LEP, 
because they would contribute 2
(triplet) or 0.5 (doublet) \neu flavours. Several new majoron-models evolved during the last
years
\cite{bur94,hir96c}.
In consequence different spectral shapes for the sum electron spectrum are predicted 
which can be
written as
\be
\frac{dN}{dE} \propto (Q-E)^n \cdot F(E,Z)
\ee
where F(E,Z) is the Fermi-function and the spectral index $n$ is 1 for the 
classical majoron, n=3 for
lepton number carrying majorons, n=5
for \zbb and n=7 for several other majoron models. A different shape is obtained 
in the vector majoron
picture of Carone \cite{car93}. It should be noted that supersymmetric Zino-exchange allows the
emission of two majorons, which also results in n=3, but a possible bound on a 
Zino-mass is less stringent
than
direct accelerator \exps \cite{moh88}. In the n=1 model the effective
\neu-majoron coupling
$\langle g_{\nu\chi}
\rangle $ can be deduced from
\be
(T_{1/2}^{0\nu\chi})^{-1} = \mid M_{GT} - M_F\mid ^2 G^{0\nu\chi} \mid 
\langle g_{\nu\chi} \rangle \mid^2
\ee
where $\langle g_{\nu\chi} \rangle $ is given by
\be 
\langle g_{\nu\chi} \rangle  = \sum_{i,j} g_{\nu\chi} U_{ei}U{ej} 
\ee
Present half-life limits for this decay (n=1) and the deduced coupling
constants are given in Table 6.
A first half-life limit for the n=3 mode was given in \cite{zub92}, a evaluation for 
\moeh is given in
\cite{tan93}. A more recent extensive study of all modes can be
found in \cite{hir96c}. Limits obtained by the Heidelberg-Moscow \expe with a
statistical significance of 4.84
kg$\cdot$y are \cite{gue96}
\bea
T_{1/2}^{0\nu\chi} > 5.85 \cdot 10^{21} y \quad \mbox{(n=3)} \\
T_{1/2}^{0\nu\chi} > 6.64 \cdot 10^{21} y \quad \mbox{(n=7)} 
\eea
\begin{table}
\begin{center}
\label{tab:majomode}
\caption{Compilation of limits on half-lives and effective majoron-neutrino
couplings from different isotopes. $^{\ast}$ corresponds to a geochemical
measurement, which has no power to discriminate between different decay modes.}
\begin{tabular}{cccc}
\hline
Isotope & Experiment & $T_{1/2} (10^{21}y)$ & $\langle g_{\nu \chi} \rangle
(10^{-4})$ \\
\hline
$^{48}$Ca & ITEP & 0.72 (90 \%) & 5.3\\
\hline
$^{76}$Ge & MPIK-KIAE & 7.91 (90 \%) & 2.3 \\
$^{76}$Ge & ITEP & 10 (68 \%) & 2.2 \\
$^{76}$Ge & UCSB-LBL & 1.4 (90 \%) & 5.8 \\
$^{76}$Ge & PNL-USC & 6.0 & 2.8\\
$^{76}$Ge & Cal.-PSI-Neu & 1.0 (90 \%) & 6.9\\
\hline
$^{82}$Se & NEMO 2 & 2.4 (90 \%) & 1.4\\
\hline
$^{100}$Mo & ELEGANT V& 5.4 (68 \%) & 0.7\\
$^{100}$Mo & NEMO 2 & 0.5 (90 \%) & 2.3\\
$^{100}$Mo & UCI & 0.3 (90 \%) & 3 \\
\hline
$^{128}$Te$^{\ast}$ & Wash. Uni-Tata & 7700 & 0.3\\
\hline
$^{136}$Xe & Cal.-PSI-Neu & 14 (90 \%) & 1.5\\
\hline
$^{150}$Nd & UCI & 0.28 (90 \%) & 1\\
\hline
\end{tabular}
\end{center}
\end{table}

Additionally the $\beta^+\beta^+$-decay in combination with EC can be 
observed via the following
decay modes
\bea
(Z,A) \ra (Z-2,A) + 2 e^+ + (2 \nel) \\
e_B^- + (Z,A) \ra (Z-2,A) + e^+ + (2 \nel)  \\ 
2 e_B^- + (Z,A) \ra (Z-2,A) + 2 \nel \\
\label{eq:ecexc}2 e_B^- + (Z,A) \ra (Z-2,A)^{\ast} \ra (Z- 2,A) + \gamma +
2 \mbox{X-rays} 
\eea
$\beta^+\beta^+$ is always accompanied by EC/EC or $\beta^+$/EC-decay.
Because of the Coulomb-barrier
and the reduction of the Q-value by 4 $m_ec^2$, the rate for
$\beta^+\beta^+$ is small and energetically only
possible for seven nuclides. Predicted half-lives for $\beta^+\beta^+$ are of 
the order $10^{26}$ y while
for 
$\beta^+$/EC this can be reduced by orders of magnitude down to $10^{22-23}$ y 
making an experimental detection
more realistic. The experimental signature of the decay modes is rather 
clear because of the two or
four 511 keV photons. The last mode (eq.\ref{eq:ecexc}) to an excited state is giving a
characteristic gamma associated
with X-ray
emission. Half-lives obtained with $^{106}$Cd and $^{96}Ru$ are of the order
$10^{18}$ y \cite{nor84,bar96a}. Extracted \neu mass limits are orders
of magnitude worse than the \obb
limits, but if there is any positive observation of the \obb mode, the $\beta^+$/EC-mode
can be used to distinguish whether this is dominated by
the \neu mass mechanism or \rh currents \cite{hir94}. 
\paragraph{Future}
Several upgrades are planned to improve the existing half-life limits. Because of the enormous
source strength after 
additional years of running the dominant project will still be the Heidelberg-Moscow \expe
probing \neu masses
down to 0.2 eV. A new \expe to improve the sensitivity on \caav is ELEGANT VI, 
using 25 modules of $CaF_2$ with
a total amount of 31 g of \caav within a CsI detector \cite{eji97}. 
A different approach might be
the use of
$CaF_2(Eu)$ as a cryogenic bolometer and to measure simultaneously the scintillation light
\cite{ale98}. \caav is interesting because it can
be treated with nuclear shell model calculations. The building up of NEMO-3, 
which should start operation in
1999, will allow to use up to 10 kg of material in form of foils
for several isotopes like \moeh \cite{bar97b}.
Even more ambitious would be the usage of
large amounts of materials (in the order of several hundred kg to tons)
like enriched \xehsd added to
scintillators
\cite{rag94}, 750 kg $TeO_2$ in form of cryogenic bolometers (CUORE) \cite{fio98} or a
huge cryostat containing several hundred detectors of
enriched \gess with a total mass of 1 ton (GENIUS) \cite{kla98}.  
Further, ideas to use a large amount of \xehsd and
detect the created daughter $^{136}$Ba with atomic traps and resonance ionisation
spectroscopy exist. This will allow no information on the decay mode and will be
dominated by \zbb \cite{moe91,mit96}.

\subsection{Magnetic moment of the neutrino}
Another possibility to check the \neu character is the search for its \mamo .
In the present standard model both types of \neus have no \mamo because \neus 
are massless and a
\mamo would require a coupling of a left-handed 
state with a right-handed one which is absent. A simple extension by including 
right-handed singlets
allows
for Dirac-masses. In this case, it can be shown that \neus can get a \mamo
due to loop diagrams which is proportional to
their mass and is given by \cite{lee77,mar77}
\be
\munu = \frac{3 G_F e}{8 \sqrt{2} \pi^2} m_{\nu} = 3.2 \cdot 10^{-19} (\frac{m_\nu}{eV}) \mub
\ee
In case of \neu masses in the eV-range, this is far to small to be observed
and to have any significant
effects in
astrophysics. Nevertheless
there exist GUT-models, which are able to increase the \mamo without increasing the mass
\cite{pal92}. However
\majo \neus still have a vanishing static moment because of CPT-invariance.
The existence of diagonal terms in the \mamo matrix would therefore prove 
the
Dirac-character of \neus.
Non-diagonal terms in the moment matrix are possible for both types of \neus
allowing transition moments of the form \nel - $\bar{\nu}_\mu$.\\ 
Limits on magnetic moments arise from \nel $e$ - scattering \exps and 
astrophysical considerations. The 
differential cross section for \nel $e$ -
scattering in presence of a \mamo is given by
\bea
\frac{d \sigma}{dT} = \frac{G_F^2 m_e}{2 \pi}
[(g_V + x+g_A)^2 +
(g_V + x- g_A)^2 (1-\frac{T}{E_\nu})^2 \\
+ (g_A^2 -
(x+g_V)^2)\frac{m_e T}{E_\nu^2}] + \frac{\pi \alpha^2 \munu^2}{m_e^2}
\frac{1-T/E_\nu}{T}
\eea
where T is the kinetic energy of the recoiling electron and 
\be
g_V = 2 sin^2 \theta_W + \frac{1}{2} \quad g_A = \pm  \frac{1}{2} \quad
(+ (-) \quad \mbox{for} \quad \nel (\bar{\nel}))
\ee
and $x$ denotes the \neu form factor related to its square charge radius $\langle r^2 \rangle$
\be 
x=\frac{2 m_W^2 }{3} \langle r^2 \rangle sin^2 \theta_W \quad x \ra -x \quad for 
\quad \bar{\nel}
\ee
The contribution associated with the charge radius can be neglected in the case $\mu_\nu
\stackrel{>}{\sim} 10^{-11} \mub$. 
As can be seen, the largest effect of a \mamo can be observed in the low
energy region, and because of
destructive interference
of the electroweak terms, searches with antineutrinos would be preferred. The obvious sources
are therefore nuclear
reactors. Experiments done so far result in a bound of $\munu < 1.52 \cdot 10^{-10}
\mub$ for \bnel \cite{kra90}. 
Measurements based on \nel$ e \ra \nel e$ and \nmu $e \ra \nmu e$
scattering were done at LAMPF and BNL yielding bounds for \nel and \nmu of 
$\mu_{\nu} \stackrel{<}{\sim}
10^{-9} \mub $ \cite{abe87,kim88}.\\
Astrophysical limits are somewhat more stringent but also more model dependent. 
An explanation of the \snp by spin precession
of $\nu_L$ into $\nu_R$ done by the magnetic field of the solar convection 
zone requires a \mamo of the order 
\munu $\approx 10^{-10} -10^{-11} \mub$ \cite{vol86}. Observation of 
\neus from Supernova 1987A
yield a somewhat model dependent
bound of \munu $< 10^{-12} \mub$ \cite{bar88,lat88}. Also the \neu
emissivity of globular
cluster stars done by
excessive plasmon decay
$\gamma \ra \nu \bar{\nu}$ is only consistent with observation for a \mamo of the same order
\cite{raf90}. This last bound applies to \neus lighter than 5 keV.\\
To improve the experimental situation and especially check the
region relevant for the
\snp new experiments are under construction. The most advanced is the
NUMU \expe \cite{ams97} currently installed near
the Bugey
reactor. It consists of a 1 m$^3$ TPC loaded with CF$_4$ under a pressure of 5 bar. The usage 
of a TPC will not only
allow to measure the electron energy but for the first time in such \exps also the 
scattering angle, therefore allowing the
reconstruction of the neutrino energy. The \neu energy spectrum at reactors in 
the energy region 1.5
$< E_{\nu} <$ 8 MeV is
known at the 3 \% level. 
To suppress background, the TPC is surrounded by 50 cm anti-Compton
scintillation detectors as
well as a passive shielding 
of lead and polyethylene. In case of no \mamo the expected count rate is 9.5 per 
day increasing to 13.4 per day if 
$\munu= 10^{-10} \mub$ for an energy threshold of 500 keV. The estimated 
background is 6 events per day. The expected
sensitivity level is down to $\munu = 3 \cdot 10^{-11} \mub$ . The usage 
of a low background Ge-NaI
spectrometer in a shallow depth near a reactor has
also been considered \cite{bed97}. The usage of large low-level detectors
with a low-energy threshold
of a few keV in underground laboratories is also under investigation. The reactor 
would be replaced
by a strong $\beta$-source. Calculations for a scenario of a 1-5 MCi $^{147}$Pm 
source (endpoint
energy of 234.7 keV) in combination with a 100 kg low-level NaI(Tl) detector with a 
threshold of about 2
keV
can be found in \cite{bar96}. 

\subsection{Search for heavy \majo \neus}
For the \ssm to work heavy \majo \neus N are necessary.
The required lightness of \neu masses makes a detection of the 
corresponding heavy state impossible. The mixing of a heavy \neu $m_H$ to the
light state $m_L$ is ruled by $\theta = \sqrt{m_L/m_H} \ll 1$. However
there exist theoretical models which decouple the mixing from any mass
relation \cite{jar90,tom94}. Assuming that in eq.(\ref{eq:massma})
$m_L^M \neq 0$ and that by
an internal symmetry at tree level the relation $m_L^M m_R^M = (m^D)^2$
is valid, the mixing is decoupled from the ratio $m_1/m_2$ and can be close
to one in case that $m_L^M \approx m_R^M$. Masses for light \neus vanish
at tree level and will be generated at higher orders.\\
From the experimental point of view, heavy \majo \neus can be searched for
at
accelerators. The LEP-data on the $Z^o$-width already exclude any
additional \neu lighter than 45 GeV. Searches for heavier \neus have been
done at LEP1.5. The search for \majo \neus heavier than the $Z^o$ focusses
on the N-decay channels 
\be
N \ra e^{\pm} W^{\mp} \quad and \quad N \ra \nu Z^o
\ee
which is identical to signatures looked for in searches of excited
fermions. A detailed description of pair production of heavy Dirac 
and Majorana \neus in $e^+e^-$collisions can be found in \cite{hof96}.
Pair 
production of \majo \neus would result in two like-sign charged
leptons. Furthermore, HERA
offers the chance to search for heavy \majo masses in ep-collisions
\cite{buc91}. For accumulated 200 $pb^{-1}$ a discovery limit up to 160
GeV is
possible. Also future high energy $e^+e^-$ machines allow an extended
search for heavy \neus via reactions
\be
e^+e^- \ra \nu N \quad N \ra e^{\pm} W^{\mp}, N \ra \nu Z , N
\ra \nu H
\ee
The dominant background will be $W^+W^-$production \cite{glu97}. LHC
offers searches
either in the pair-production or single \majo \neu production mode
\cite{ma89,dic91,alm97}. The
advantage of single \majo production is that it depends only linearly on
the \neu mixing. The single production channel via
\be
pp \ra e^- N X \ra e^- e^- W^+ X
\ee
offers a signal of two same sign leptons, two jets with the invariant
mass of $m_W^2$ and no missing energy. For an assumed luminosity of 10
fb$^{-1}$ the discovery potential goes up to 1.4 TeV (0.8 TeV) for an
assumed mixing of $sin \theta \approx 10^{-2} (10^{-3})$.

\section{Neutrino \oszs}
In case of massive \neus the mass eigenstates do not have to be identical with 
the flavour eigenstates,
similar
to the CKM-mixing in the quark sector. This offers the chance for \neu \oszsp Oscillations 
might be
the only chance to see effects of \nmu and \ntau in the eV mass range which is not
accessible in direct
\expsp
\subsection{General formalism}
The concept of \neu \oszs has been introduced by \cite{pon57}. The weak
eigenstates $\nu_\alpha$ are related
to the mass
eigenstates $\nu_i$ via a unitary matrix U 
\be
\nu_\alpha = \sum_i U_{\alpha i} \nu_i
\ee
which is given for Dirac \neus as
\be
U = \left( \begin{array}{ccc}
c_{12} c_{13} & s_{12} c_{13} & s_{13} e^{- i \delta}\\
- s_{12} c_{23} - c_{12}s_{23}s_{13} e^{i \delta}& c_{12}c_{23} - 
s_{12}s_{23}s_{13}e^{i \delta} &
s_{23}c_{13}\\
s_{12}s_{23}- c_{12}s_{23}s_{13} e^{i \delta} & -c_{12}s_{23} - s_{12}c_{23}s_{13}e^{i
\delta} &  c_{23}c_{13} 
\end{array} \right)
\ee
and in the \majo case as
\begin{flushleft}
\begin{equation}
U = \left( \begin{array}{ccc}
c_{12} c_{13} & s_{12} c_{13}e^{- i \delta_{12}} & s_{13} e^{- i \delta_{13}}\\
- s_{12} c_{23} e^{i \delta_{12}} - c_{12}s_{23}s_{13} e^{i(\delta_{13} + \delta_{23})}&
c_{12}c_{23} - s_{12}s_{23}s_{13}e^{i (\delta_{23}+\delta_{13}-\delta_{12})} &
s_{23}c_{13}e^{i \delta_{23}}\\
s_{12}s_{23}e^{i(\delta_{13} + \delta_{23})} - c_{12}s_{23}s_{13} e^{i(\delta_{13}+\delta_{23})}
& -c_{12}s_{23} e^{i \delta_{23}}- s_{12}c_{23}s_{13}e^{i
(\delta_{13}-\delta_{12})} &  c_{23}c_{13} 
\end{array} \right)
\end{equation}
\end{flushleft}
with $c,s = cos \theta, sin\theta$.
The quantum mechanical transition probability can be derived (assuming
relativistic \neus and CP-conservation)
as \cite{bil87} 
\be
P(\nu_\alpha \ra \nu_\beta) = \sum_i \mid U_{\beta i} \mid^2  \mid U_{\alpha i} \mid^2 + Re
\sum_{i \neq j}
U_{\beta i} U^{\ast}_{\beta j} U^{\ast}_{\alpha i} U_{\alpha j} exp (-it
\delm_{ij} /2E)
\ee
with $\delm_{ij} = \mid m_i^2 - m_j^2 \mid$.
In the simple two-flavour mixing the probability to find $\nu_\beta$ in a distance $x$ 
with respect to
a source of $\nu_\alpha$ is given by
\be
P(\nu_\alpha \ra \nu_\beta) = \sint sin^2 \frac{\pi x }{L}
\ee
giving the \osz length L in practical units as
\be
L = \frac{4 \pi E \hbar}{\delm c^3} = 2.48 (\frac{E}{MeV}) (\frac{eV^2}{\delm}) \quad m
\ee
For a more extensive review on N flavour mixing, wave-packet treatment 
and coherence considerations
see \cite{bil87,kay89,kim93}.
Terrestrial \exps are done with nuclear reactors and accelerators. The discussed \osz searches
involve the three known \neus as well as a possible sterile \neu $\nu_S$.

\subsection{Reactor \exps }
Reactor experiments are disappearance
experiments looking for \bnel $\ra \bar {\nu}_X$.
\subsubsection{Principles}
Reactors are a source of MeV \bnel due to the fission of nuclear fuel. The main isotopes
involved are
$^{235}$U,$^{238}$U,$^{239}$Pu and $^{241}$Pu. The neutrino rate per fission has been measured
\cite{sch85}
for all isotopes except $^{238}$U and is in good agreement with
theoretical calculations \cite{kla82}. 
Experiments typically 
try to measure the positron spectrum which can be deduced from the \bnel - spectrum
and either compare it directly to the theoretical predictions
or measure it at several distances from the reactor and search for spectral changes. Both
types of experiments were done in the past. The \bnel cross section is known to about 1.4 \%
\cite{dec94}. The detection reaction
is 
\be
\label{gl1}
\bnel + p \ra e^+ + n
\ee
with an energy threshold of 1.804 MeV.
The detection reaction (\ref{gl1}) is always the same, resulting in different strategies for
the detection of the
positron and the neutron. Normally coincidence techniques are used between the annihilation
photons and the neutrons 
which diffuse and
thermalise within 10-100 $\mu$s.
The main background are cosmic ray muons producing neutrons 
in the surrounding of the detector.

\subsubsection{Experimental status}
Several reactor experiments have been done in the past (see Table 7). All
these experiments
had a fiducial mass of less than 0.5 t and the distance to the reactor was never
more than 250 m.
\begin{table}
\begin{center}
\label{tab:reactorex}
\caption{List of finished reactor experiments. Given
are the power of the reactors and the distance of
the experiments with respect to the reactor.}
\begin{tabular}{ccc}
\hline
reactor & thermal power [MW] & distance [m]\\
\hline
ILL-Grenoble (F) & 57 & 8.75 \\
Bugey (F) & 2800 & 13.6, 18.3 \\
Rovno (USSR) & 1400 & 18.0,25.0 \\
Savannah River (USA) & 2300 & 18.5,23.8\\
G\"osgen (CH) & 2800 & 37.9, 45.9, 64.7 \\
Krasnojarsk (Russia) & ? & 57.0, 57.6, 231.4 \\
Bugey III (F) & 2800 & 15.0, 40.0, 95.0 \\
\hline
\end{tabular}
\end{center}\end{table}

\begin{figure}
\begin{center}
\epsfig{file=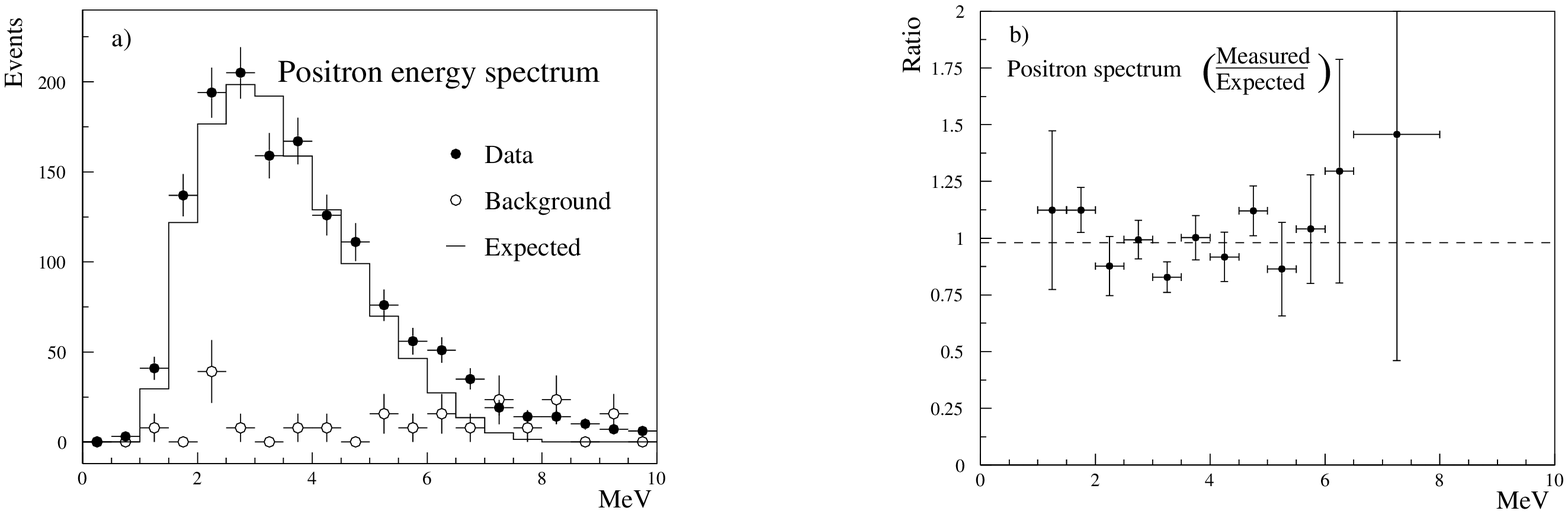,width=13cm,height=6cm}
\caption{The measured positron
spectrum of the CHOOZ-\expe (left). On the right the ratio with the
expectation is shown. No oscillation signal is visible (from \protect \cite{apo98}).}
\label{pic:choozspec}
\end{center}
\end{figure}

Two new reactor experiments producing data are
CHOOZ and Palo Verde. The
CHOOZ-experiment in France \cite{chooz}
has some advantages with respect to previous experiments. First of all the
detector is located underground with
a shielding of 300 mwe, reducing the background due to atmospheric muons by a factor of 300.
Moreover,
the detector 
is about 1030 m away
from two 4.2 GW reactors (more than a factor 4 in comparison to previous experiments)
enlarging the sensitivity to smaller \delm . In addition
the main target has about 4.8 t and is therefore much larger than those used before.
The main target consists of a specially developed
Gd-loaded scintillator. This inner detector is surrounded by an additional
detector containing 17 t of scintillator
without Gd and 90 t of scintillator as an outer veto. The signal in the inner detector is
the detection of the
annihilation photons in coincidence with n-capture on Gd, the latter producing gammas with a
total sum of 
up to 8 MeV. 
The first published positron spectrum \cite{apo98} is shown in Fig. 
\ref{pic:choozspec} and shows no
hints
for \oszsp The resulting exclusion plot is shown in Fig.\ref{pic:exclplatm}.
\\
The second experiment is the Palo Verde (former San Onofre) experiment \cite{pave} near
Phoenix, AZ (USA). It consists of 12 t liquid scintillator
also loaded with Gd. The scintillator is filled in 66 modules arranged in an 11$\times$6 array.
The coincidence of three modules serves as a signal.
The experiment is located under a shielding of 46 mwe in a distance
of about 750 (820) m to three reactors with a total power of 10.2 GW.\\
A further project plans a 1000 t liquid scintillator detector (KamLAND)
\cite{kamland}.
It is approved by the Japanese Government and will be constructed at the Kamioka site.
Having a distance of 160 km to the next reactor, it will probe \delm down to $10^{-5}
eV^2$.

\subsection{Accelerator \exps }
The second source of terrestrial \neus are high energy accelerators. Experiments
can be of either appearance or
disappearance type \cite{zub97a}. 
\subsubsection{Principles}
Accelerators typically produce \neu beams by shooting a proton beam on a fixed target.
The produced secondary pions and kaons decay and create a
\neu beam dominantly consisting of \nmu. 
The detection mechanism is via charged weak currents
\be
\nu_i N \ra i + X  \quad i= e, \mu, \tau
\ee
where N is a nucleon and X the hadronic final state.
Depending on the intended goal, the search for
\oszs therefore requires a detector
which is capable of detecting electrons, muons and $\tau$ - leptons in the final
state.

\subsubsection{Experimental status}
\begin{figure}
\begin{center}  
\epsfig{file=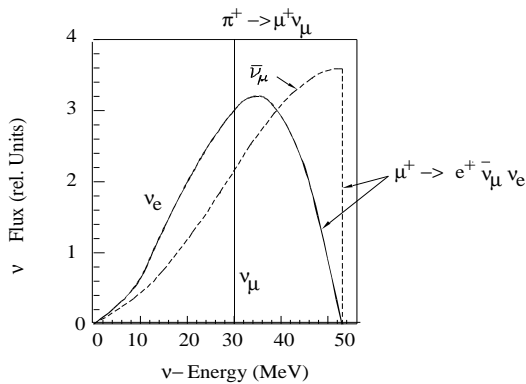,width=7cm,height=5cm}
\caption{Neutrino spectrum from $\pi$-decay. Beside
the monoenergetic line from $\pi^+$-decay at rest, the continuous spectra
of \nel and \bnmu are shown.}
\label{pic:pispek}
\end{center}
\end{figure}

\begin{figure}
\begin{center}
\begin{tabular}{cc}
\epsfig{file=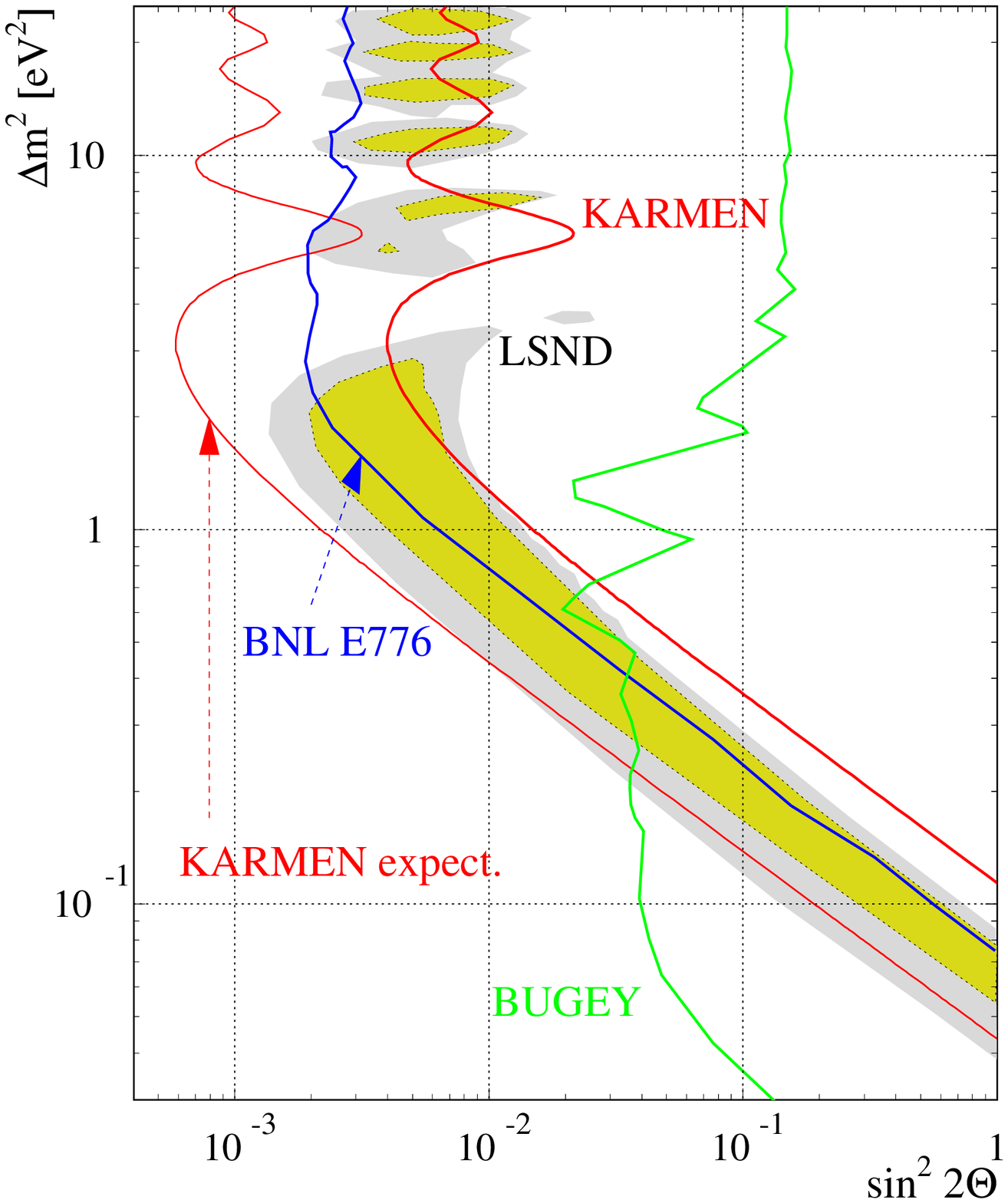,width=6cm,height=6cm} &
\epsfig{file=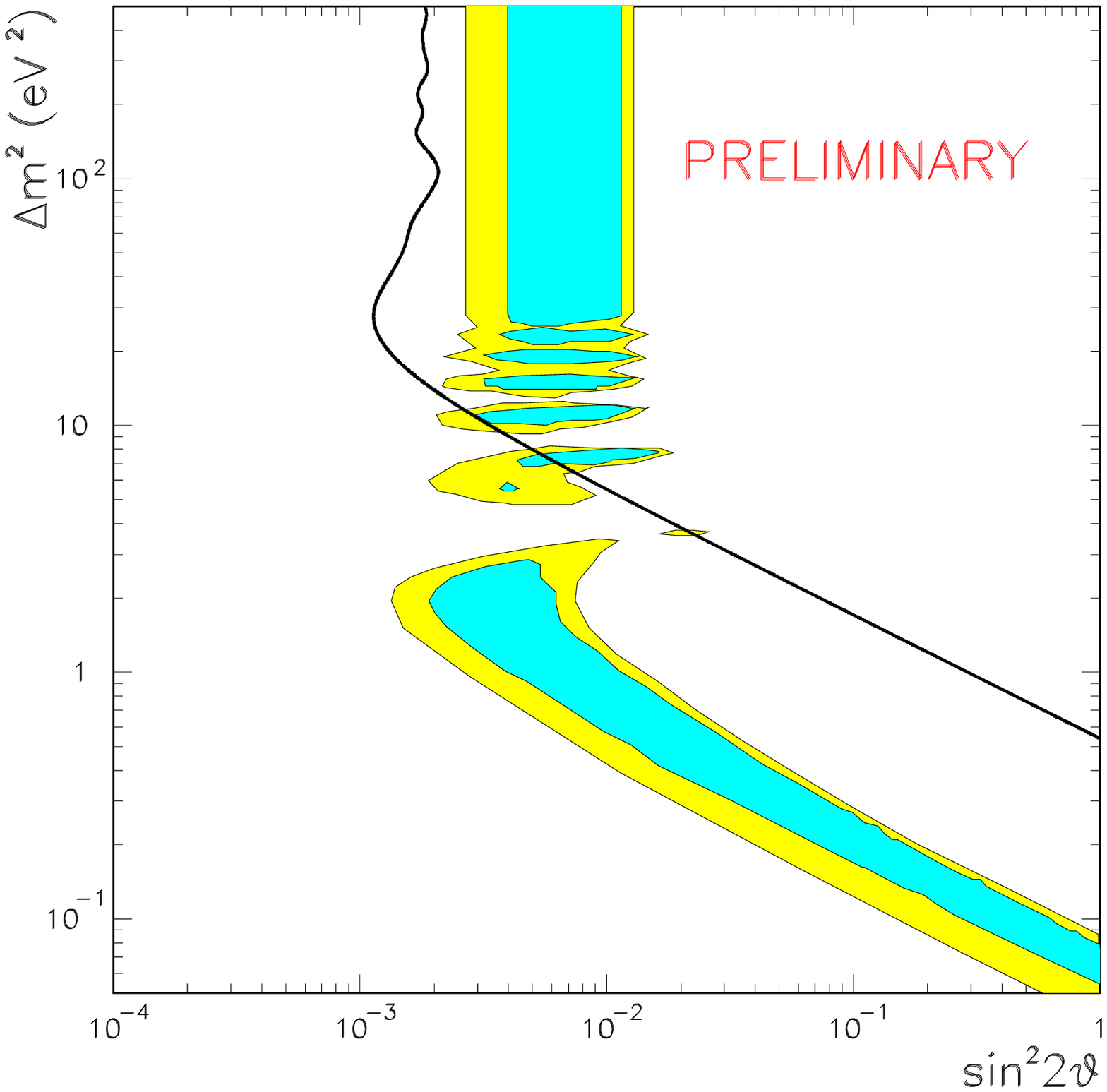,width=6cm,height=6cm}
\end{tabular}
\end{center}
\caption{\nel - \nmu parameter plot. Shown is the
region of
evidence from the \lsnd $\pi$-decay at rest analysis (grey areas) and different exclusion curves
from Bugey, KARMEN, E776 (left) and the preliminary NOMAD result (right).}   
\label{pic:lsndev} 
\end{figure}

\begin{figure}
\begin{center}  
\epsfig{file=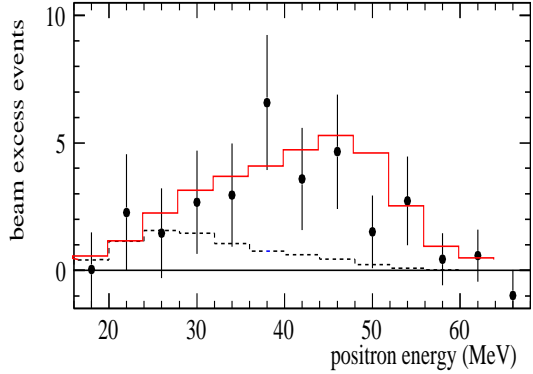,width=7cm,height=5cm}
\caption{Positron spectrum of beam-excess data. The dashed curve corresponds
to the estimated \neu background while the solid line indicates \neu \oszs for large \delm plus
the estimated background (from \protect \cite{lsnd}).}
\label{pic:lsndex}
\end{center}
\end{figure}

\paragraph{Accelerators at medium energy}
At present there are two experiments running with \neus at medium energies
($E_\nu \approx $ 30 - 50 MeV) namely KARMEN and \lsnd. Both experiments use
800 MeV proton beams on a beam dump to produce pions. The expected \neu
spectrum from pion and $\mu$-decay is shown in Fig. \ref{pic:pispek}. The
beam contamination 
of \bnel is in the order of $10^{-4}$.
The KARMEN experiment \cite{karmen} at the neutron spallation source ISIS at Rutherford
Appleton Laboratory is using 
56 t of a segmented liquid scintillator. The main
advantage of this experiment is the known time structure of the two proton pulses hitting the
beam dump 
(two pulses of 100 ns with a separation of 330 ns and a repetition rate of 50 Hz).
Because of the pulsed beam, positrons are expected within 0.5-10.5
$\mu$s after beam on target.
The signature for detection is a delayed coincidence of a positron in the 10 - 50 MeV region
together with
$\gamma$-emission from either p(n,$\gamma$)D or Gd(n,$\gamma$)Gd reactions. The first results
in 2.2 MeV photons
while the latter allows gammas up to 8 MeV. 
The limit reached so far is shown in Fig. \ref{pic:lsndev}. 
Recently KARMEN published a 2- and 3- flavour analysis of \nel -\ntau and
\nel - \nmu \oszs by comparing the energy averaged CC-cross section for
\nel interactions with expectation as well as making a detailed maximum 
likelihood analysis of the spectral shape of the electron spectrum observed from \csz
(\nel,e$^-$)\sszw$_{gs}$ reactions \cite{arm98}.
To improve the sensitivity by reducing the neutron background, a new
veto shield against atmospheric muons
was constructed which has been in operation since Feb. 1997 and is surrounding the whole
detector. 
The region which can be investigated in 2-3 years of running in the upgraded 
version is also shown
in Fig. \ref{pic:lsndev}. 
\\
The \lsnd experiment \cite{lsnddet} at LAMPF is a 167 t mineral oil based liquid
scintillation detector using
scintillation and Cerenkov light for detection. It consists of an approximately
cylindrical tank
8.3 m long and 5.7 m in diameter. The experiment is about 30 m away from a copper beam stop
under an
angle of 12$^o$ with respect to the proton beam. For the \osz search in the channel 
\bnmu -\bnel a signature of a positron within the energy range 36 MeV $<E_e<$ 60 MeV together
with an
in time and spatial correlated 2.2 MeV photon from p(n,$\gamma$)D is required. The analysis
(Fig. \ref{pic:lsndex})
\cite{lsnd} 
ends up in evidence for
\oszs in the region shown in Fig.\ref{pic:lsndev}.
Recently \lsnd  published their \nel - \nmu analysis for pion decays in flight
by looking for isolated electrons in the region 60 MeV $< E_e <$ 200 MeV coming from
\csz (\nel,e$^-$)\sszw$_{gs}$ reactions \cite{lsnd1}, which is in
agreement with the former
evidence from pion decay at rest.
Also \lsnd continues with data acquisition.
\newline
An increase in sensitivity in the \nmu - \nel \osz channel can be reached in the future
if there is a possibility for \neu physics at the planned European Spallation Source (ESS)
or the National Spallation Neutron Source (NSNS) at Oak Ridge which might have a 1 GeV proton
beam in 2004.
The Fermilab
8 GeV proton booster offers the chance for a \neu experiment as well which could start
data taking in 2001. It would use part of the \lsnd equipment and consist of 600 t mineral oil
contained and be located 500 m away from the \neu source (MiniBooNE)\cite{chu97}.
An extension using a second detector at 1000m is possible (BooNE).
At CERN
the PS \neu beam could be revived with an average energy of 1.5
GeV and two detector locations at 128 m and 850 m as it was used by the former CDHS
\cite{dyd84} and
CHARM-experiment \cite{all88}. By measuring the \nel / \nmu ratio the
complete
\lsnd region can be investigated \cite{xxx97}.
\paragraph{Accelerators at high energy}
High energy accelerators provide \neu beams with an average energy in the GeV region.
With respect to high energy experiments at present especially CHORUS and
NOMAD at CERN will provide new limits. They are running at the CERN wide band neutrino beam
with an average energy
of around 25 GeV,
produced by 450 GeV protons accelerated in the SPS and then hitting a beryllium beam dump.
To reduce the uncertainties in the \neu flux predictions, the NA56 - \expe
measured the resulting pion and kaon spectra \cite{amb98}.  
The experiments are 823 m (CHORUS) and 835 m (NOMAD) away from the beam
dump and designed to
improve the existing limits on \nmu - \ntau \oszs by an order of magnitude. 
The beam contamination of prompt \ntau from $D_s^{\pm}$-decays is of the
order 2-5 $\cdot 10^{-6}$ \cite{van96,gon96}.
Both experiments differ in their detection technique. While CHORUS relies on 
seeing the track of the $\tau$ - lepton and the associated decay vertex with the
kink because of the $\tau$-decay,
NOMAD relies on kinematical criteria.
\newline
 The CHORUS experiment \cite{chorusdet} uses
emulsions with a total mass of 800 kg segmented into 4 stacks, 8 sectors each as a main
target. To determine
the vertex within the emulsion as accurate as possible, systems of thin emulsion sheets
and
scintillating fibre trackers
are used. Behind the tracking devices follows a hexagonal air core magnet for momentum
determination of hadronic tracks, 
an electromagnetic lead-scintillating fibre calorimeter with an energy 
resolution of $\Delta E / E = 13 \%
/ \sqrt{E}$ for electrons as well
as a muon spectrometer.
A $\tau$ - lepton created in the emulsion by a charged current reaction is producing a track
of 
a few hundred $\mu$m.
After the running period the emulsions are scanned with automatic microscopes coupled to CCDs.
The \expe searches for the muonic and hadronic decay modes of the
$\tau$ and took data from 1994 to 1997. 
The present limit (Fig.\ref{pic:tosca}) provided by CHORUS for the \nmu -
\ntau channel for large \delm
is
\cite{chores}
\be
\sint < 1.2 \times 10^{-3} \quad (90 \% CL) 
\ee
The final goal is to reach a sensitivity down to \sint $\approx 2 \times 10^{-4}$ for large
\delm .
\newline
The NOMAD experiment \cite{nomaddet} on the other hand relies on the
kinematics. It has as a main active target 45 drift chambers
representing a total mass of 2.7 tons followed by transition radiation
and preshower detectors for e/$\pi$ separation. After an
electromagnetic calorimeter with an energy resolution of $\Delta E / E = 3.22 \% / \sqrt{E}
\oplus 1.04 \%$ 
and a hadronic calorimeter five muon
chambers follow. Because most of the devices are located within a magnetic field of 0.4 T a
precise momentum determination due
to the curvature of tracks is possible.
The $\tau$-lepton cannot be seen directly, the signature is determined by the decay
kinematics.
The main background for the $\tau$-search are regular charged and neutral current reactions.
In normal \nmu charged current events the muon balances the hadronic final
state in transverse momentum $p_T$ with respect to the \neu beam. Hence the value for missing
transverse momentum is small.
The angle $\Phi_{lh}$ between the outgoing lepton and the hadronic final state is close to
180$^o$ while
 the angle
$\Phi_{mh}$ between the missing momentum and the hadronic final state is more or less equally
distributed.
In case of a $\tau$ - decay there is significant missing $p_T$ because of the escaping \neus
as well as a 
concentration of $\Phi_{mh}$ to larger angles because of the kinematics.
In the \nmu - \ntau channel for large \delm NOMAD gives a limit of
\cite{nomadres}
\be
\sint < 1.2 \times 10^{-3} \quad (90 \% CL)  \\
\ee
Both limits are now better than the limit of E531 (Fig.\ref{pic:tosca}). 
Having a good electron identification, NOMAD also offers 
the possibility to search for \oszs in the \nmu - \nel channel.
A preliminary limit (Fig.\ref{pic:lsndev}) on \nmu - \nel is
available as (for large $\Delta m^2$) 
\be 
\sint < 2 \times 10^{-3} \quad (90 \% CL)\ee
This and a recently published CCFR result \cite{ccfr} seem to rule out the large
\delm region of the LSND evidence. 
NOMAD will continue data taking until Sept. 1998.

\subsubsection{Future \exps }
\begin{figure}
\begin{center}
\epsfig{file=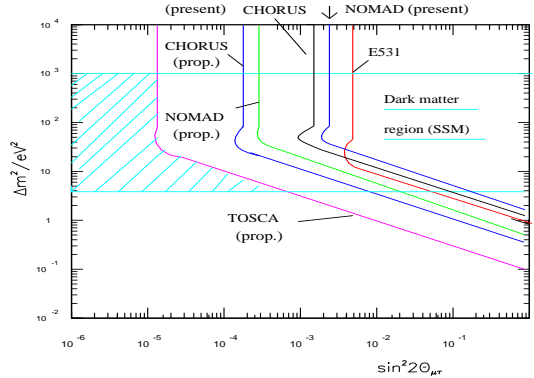,width=7cm,height=5cm}
\caption{Status and proposed curves for \nmu - \ntau
\oszsp Shaded are the regions expected for an eV-\ntau as dark matter candidate
according to the see-saw-model, where \nel - \nmu \oszs are used to explain the
solar \neu deficit. Meanwhile the CHORUS and NOMAD exclusion curve moved down to
1.2 $\cdot 10^{-3}$ for large \delm .}
\label{pic:tosca}
\end{center}
\end{figure}

Possible future ideas split into two groups depending on the physics goal. One group is
focussing on improving the
existing bounds in the eV-region by another order of magnitude with respect to CHORUS and NOMAD.
This effort is
motivated by the classical
\ssm which offers a \ntau in the eV-region as a good candidate for \hdm by assuming that the
solar \neu problem can be solved 
by \nel - \nmu \oszsp The second motivation is to check the \lsnd evidence. 
The other group plans to
increase
the source - detector distance to probe smaller \delm and to be directly comparable to
atmospheric scales (see chapter \ref{cha54}).
\paragraph{Short and medium baseline experiments} 
Several ideas exist for a next generation of short and medium baseline experiments. At CERN the 
proposed follow up
is TOSCA, a
detector combining features of NOMAD and CHORUS \cite{tosca}. The idea is
to use 2.4 tons
of emulsion 
within the NOMAD magnet in
form of 6 target modules. Each module contains an emulsion target consisting of 72 emulsion
plates, as well as a set of  
large silicon microstrip detector planes and a set of honeycomb tracker planes. Both
will allow a precise determination of the interaction vertex improving significantly the
efficiency. To verify the
feasibility 
of large silicon
detector planes maintaining excellent spatial resolution over larger areas, NOMAD
included a
prototype (STAR) in the 1997 data taking.
Moreover options to extract a \neu beam at lower energy of the proton beam
(350 GeV) at 
the CERN SPS to reduce the prompt \ntau background are discussed.
The proposed sensitivity in the \nmu - \ntau channel is around 2$\times 10^{-5}$ for large \delm
(\delm
$>100 eV^2$) (Fig. \ref{pic:tosca}). 
Also proposals for a medium baseline search exist
\cite{rev97,aut97}. The CERN \neu beam used
by CHORUS and NOMAD is coming up to the surface again in a distance of about 17 km
away from the beam dump. An installation of an ICARUS-type detector (liquid Ar TPC) \cite{rev97}
could be made here. In a smaller version, two fine grained calorimeters located at CERN and
in 17 km distance might be used as well \cite{aut97}.
\paragraph{Long baseline \exps}
Several accelerators and underground laboratories around the world offer the possibility to
perform \lbls .
This is of special importance to probe the region of atmos\-pheric
\neus directly.\\
{\it KEK - \sk :}
The first of these experiments will be the KEK-E362 experiment (K2K) \cite{keksk} in
Japan
sending a \neu beam from KEK to \sk. The distance is 235 km. A 1 kt near detector, about 1 km
away
from the
beam dump will
serve as a reference and measure the \neu spectrum. The \neu beam with
an average energy of 1 GeV is produced by a 12 GeV proton beam dump. The detection method
within \sk will 
be identical to that of their atmospheric \neu detection.
The beamline should be finished by the end of 1998 and the experiment will start data taking in
1999.
In connection with the JHC-project an upgrade of KEK is planned to a 50 GeV proton beam, which
could start producing data around
2004 and would make a \ntau appearance search possible.\\
{\it Fermilab - Soudan:}
A \neu program is also associated with the new Main Injector at Fermilab. The long
baseline project will
send a \neu beam to the Soudan mine about 735 km away from Fermilab. Here the MINOS experiment
\cite{minos} will be
installed. It also consists of a near detector located at Fermilab
and a far detector at Soudan. The far
detector will be made of 8 kt magnetized Fe toroids in 600 layers with 2.54 cm thickness
interrupted by about 32000
m$^2$ active detector planes in form of plastic scintillator strips with x and y readout to get the
necessary tracking 
informations. An additional hybrid emulsion detector for $\tau$-appearance is also under
consideration. The project could start
at the beginning of next century.\\
{\it CERN - Gran Sasso:}
A further program considered in Europe are \lbls using a \neu beam from CERN down to Gran
Sasso Laboratory.
The distance is 732 km. Several experiments have been proposed for 
the \osz search. The first proposal is the ICARUS
experiment \cite{icarus} which will be installed in Gran Sasso anyway for
the search of
proton decay and 
solar neutrinos.
This liquid Ar TPC can also be used for long baseline searches. A prototype  of 600 t is
approved for installation which will happen in 1999.
A second proposal, the NOE experiment \cite{noe}, plans
to build a giant lead-scintillating fibre detector with a total mass of 4.3 kt. 
The calorimeter modules will be interleaved with transition radiation 
detectors with a total of 2.4 kt. The complete detector will
have twelve modules, each 8m$\times$8m$\times$5m, and
a module for muon identification at the end. A third proposal is the building
of a 125 kt water-RICH detector (AQUA-RICH) \cite{tom}, which could be installed 
outside the Gran 
Sasso tunnel. The readout will be done by 3600 HPDs with
a diameter of 250 mm and having single photon sensitivity.
Finally there exists a proposal for a 750t iron-emulsion sandwich detector
(OPERA) \cite{niwa}
which could be installed either at the Fermilab-Soudan or
the CERN-Gran Sasso project. It could consist of 92 modules, each would have 
a dimension orthogonal to the beam of
3$\times$3 m$^2$ and would consist out of 30 sandwiches. One sandwich is
composed out of 1 mm iron, followed by two 50 $\mu$m 
emulsion sheets, spaced by 100 $\mu$m. After a gap of 2.5 mm, which could be 
filled by low density material, two additional
emulsion sheets are installed. 
The $\tau$, produced by CC reaction in the iron, decays in the gap region,
and the emulsion sheets
are used to verify the kink of the decay.\\
A project in the very far future could be \osz \exps involving a 
$\mu^+ \mu^-$-collider currently 
under investigation. The
created \neu beam is basically free of \ntau and can be precisely determined to be 50 \%
$\nmu (\bar{\nu}_\mu)$ and 50\% \bnel ($\nu_e$) for $\mu^- (\mu^+)$.
Because the $\mu^+
\mu^-$-collider would
be a high luminosity machine, one even can envisage very \lbls e.g. from Fermilab to Gran
Sasso
with a distance of 9900 km \cite{gee97}.

\subsection{Atmospheric \neus}
\label{cha54}
\begin{figure}
\begin{center}
\epsfig{file=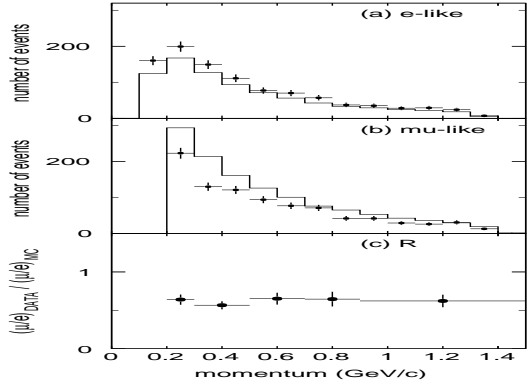,width=7cm,height=5cm}
\caption{Momentum spectra of \sk sub-GeV sample a) for the e-like events b)
for the $\mu$-like events and c) the obtained $R$-ratio. $R$ stays essentially flat over all
five bins (from \protect \cite{fuk98}).} 
\label{pic:skpspek}
\end{center}
\end{figure}

A different source of \neus are cosmic ray interactions within the
atmos\-phere. 
A detailed prediction of the expected flux depends on three main ingredients, namely the cosmic
ray spectrum and composition, the geomagnetic cutoff and the \neu yield of the hadronic
interaction in the
atmosphere. At lower energies ($E \lsim$1 GeV) \neus basically result from pion- and muon-decay
leading to rough expectations for the fluxes like $\nmu \sim \bnmu \sim 2 \nel$ or $\nel / \bnel
\sim \mu^+ / \mu^-$.
The ratio \nel / \nmu drops quickly above 1 GeV, because for $E_{\mu} > 2.5$ GeV the path length
for muons
becomes larger than the production height. At even higher energies the main source of \nel are
$K_{e3}$-decays ($K^0_L \ra \pi e \nel$). Contributions of prompt \neus from charm decay are
negligible and
might become important in the region above 10 TeV \cite{thu95}.  
\begin{figure}
\begin{center}
\epsfig{file=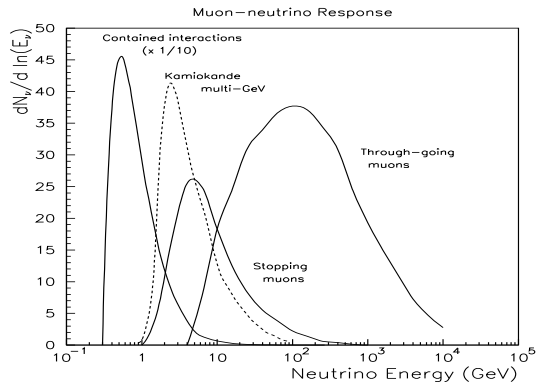,width=7cm,height=5cm}
\caption{Different classes of observable
muons as a function of \neu energy. The peak corresponding to contained events is reduced by a
factor 10 (after \protect \cite{gai95}).}
\label{pic:atmmuons}
\end{center}
\end{figure} 
For \neus in the energy range 300 MeV \lsim \enu
\lsim 3 GeV the energy of the primary typically lies in the region 5 GeV 
\lsim $E_P$ \lsim 50 GeV.
This region of the spectrum is affected by the geomagnetic cutoff, which 
depends on the gyroradius of the
particles, introducing a factor A/Z between nuclei and protons of the same
energy. However \neu
production depends on the energy per nucleon E/N.
Furthermore the energy range below about 20 GeV is also affected by the 11-year activity 
cycle of the sun, which is in the maximum phase
preventing low energy cosmic rays to penetrate into the inner solar system.  
Neutrinos in the region well beyond 1 GeV can be detected via horizontal or \ugm produced by
CC
reactions. The dominant part is given from events between 10 GeV $< \enu < 10^4$ GeV.
The contribution of primaries
with energies larger than $10^5$ GeV/nucleon to the upward going muon flux is only about 15 \%.
Several authors made calculations for the \neu flux for
different
detector sites covering the energy region from
100 MeV to $10^4$ GeV \cite{bar89,bug89,hon90,per94,agr96,gai96}. The
absolute predictions differ by about 30 \% due to different assumptions on 
the cosmic ray spectrum and composition and
the description of the hadronic interaction.
Absolute atmospheric \neu spectra in the interval 320 MeV $< E_{\nel} <$ 30 GeV for \nel
and 250 MeV $<
E_{\nmu} <$ 10 TeV are measured by the Frejus-experiment \cite{dau95}.
The observed \neu event types can be divided by their
experimental separation into contained, stopped and throughgoing 
events (Fig.\ref{pic:atmmuons}).
For \neu \osz searches it is convenient to use the ratio $\mu$/e or even the double ratio R
of experimental values versus Monte Carlo prediction 
\be
R= \frac{(\mu /e)_{exp}}{(\mu /e)_{MC}}
\ee
where $\mu$ denotes muon-like and $e$ electron-like events. Here 
a large number of systematic effects cancel out.
The above mentioned calculations agree for $R$ within 5 \% for \enu between 400 MeV 
and 1 GeV but
show a
significant difference in normalisation and spectral shape. This effect can mainly 
be traced back to 
different assumptions on the production of low energy pions from 10
- 30 GeV p-Air interactions. This might be improved by the results of
the recent NA56 measurements \cite{amb98}. Furthermore the predictions can be 
cross-checked with atmospheric muon flux measurements which are closely related
\cite{bel96,cir97,bug98}.\\
The purest sample to investigate are the contained events corresponding to
\enu \lsim 1 GeV.
The events are basically due to
quasielastic CC and single pion NC reactions \cite{lip95}. Unfortunately the
relevant cross sections for these processes have a relatively large uncertainty 
in the energy range of
interest.
By far the highest statistics for the sub-GeV region
($E_{vis} < 1.33$ GeV, where $E_{vis}$ is the energy of an electromagnetic shower producing a
certain amount of Cerenkov-light) is given by \sk. With a significance of 33 kt$\times$y they 
accumulated 1158 $\mu$-like and 1231 e-like events in their contained single ring sample
\cite{fuk98}. The capability to distinguish e-like and $\mu$-like events in water
Cerenkov-detectors was verified at KEK \cite{kas96}. The
momentum spectra are shown in Fig.\ref{pic:skpspek}. The value
obtained with two independent analyses is given by $R=0.61 \pm 0.03 (stat.) \pm 0.05 (sys.)$.
\begin{figure}
\begin{center}
\epsfig{file=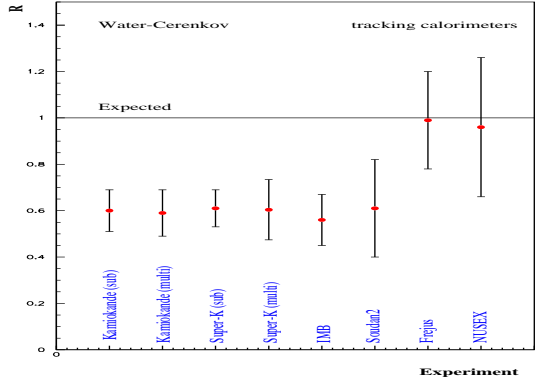,width=7cm,height=5cm}
\caption{Observed R-values by different \exps . No \osz corresponds to R=1.}
\label{pic:rvalues}
\end{center}
\end{figure}
A compilation of experimental results is shown in Table 8 (Fig. \ref{pic:rvalues}).
\begin{table}
\begin{center}
\label{tab:rvalues}
\caption{Compilation of existing R ratio measurements. The statistics is now
clearly dominated by \sk . The no \osz case corresponds to R = 1.}
\begin{tabular}{ccc}
\hline
Experiment & R & stat. significance (kT $\times$ y) \\
\hline
\sk (sub GeV) & $0.63 \pm 0.03 \pm 0.05$ & 33.0\\
\sk (multi GeV) & $0.65 \pm 0.05 \pm 0.08$ & 33.0 \\
Soudan2 & $0.61 \pm 0.15 \pm 0.05$ & 3.2\\
IMB & $0.54 \pm 0.05 \pm 0.11$ & 7.7\\
Kamiokande (sub GeV) & $0.60^{+0.06}_{-0.05} \pm 0.05$ & 7.7\\
Kamiokande (multi GeV) & $0.57^{+0.08}_{-0.07} \pm 0.07$ & 7.7\\
Frejus & $1.00 \pm 0.15 \pm 0.08$ & 2.0 \\
Nusex & $0.96^{+0.32}_{-0.28}$ & 0.74 \\
\hline
\end{tabular}
\end{center}
\end{table}

\begin{figure}
\begin{center}
\begin{tabular}{cc}
\epsfig{file=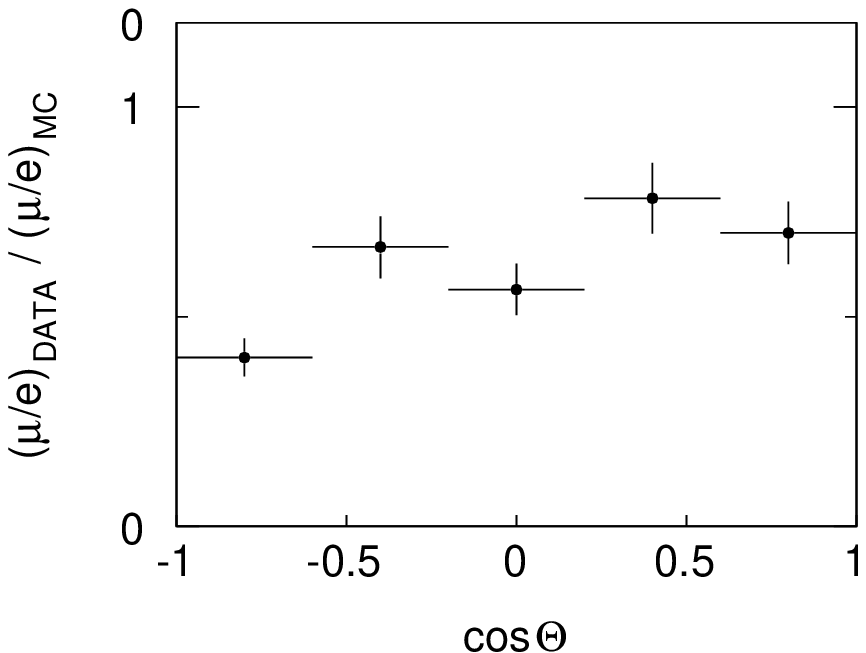,width=7cm,height=5cm} &
\epsfig{file=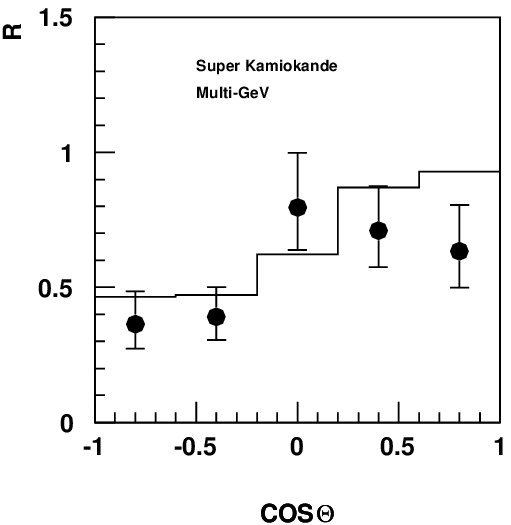,width=7cm,height=5cm}
\end{tabular}
\caption{Zenith angle dependence of R for the sub- (left) and multi-GeV
(right) sample of \sk . The line in the multi-GeV sample corresponds to \nmu - \ntau \oszs with
\sint =1 and \delm = $5 \cdot 10^{-3} eV^2$ (from \protect \cite{fuk98,fuk98a}).}
\label{pic:skrratiosubmul}
\end{center}
\end{figure}

\begin{figure}
\begin{center} 
\epsfig{file=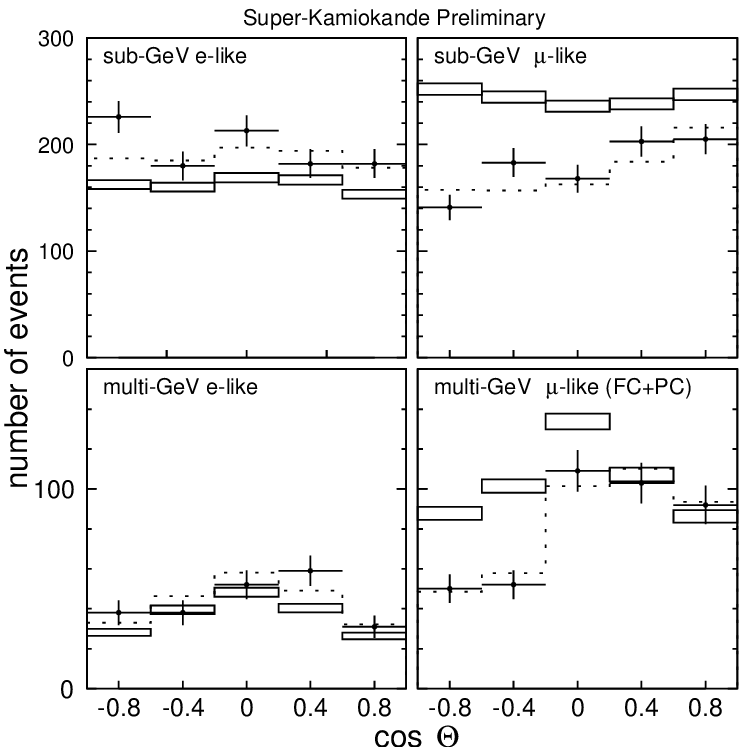,width=7cm,height=9cm}
\caption{Zenith angle distribution of electron- (left) and
muon like (right) events for the sub- and multi GeV sample of \sk . The bars correspond to MC
expection without \neu \oszs and are reflecting the statistical uncertainty. The $\pm$ 20 \%
normalisation uncertainty is not shown. The normalisation is adjusted to the upward going
events. The dotted curve corresponds to a \nmu -\ntau \osz scenario with \sint = 1 and \delm
= $5 \cdot 10^{-3} eV^2$. A clear $\mu$-deficit coming from below can be seen. The ratio 
R obtained from the histograms is shown in Fig. \protect \ref{pic:skrratiosubmul} 
(from \protect \cite{fuk98a}).} 
\label{pic:zenith}
\end{center} 
\end{figure}

\begin{figure}
\begin{center}
\begin{tabular}{cc}
\epsfig{file=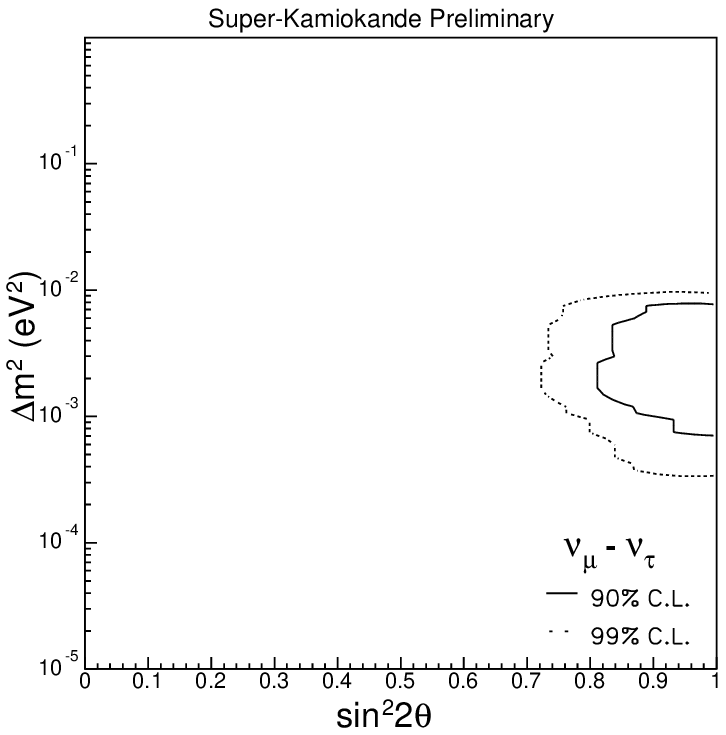,width=6cm,height=6cm} &
\epsfig{file=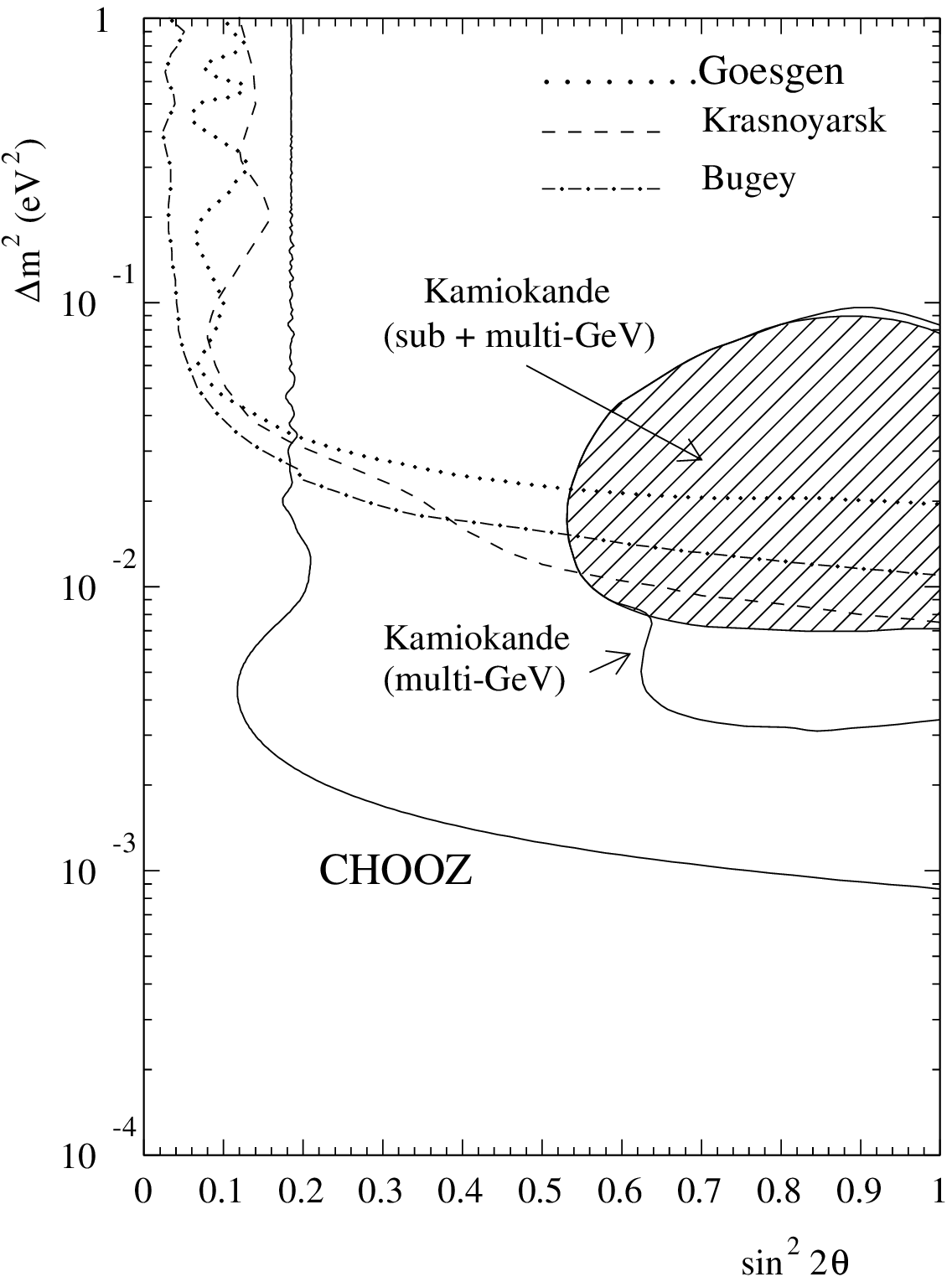,width=6cm,height=6cm}
\end{tabular}
\caption{a) \nmu - \ntau allowed region of \sk analysis of zenith angle
distribution (from \protect \cite{kea98}). b) \nel - \nmu exclusion plot as given
by
reactor \exps especially CHOOZ. This excludes any interpretation of
atmospheric \neu data as observed by Kamiokande by \oszs in this channel
(from \protect \cite{apo98}).}
\label{pic:exclplatm}
\end{center}
\end{figure}

\begin{figure}
\begin{center}
\begin{tabular}{cc}
\epsfig{file=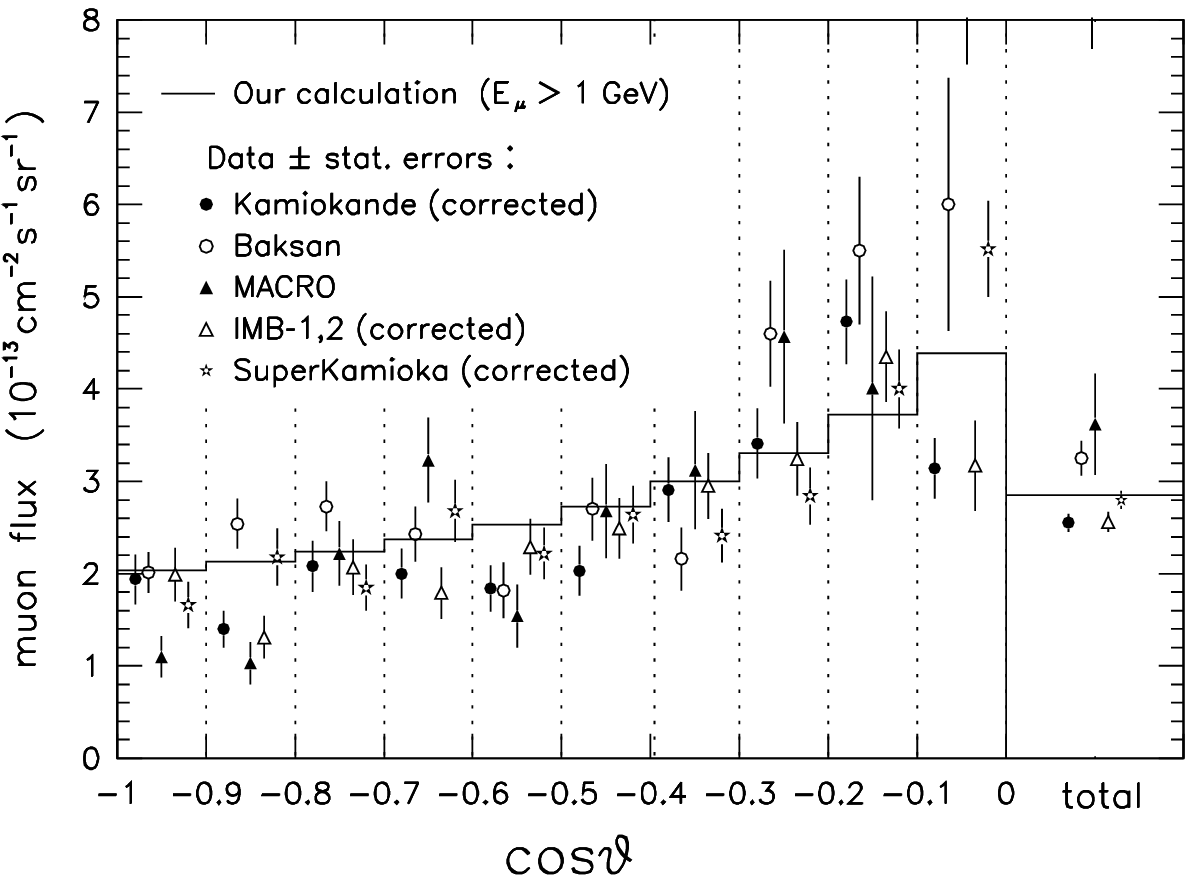,width=7cm,height=5cm} &
\epsfig{file=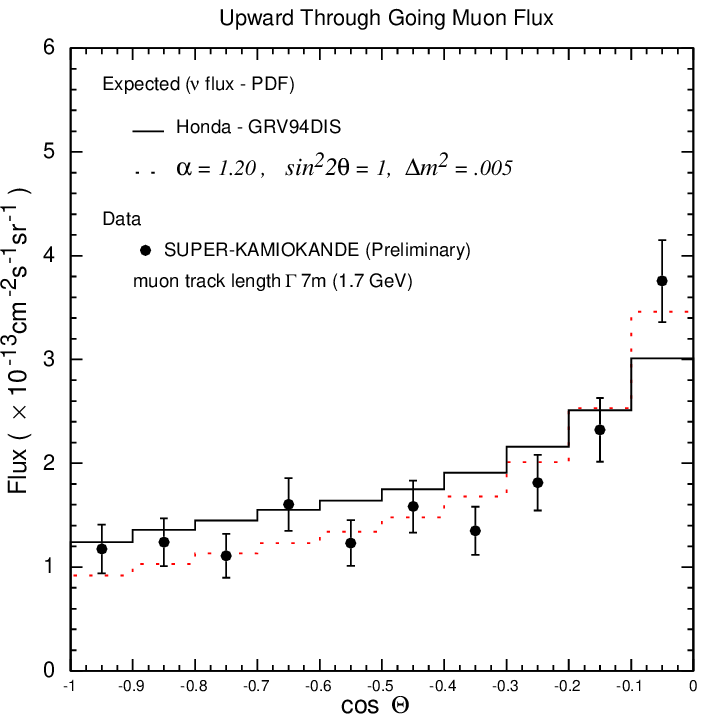,width=5cm,height=7cm}
\end{tabular}
\caption{Zenith angle distribution of \ugm compiled by Fogli et
al. \protect \cite{fog97} as of Aug. 1997 (left). The up-to-date spectrum as observed
by \sk is shown on the right.
The line corresponds to the flux predictions of Honda et al. \protect \cite{hon95} and
using the GRV 94
DIS structure functions \protect \cite{glu94}. An
\osz scenario with \sint = 1 and \delm = $5 \cdot 10^{-3} eV^2$ including a factor $\alpha=1.2$ in the
normalisation is shown as a dotted curve (from \protect \cite{kea98}).}
\label{pic:ugmzenit} 
\end{center}
\end{figure}

While Frejus and NUSEX are in agreement with
expectations, it can be
seen that the water
Cerenkov detectors and Soudan2 show a significant reduction. 
Besides looking on the R-ratio for \osz searches, the
zenith angle distribution can be used (Fig.\ref{pic:skrratiosubmul}).
Because the
baselines are quite different for downward (L $\approx$ 20 km) and \ugm (L
$\approx 10^4$ km), any \osz
effect should show up in a zenith angle dependence. The recent distributions from \sk also for
the multi-GeV sample ($E_{vis} > 1.33$ GeV), consisting of contained and partially contained
events, are shown in Fig.
\ref{pic:zenith} showing an up-down asymmetry which could be explained by \neu \oszs
\cite{fuk98a,kea98}. To verify
this assumption an L/E analysis for fixed \delm , as the one proposed for
the LEP-experiments
in \cite{man98a}, is done, which shows a characteristic
oscillation pattern. From the zenith angle distribution and the momentum spectra it seems
evident that there is a deficit in muon-like
events, which might be explained by $\nmu - \ntau$ or $\nmu - \nu_S$ \oszsp The region allowed 
by $\nmu - \ntau$ \oszs is shown in Fig.\ref{pic:exclplatm}. An independent three flavour
analysis results in a best fit value of \delm $\approx 7 \cdot 10^{-3} eV^2$ for
maximal mixing \cite{yas98}. Additionally the
CHOOZ-result excludes all Kamiokande data to be due to $\nmu - \nel$ \oszs and are shown for 
comparison in Fig.\ref{pic:exclplatm} as well. Moreover in a recent analysis of all atmospheric
data including the earth matter effect (see chapter \ref{cha632}), the CHOOZ-result rules out
the \nmu - \nel solution for Super-K at 90 \% CL \cite{gon98}. Furthermore different \osz channels
might be
distinguished by a detailed investigation of up-down asymmetries
\cite{foo97} or by measuring
the NC/CC ratio \cite{vis97}.\\ 
Neutrino events at higher energies are detected via their CC
reactions producing \ugm . The effective detector area can be increased because of the
muon range allowing \nmu CC in the surrounding rock (see chapter \ref{cha722}). 
The corresponding
muon flux of
the used horizontal and \ugm has to
be compared with absolute predictions. One also has to take care of the angular dependent
acceptance of the detector. Here the main uncertainty for the \neu flux stems from kaon
production and the knowledge of the involved structure functions. 
The behaviour of low-energy cross sections is dicussed in \cite{lip95}. Also here the models can
be adjusted to recent muon flux measurements in the
atmosphere
\cite{bel96} even though one has to take into account that for E$>$100 GeV relatively
more \neus are
produced by
kaon-decays while the muon-flux is still given dominantly by pion-decay. 
The observations of \ugm are compiled in Fig.\ref{pic:ugmzenit}. A zenith angle 
distribution from
\ugm
as measured with \sk is shown in Fig. \ref{pic:ugmzenit}. Two independent ways
of verifying the
\osz solution are the ratio of stopped/throughgoing muons and the shape of the
zenith
angle distribution \cite{lip97}. Both were done by \sk and support their oscillation evidence
\cite{fuk98a}.

\section{Solar neutrinos}
\begin{figure}
\begin{center}  
\begin{tabular}{cc}
\epsfig{file=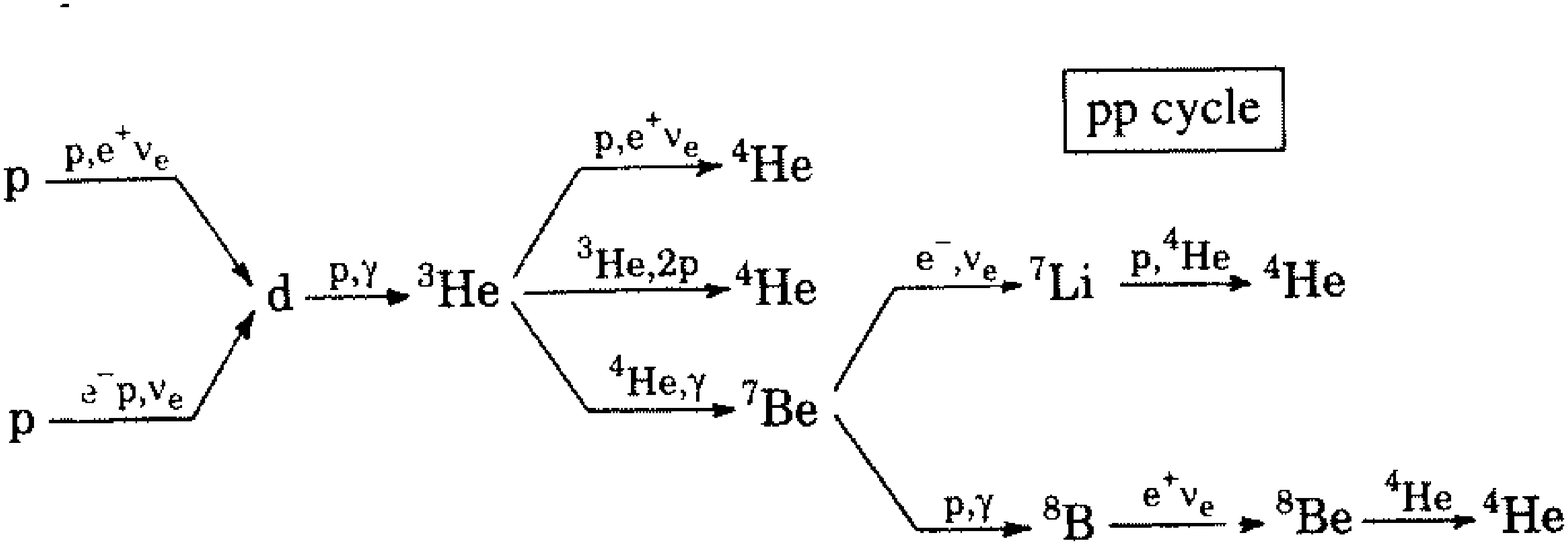,width=7cm,height=5cm} &
\epsfig{file=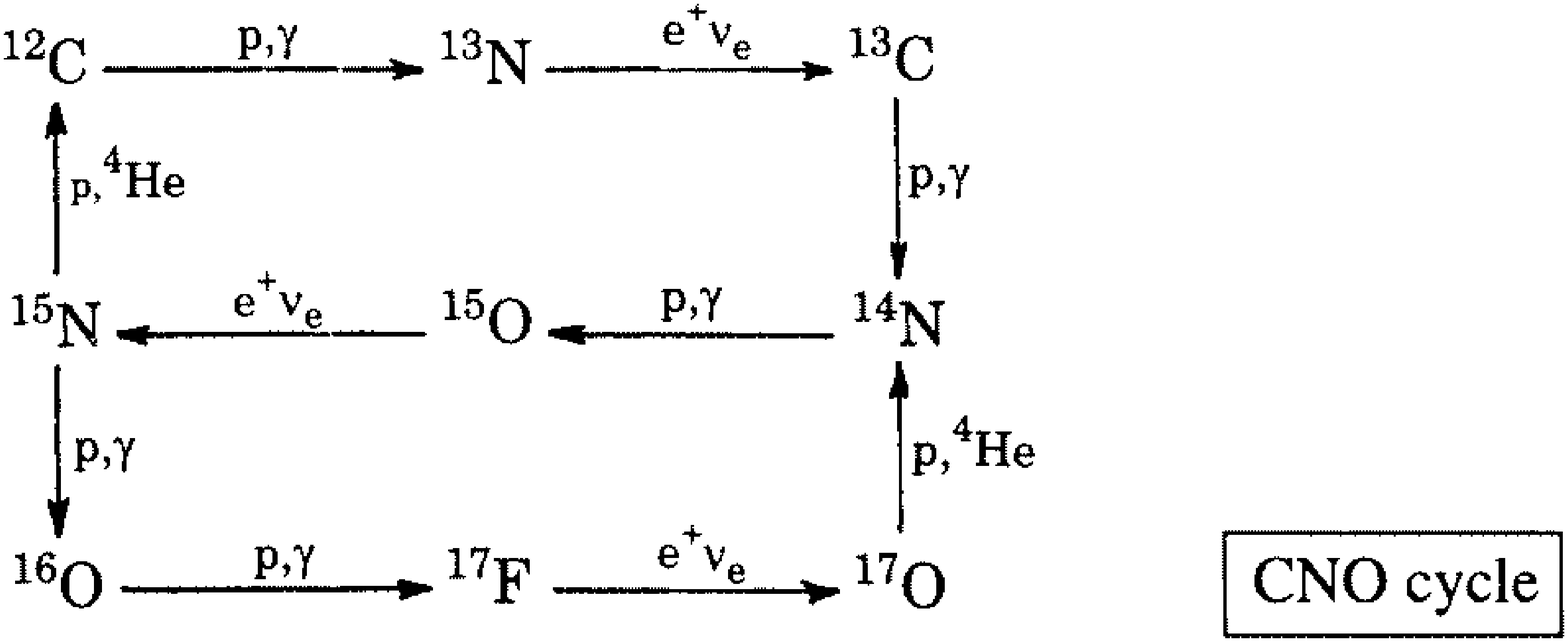,width=7cm,height=5cm}
\end{tabular}
\caption{Energy generation in the sun via the pp und CNO
cycles.}
\label{pic:pp}
\end{center}
\end{figure}

The closest astronomical neutrino source is the sun. 
The investigation and understanding of the sun as a typical main sequence star is of outstanding
importance for an understanding of stellar evolution. 
Stars are producing their energy via nuclear reactions. The hydrogen
burning is done in two ways as shown in Fig.\ref{pic:pp}, the pp-chain and
the
CNO-chain. The net result is the same giving
\be
4 p \ra \hev + 2 \nel + 26.73 MeV
\ee
The prediction of the expected
neutrino flux depends on detailed calculations of the solar structure
resulting in temperature,
pressure and density profiles and the knowledge of nuclear cross sections
for determining the energy
generation. Once the flux is in hand, it is still a matter of detecting
this low-energy \neus typically below 15 MeV with the main component below 500 keV.
The principle methods are radiochemical detectors using inverse $\beta$-decay and
real-time experiments looking for neutrino-electron scattering. Because of
the low cross-sections
involved, it is convenient to introduce a new unit for the expected event
rates in
radiochemical
detectors called SNU (solar
\neu unit) given by
\be
1 SNU = 10^{-36} \quad \mbox{captures per target atom per second}
\ee
The fundamental equations and ingredients of standard solar models are
discussed first. For more detailed reviews see
\cite{cla68,bah89,bah95}.
\subsection{Standard solar models (SSM)}
The sun as a main sequence star is producing its energy by hydrogen fusion and its
stability is ruled by thermal and hydrodynamic equilibrium. Modelling of
the sun as well as the prediction of the expected 
\neu flux requires the basic
equations of stellar evolution:\\
Mass conservation
\be
\frac{dM(r)}{dr} = 4 \pi r^2 \rho(r)
\ee
where $M(r)$ denotes the mass within a sphere of radius $r$.\\
Hydrostatic equilibrium (gravity is balanced by gas and radiation pressure)
\be
\frac{dp(r)}{dr}= - \frac{GM(r)}{r^2} \rho(r)
\ee
Energy balance, meaning the observed luminosity L is generated by an energy
generation rate $\epsilon$
\be
\frac{dL(r)}{dr} =  4 \pi r^2 \rho(r) \epsilon
\ee
Energy transport dominantly by radiation and convection which is given in the radiation
case by
\be
\frac{dT(r)}{dr} = - \frac{3}{64 \pi \sigma} \frac{\kappa \rho(r)
L(r)}{r^2 T^3}
\ee
with $\sigma$ as the Stefan-Boltzman constant and $\kappa$ as absorption coefficient.
These equations are governed by additional three equations of state for the pressure $p$,
the absorption coefficient $\kappa$ and the energy generation rate $\epsilon$:
\be
p=p(\rho,T,X), \quad \kappa = \kappa (\rho,T,X), \quad \epsilon = \epsilon (\rho,T,X)
\ee
where $X$ denotes the chemical composition.
The Russell-Vogt theorem then assures, that for a given $M$ and $X$ an 
unique equilibrium
configuration will evolve, resulting in certain radial pressure, temperature and density
profiles. Under these assumptions, solar models can be calculated as an
evolutionary
sequence from an
initial chemical composition. The boundary conditions are that the model
has to reproduce the age, luminosity, surface temperature and mass of the present sun.
The two typical adjustable parameters are the 
\hev abundance and the relation of the convective mixing length to the
pressure scale height.
This task has been done by several groups
\cite{bah95,iau90,tur93a,tur93b,ric96,dar96,cas97}. 
Nevertheless there remain sources of
uncertainties. Some will be discussed in a little more detail.
\subsubsection{Diffusion}
Several experimental evidences strongly suggest a significant mixing and
gravi\-tational settling of He
and the heavier elements in the sun. The long standing problem of \lis - depletion in the solar
photosphere can
be explained if \lis is destroyed by nuclear burning processes, which on the 
other hand requires
temperatures of about 2.6 $\cdot 10^6$ K. Such temperatures exist nowhere at the base of the
convection zone, therefore \lis has to be brought to the inner regions. Also the 
measured sound speed
profiles in the solar interior obtained by helioseismological data can be better reproduced by
including diffusion processes.
Therefore these effects were included in the latest solar models.
\subsubsection{Initial composition}
The chemical abundances of the heavier elements (larger than helium) is an 
important ingredient for solar
modelling. Their abundances influence the radiative opacity and
therefore the temperature profile within the sun. Under the assumption
of a homogeneous sun, it is assumed
that the element abundance in the solar photosphere still corresponds to the 
initial values. The relative
abundances of the heavy elements are best determined in certain kind of meteorites, the
type I carbonaceous chondrites, 
which can be linked and found in good agreement with the photospheric abundances
\cite{gre93,gre93a}.
Abundances of C,N and O are
taken from photospheric values, the \hev abundance cannot be measured 
and is used as an adjustable
parameter.
\subsubsection{Opacity and \eos}
The opacity or Rosseland mean absorption coefficient $\kappa$ is defined
as a harmonic mean integrated over
all frequencies $\nu$
\be
\frac{1}{\kappa} = \frac{\int_0^{\infty}\frac{1}{\kappa_\nu} \frac{dB_\nu}{dT} d
\nu}{\int_0^{\infty} \frac{dB_\nu}{dT} d \nu} 
\ee
where $B_\nu$ denotes a Planck-spectrum.
The implication is that more energy is transported at frequencies at which the 
material is more transparent
and at which the radiation field is more temperature dependent. The 
calculation of the Rosseland mean
requires a knowledge of all involved absorption and scattering cross 
sections of photons on atoms, ions and
electrons. The calculation includes bound-bound (absorption), bound-free 
(photoionization), free-free
(inverse bremsstrahlung) transitions and Thomson-scattering. 
Corrections for electrostatic interactions of
the ions with electrons and for stimulated emissions have to be taken 
into acount. The number densities
$n_i$
of the absorbers can be extracted from the Boltzmann and Saha equations. 
The radiative opacity per unit
mass can then be
expressed as (with the substitution u= $h \nu / k T$)
\be
\frac{1}{\kappa} = \rho \int_0^{\infty} \frac{15 u^4 e^u / 4 \pi^4 
(e^u -1)^2}{(1-e^u) \sum_i \sigma_i n_i + 
\sigma_s n_e} du
\ee
where $\sigma_s$ denotes the Thomson scattering cross section.\\
The most comprehensive compilation of opacities is given by the Livermore group (OPAL)
\cite{ale94,igl96}. It includes
data of 19 elements in a wide range of temperature, density and composition.
The main contribution to the opacity in the centre of the sun is 
given by inverse bremsstrahlung with a few
per cent contribution of Thomson scattering. A detailed study on opacity
effects on the solar interior can
be found in Tripathy et al. \cite{tri97}.\\
A further ingredient for solar model calculations is the equation of state, meaning the
density as a function of
$p$ and $T$
or as widely used in the calculations, the pressure expressed as 
a function of density and temperature.
Except for the solar atmosphere, the gas pressure outweighs the radiation
pressure anywhere in the sun. The
gas pressure is given by the perfect gas law, where the mean molecular weight 
$\mu$ must be determined by
the corresponding element abundances. The different degrees of ionisation 
can be determined using the Saha
equations. An \eos in the solar interior has to consider plasma effects 
(normally via Debye-H\"uckel
treatment) and the partial electron degeneracy deep in the solar
core. The latest \eos is given by Rogers et al. \cite{rog96}.
\subsubsection{Nuclear reaction rates} 
\begin{figure}
\begin{center}
\epsfig{file=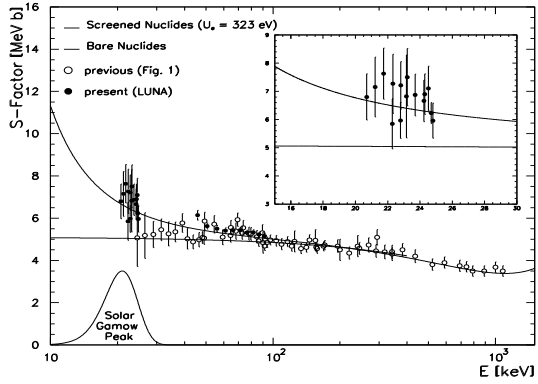,width=7cm,height=5cm}
\caption{Cross section for the \hed(\hed,2p)\hev at
low energies as obtained with the LUNA \expe . For the first time data points
directly in the Gamow peak are experimentally obtained (from \protect
\cite{arp97}).} 
\label{pic:he3he3}
\end{center}
\end{figure}

A detailed prediction of the solar structure and the corresponding \neu 
flux relies on the nuclear
reaction rates 
\cite{rol88,adl98}.
Their precise knowledge determines the branching in the complex
network of reactions and the yields of all isotopes. In contrast to typical
energies in the solar interior, which are in the keV region (Gamow-region),
laboratory measurements are normally at about MeV and
one has to extrapolate down. Because the cross section for non-resonant
charged particle interactions is steeply falling, it is usually
parametrized as 
\be 
\sigma (E) = \frac{S(E)}{E} exp (-2 \pi \eta)
\ee
where the Sommerfeld-parameter $\eta$ is given by $\eta=Z_1 Z_2 e^2 / \hbar v$. In cases of no
resonances,
the nuclear or astrophysical S-factor S(E) should show a rather smooth behaviour. 
It is therefore
typically the S-factor
which is extrapolated down to solar energies. Since the energy of the Gamow-Peak is temperature
dependent, S(E) is for ease of computation expanded in a Taylor series with respect to energy
\be
S(E) = S(0) + \dot{S}(0) E + \frac{1}{2} \ddot{S}(0) E^2 + ...
\ee
where $S(0),\dot{S}(0)$ etc. are obtained by a fit to the experimental data. A
compilation of
S(0) values for all relevant reactions of the pp-cycle can be found in Table 9. 
Because such
extrapolations contain some uncertainties, the idea is to measure the cross section
directly in the
relevant energy range. To reach this goal, several additional requirements
have to be fulfilled, e.g.
going underground. In a first step this is done by the LUNA collaboration
building a 50 kV accelerator at Gran Sasso Laboratory to investigate 
the \hed(\hed,2p)\hev reaction 
as the final step in the ppI-chain \cite{arp97}. As can be seen in Fig.
\ref{pic:he3he3}, the
experimental data exceed the theoretical
expectation of bare nuclei which is due to a still restricted knowledge of screening effects.
Further activities will include an upgrade to a 200 kV and even 2 MV accelerator making the
additional measurements of the \bes(p,$\gamma$)\ba and $^{14}N(p,\gamma$)\sfz cross sections
possible. New measurements for the \bes(p,$\gamma$)\ba cross section at cms
energies between 350 and 1400 keV exist \cite{ham98}. Earlier measurements of Kavanagh et
al. \cite{kav69} and Fillipone et al. \cite{fil83} showed a 30 \% difference in the absolute
value of S(E) in this region. The new measurement seems to support the lower
S(E) values of \cite{fil83}. A further proposal to investigate this important
cross-section exists by using ISOLDE at CERN \cite{bro97}.
\begin{table}
\begin{center}
\label{tab:S0werte}
\caption{Compilation of S(0) values relevant for the pp-process. For most reactions the 1$\sigma$ error is given.}
\begin{tabular}{cc}
reaction & S(0) (keV $\cdot$b)  \\
\hline
pp$\ra d e^+ \nu$ & $4.00(1 \pm 0.007^{+0.020}_{-0.011}) \cdot 10^{-22}$  \\
d(p,$\gamma$)\hed & 0.25 $\cdot 10^{-3}$  \\
\hed(\hed,2p)\hev & 5400 $\pm$ 400\\
\hed(p,$e^+$ \nel)\hev & 2.3 $\cdot 10^{-20}$\\
\hed($\alpha,\gamma$)\bes & 0.53 $\pm$ 0.05  \\
\bes(p,$\gamma$)\ba & 0.019$^{+0.004}_{-0.002}$  \\
\hline
\end{tabular}
\end{center}
\end{table}

\subsubsection{Neutrino flux predictions}
\begin{figure}
\begin{center}
\epsfig{file=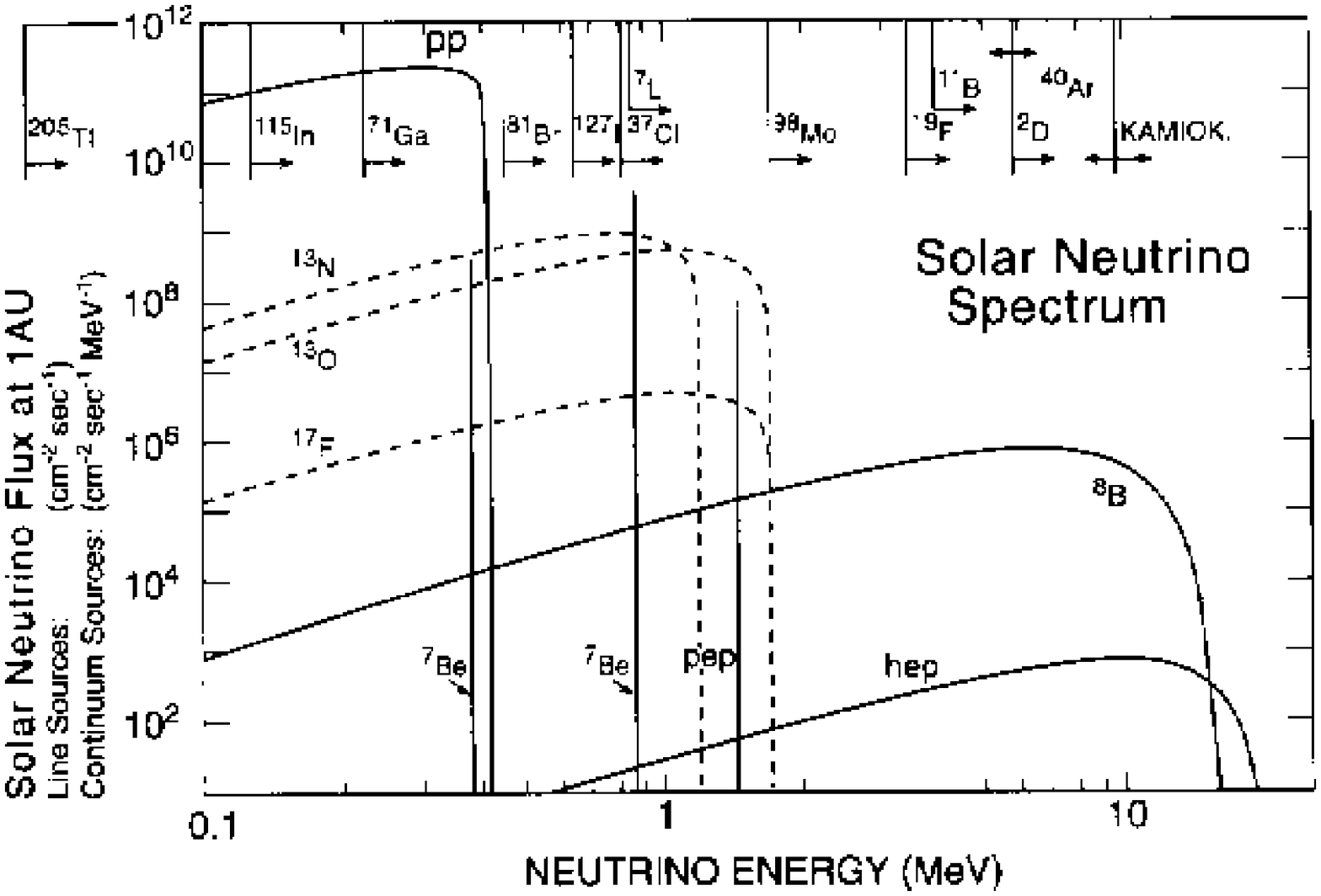,width=7cm,height=5cm}
\caption{Predicted solar \neu spectrum including thresholds for various running
and planned \exps (with kind permission of T. Kirsten).}
\label{pic:solarnspek}
\end{center}   
\end{figure}

Several groups are working on the detailed modelling of the sun in order to predict
accurately the
solar \neu flux and to reproduce the sound speed profile measured with helioseismology.
A comparison of the different predicted \neu fluxes at the position of the
earth is given in Table 10 and Fig.\ref{pic:solarnspek}.
\begin{table}
\begin{center}
\label{tab:solneuflux}
\caption{Flux predictions from four different solar model calculations.}
\begin{tabular}{ccccc}
\hline
Flux (\cms) & BP95 \cite{bah95} & RVCD \cite{ric96}& CDFLR \cite{cas97} & DS96 \cite{dar96}\\
\hline
${\phi }_{\nu }(pp)~[{10}^{10} \cms]$
& 5.91 & 5.94 & 5.99 & 6.10 \\
${\phi }_{\nu }(pep)~[{10}^{8}$\cms]
& 1.40 & 1.38 & 1.40 & 1.43 \\
${\phi }_{\nu }(^{7}{Be})~[{10}^{9}$\cms] &
5.15 & 4.80 & 4.49 & 3.71 \\
${\phi }_{\nu }(^{8}{B})~[{10}^{6}$\cms] &
6.62 & 6.33 & 5.16 & 2.49 \\
${\phi }_{\nu }(^{13}{N})~[{10}^{8}$\cms] &
6.18 & 5.59 & 5.30  & 3.82 \\
${\phi }_{\nu }(^{15}{O})~[{10}^{8}$\cms] &
5.45 & 4.81 & 4.50 & 3.74 \\  
\hline
\end{tabular}
\end{center}
\end{table}

\subsection{Solar \neu experiments}
At present results of five \neu experiments are available, namely the
chlorine-experiment, the two
gallium experiments GALLEX and SAGE and the only real-time experiment Kamiokande 
and its follow-up
Super-Kamiokande. The discussion follows the historical ordering starting with the
chlorine-experiment.
\subsubsection{The chlorine-experiment}
The origin of \neu astrophysics is the chlorine solar 
\neu \expe by R. Davis in the Homestake mine in South
Dakota \cite{dav97,cle98}. The detection reaction is 
\be
\label{eq:cl37}^{37}Cl + \nu_e \rightarrow ^{37}Ar + e^- 
\ee
with a threshold of 814 keV. Therefore it is basically sensitive 
to \ba and \bes \neus with small
contributions due to pep, $^{13}N$ and $^{15}O$ \neusp All, except
the \ba \neus , lead only to the ground
state of \arsid whereas \ba is also populating excited states including the isotopic analogue
state. The cross
section for the
reaction (\ref{eq:cl37}) averaged over the \ba spectrum has been measured recently
to be \cite{auf94,bah96}
\be
1.14 \pm 0.11 \cdot 10^{-42} cm^2 
\ee
The predicted SNU-rate for the experiment due to different flux
contributions is given in Table 11.
\begin{table}
\begin{center}
\label{tab:snuradio}  
\caption{Contributions (in SNU) of the different flux components of solar neutrinos to the signal
in different radiochemical detector materials of running or planned \exps . The fluxes of BP95 are
used.}
\begin{tabular}{cccccc}
\hline
Source & \clsd & \gaes & \ihsz & \lis & \xehed \\
\hline
pp & 0 & 69.7 & 0 & 0 & 9.7 \\
pep & 0.22 & 3.0 & 1.85 & 9.17 & 1.6\\
\bes & 1.24 & 37.7 & 13.0 & 9.78 & 17.8\\
\ba & 7.36 & 16.1 & 18.4 & 25.8 & 12.7 \\
$^{13}N$ & 0.11 & 3.8 & 0.73 & 2.62 & 1.6\\
$^{15}O$ & 0.37 & 6.3 & 2.43 & 13.4 & 1.8\\
\hline
\hline
Sum & 9.3 & 136.6 & 36.5 & 60.8 & 45.2 \\
\hline
\end{tabular}
\end{center}
\end{table}

\begin{figure}
\begin{center}  
\epsfig{file=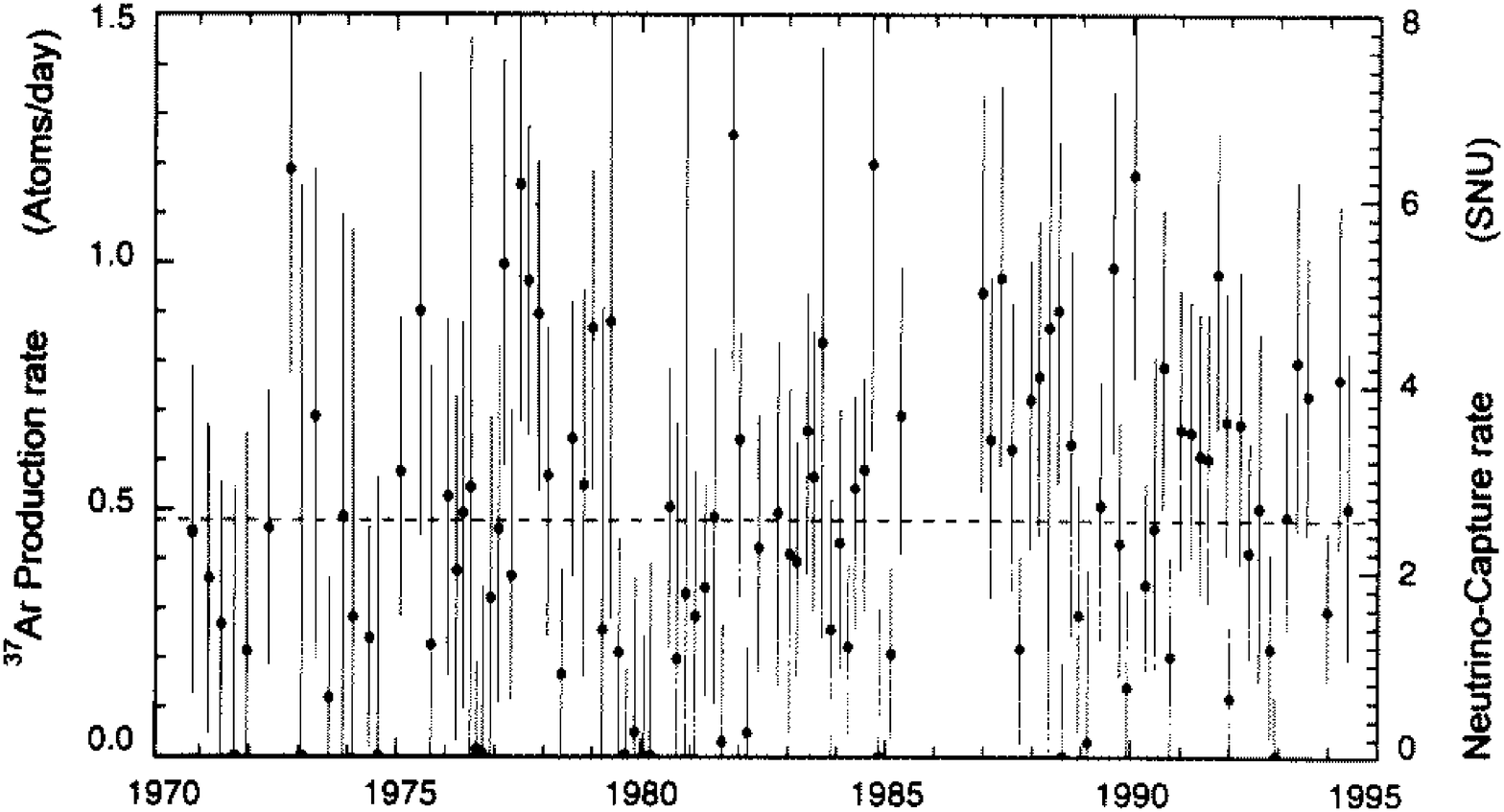,width=7cm,height=5cm}
\caption{SNU-rate as function of time for the Cl-\expe (from \protect
\cite{lan97}).}
\label{pic:cldata}
\end{center}
\end{figure}

The production of one \arsid atom/day corresponds to 5.35 SNU. The experiment consists of 615 t
tetrachloroethylene (C$_2$Cl$_4$) under a shielding of about 4000 mwe. 
The natural abundance of \clsd is 24
\% resulting in 2.2 $\cdot 10^{30}$ target atoms. An extraction of the 
produced \arsid happens roughly
every two
months, and the extraction efficiency is controlled by adding a small
amount of isotopical pure inert \arsd or \arad .
To do this, helium is flushed through the tank taking the volatile argon out of the
solution and allowing 
the collection of the argon in a cooled charcoal trap. After purification, 
the
argon is filled with the counting gas P10 into
specially developed low-level miniaturized proportional counters.
The detection reaction uses the EC of \arsid
\be
e^- + \arsid \rightarrow \clsd + \nu_e 
\ee
with a half-life of 35 days and focuses on observing the 2.82 keV Auger electrons.
To discriminate further against background energy and pulse rise time information are used and
the
counters are plugged into a special low-level shielding.
The average measured value using 108 runs after starting the experiment in 1970 is given by
\cite{lan97}
\be
2.56 \pm 0.16 (stat.) \pm 0.15 (sys.)  SNU
\ee
whereas the single runs can be seen in Fig. \ref{pic:cldata}.
The theoretical expectations are 9.3 $\pm$ 1.4 SNU \cite{bah95}, 
4.1 $\pm$ 1.2 SNU \cite{dar96} and
6.4 $\pm$ 1.4 SNU \cite{tur93b}.
This discrepancy is the origin of the solar \neu problem.

\subsubsection{Real time water Cerenkov detectors}
\begin{figure}
\begin{center}
\epsfig{file=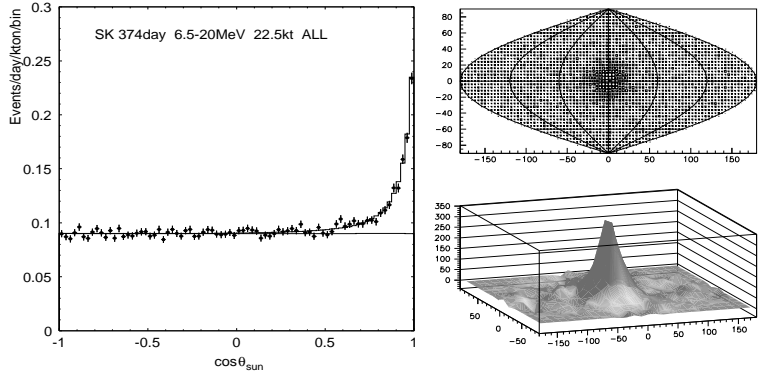,width=10cm,height=5cm}
\caption{Angular distribution of observed recoil electrons in \sk . Clearly
visible is an increase in direction of the sun (cos $\theta$ =1). The
double angle plots (right) with the sun at (0,0) show clearly that the \neus 
are coming from the sun (from \protect \cite{fuk98b}).}
\label{pic:sksolarsp}
\end{center}
\end{figure}

\begin{figure}
\begin{center}
\epsfig{file=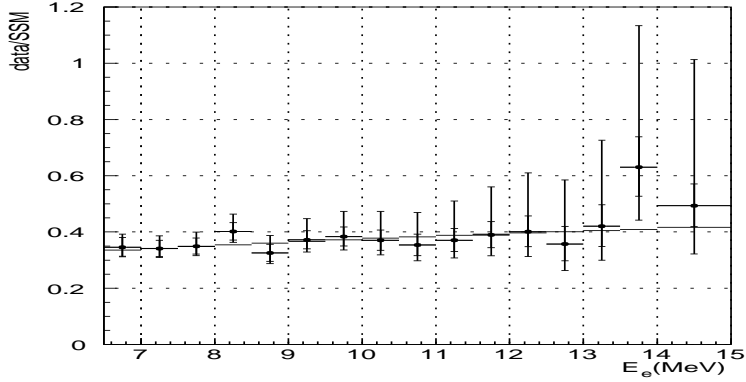,width=10cm,height=5cm}
\end{center}
\caption{Energy spectrum of \ba \neus as observed by \sk divided 
by the solar model of [184]. The solid line corresponds to a fit with a SMA
solution (from \protect \cite{tot97}).}
\label{pic:skb8}
\end{figure}

The only real time solar \neu \exps are \sk and its precessor Kamiokande. 
The detection principle 
is the Cerenkov light produced in neutrino-electron scattering within the
water. 
Energy and directional
information are reconstructed from the
corresponding number and timing of the hit photomultipliers. 
The scattering angle of the struck electron is related with the incident \neu energy as 
\be
cos \theta_e = \frac{1+\frac{m_e}{E_{\nu}}}{1+\frac{2 m_e}{T_e}}
\ee
where $T_e$ denotes the kinetic energy of the recoil electron. The directional information is
shown in Fig.\ref{pic:sksolarsp}.
While Kamiokande consisted out
of 3000t of water using only 680 tons as fiducial volume, \sk consists of 50000 t using 
22.5 kt as fiducial
volume. The detector threshold is 6.5 MeV for \sk (in the late stage of Kamiokande it was at
7.5 MeV) making these detectors only sensitive to \ba \neusp
The measured fluxes are \cite{tot97}
\bea 
\Phi (\ba) &=& 2.80 \pm 0.19 \pm 0.33 \cdot 10^6 \cms
\quad \mbox{Kamiokande (final)} \\
\Phi (\ba) &=& 2.44 \pm 0.05 (stat.) ^{+0.09}_{-0.07} (sys.) \cdot 10^6 \cms
\quad \mbox{\sk}
\eea
where the theoretical expectations are $6.62 \cdot 10^6 \cms$ (BP), $4.52 \cdot 10^6
\cms $ (TCL) and $2.49 \cdot 10^6 \cms$ (DS). The ratio of measured to expected electron recoil
spectrum is given in Fig.
\ref{pic:skb8}.
Super-Kamiokande recently implemented a low energy trigger with a threshold of 4.5 MeV.

\subsubsection{The gallium \exps }
\begin{figure}
\begin{center}  
\epsfig{file=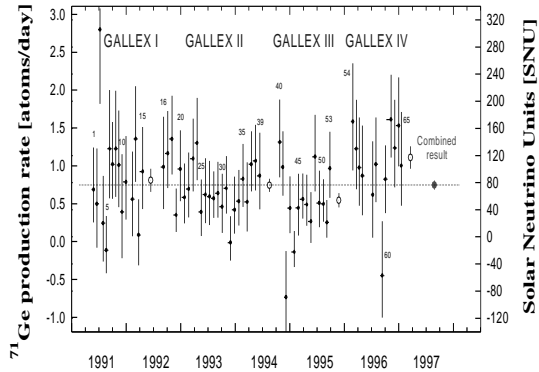,width=7cm,height=5cm}
\caption{SNU rate as a function of time for
the GALLEX \expe (from \protect \cite{kir97}).}
\label{pic:garesults}
\end{center}
\end{figure}

The only \exps which are able to measure the pp- \neus directly are GALLEX
and SAGE, using \gaes as
target material. The underlying reaction is 
\be
^{71}Ga + \nu_e \rightarrow ^{71}Ge + e^-
\ee
with a threshold of 233 keV. The main difference between the two experiments lies in
the chemical
state of the gallium and therefore also in the extraction of the
produced \gees . While GALLEX is using 30 t of gallium in form of 
a 110 t GaCl$_3$ solution, SAGE is using 
about 57 tons of metallic gallium.
After extraction, \gees is converted into GeH$_4$ and filled together with Xe into
special miniaturised proportional counters. 
The detection relies on the Auger-electrons and X-rays from K and L-capture 
in the \gees decay producing two
lines at 10.37 keV and 1.2 keV. As in the chlorine-experiment, besides the energy information
also pulse
rise
time analysis is used and the counting is done inside a special low-level shielding. Both
experiments for the first time checked their overall efficiency by
using MCi \cref sources. The present results are \cite{abd97,kir97}
\bea
76.4 \pm 6.3 (stat.)^{+4.5}_{-4.9} (sys.) \quad SNU \quad
\mbox{GALLEX }\\
66.6 ^{+6.8}_{-7.1} (stat.) ^{+3.8}_{-4.0} (sys.) \quad SNU \quad
\mbox{SAGE}
\eea
with theoretical predictions of $137^{+8}_{-7}$ SNU \cite{bah97}, $123 \pm 8$ 
\cite{tur93b} and $115
\pm 6$ SNU \cite{dar96}. Clearly the experiments are far off. The individual 
runs for GALLEX are shown in Fig.
\ref{pic:garesults}.

\subsection{Solutions to the solar neutrino problem}
\begin{figure}
\begin{center}  
\begin{tabular}{cc}
\epsfig{file=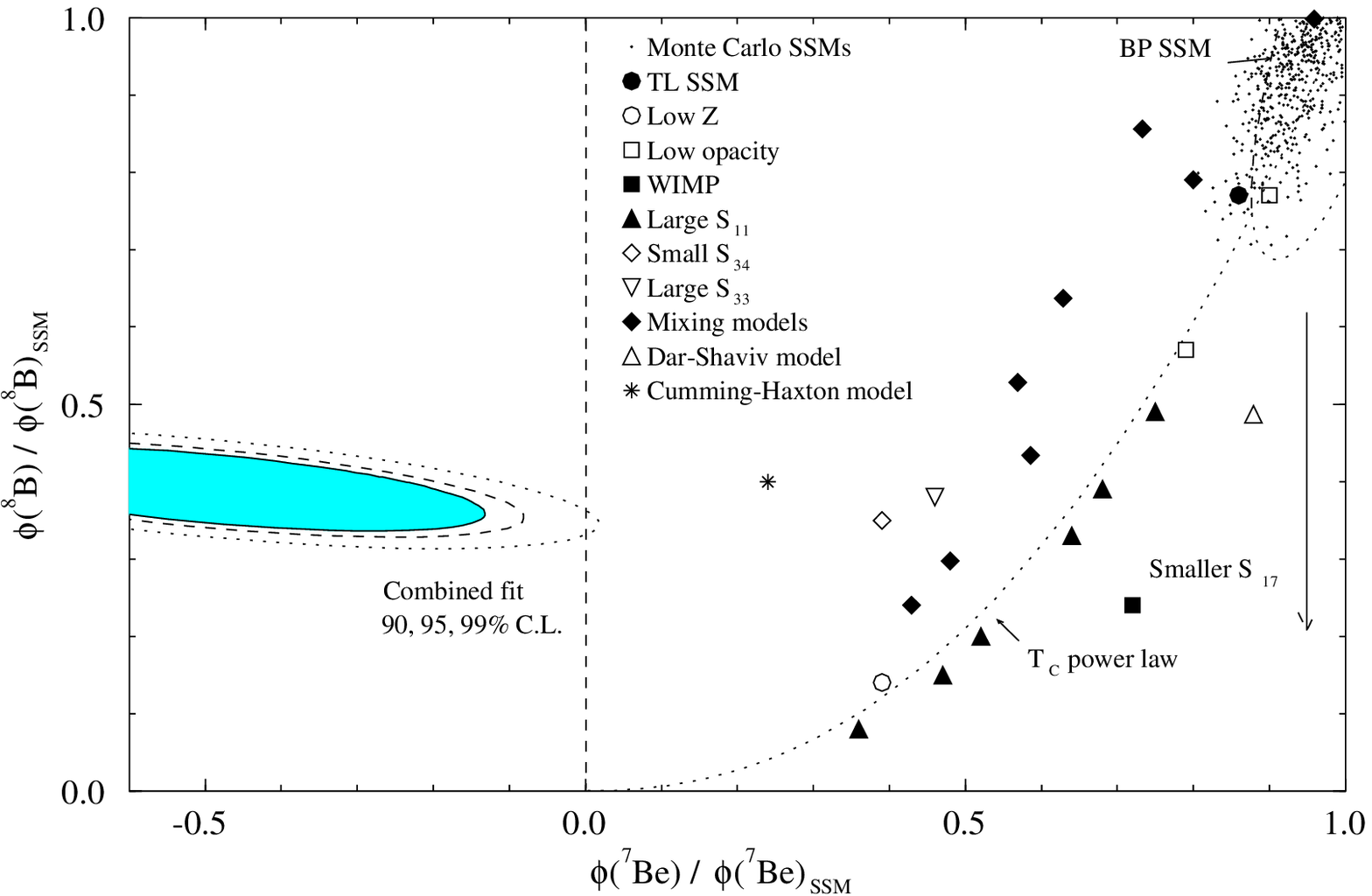,width=7cm,height=5cm} &
\epsfig{file=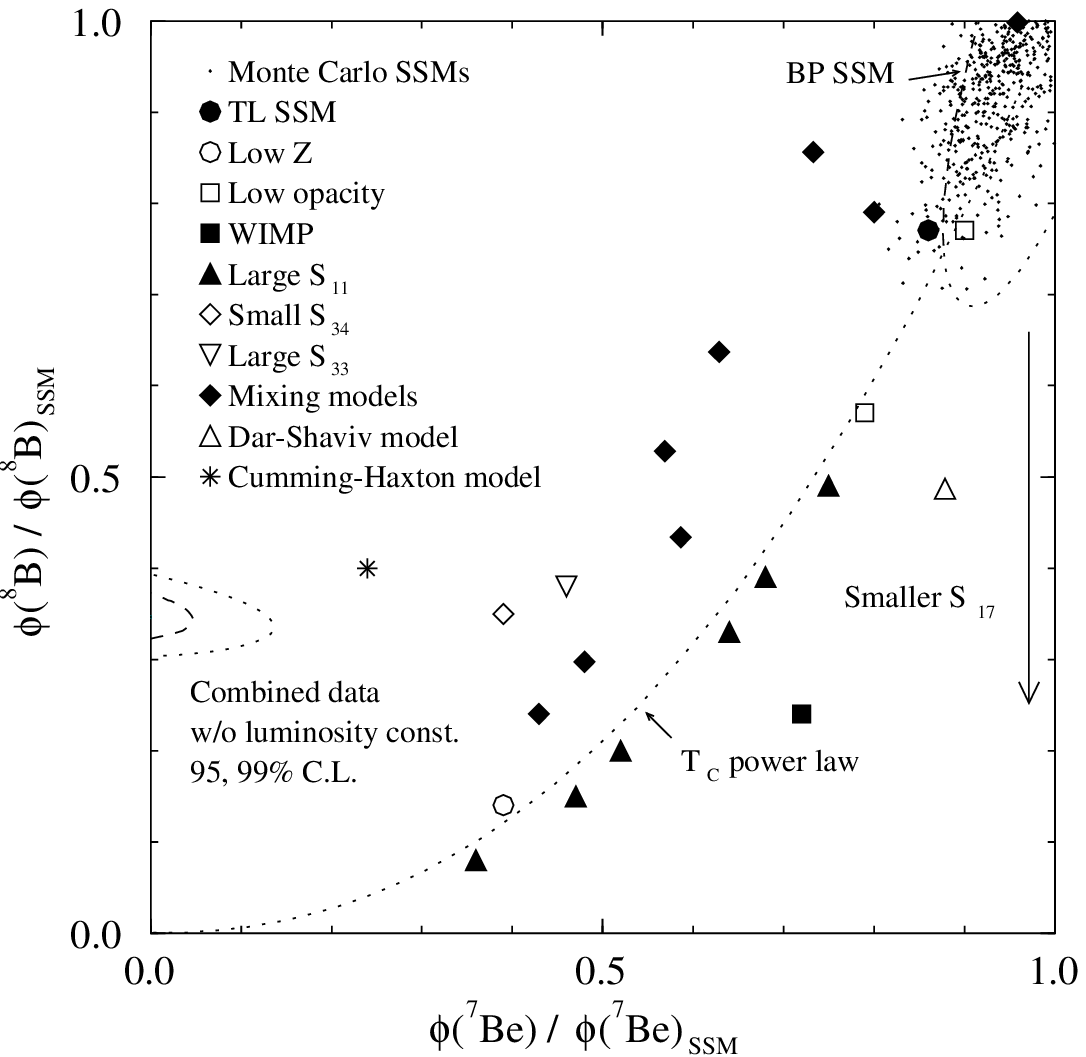,width=7cm,height=5cm}
\end{tabular}
\caption{ 
Constraints from the Cl,Ga and Cerenkov-\exps in comparison with several flux
predictions. In the upper right corner are the solar models from Bahcall and Pinnsoneault
\protect \cite{bah95}, the dashed
curve corresponds to a power law behaviour of the solar core temperature and the other
mentioned solar models are explained in Hata and Langacker \protect \cite{hat95}.
Reducing the $S_{17}$ factor
leads
to an evolution in the direction of the arrow. 
The fit to the experimental data results in negative values for 
the \bes flux (grey area). On the
right side no luminosity constraint is implemented (from \protect \cite{hat97}).}
\label{pic:fluxfit}
\end{center}
\end{figure}

The observed results split the solar \neu problem into three.
The first and original one is that the measured rate in the chlorine - experiment,
dominated by
\ba \neus , is less than the SSM prediction. 
This problem might be explained as an astrophysical solution by reducing the \ba - flux by a
temperature decrease in the solar core or by the involved
uncertainties in the nuclear cross sections. On the other hand \sk measures the \ba - flux and  
taking this value as a fact, the contribution of the \ba \neus
to the chlorine signal already exceeds the experimental value. This is independent of any solar
model. There is no astrophysical scenario which could distort the \ba beta 
spectrum in such a way
that both \exps are in agreement. Any possible deviation is shown to be at maximum at the
10$^{-5}$ level \cite{bah91}. There is almost no room for the \bes \neus , but \ba is produced
from \bes. The
third problem is, that the gallium \exps
do not allow any significant contribution beside the expected pp-value. Also here there is
almost no room
for the \bes \neus contributing more than 30 SNU in the SSM. A fit to all
available data, 
independent of a solar model,
is given by Hata et al. \cite{hat97} and shown in Fig. \ref{pic:fluxfit}. The best fit 
values achieved for the
fluxes are
$\Phi(\bes)/\Phi(\bes)_{SSM} 
= -0.6 \pm 0.4$ and  $\Phi(\ba)/\Phi(\ba)_{SSM} = 0.4 \pm 0.5$. Restricting the fluxes to
physical
regions ($\Phi>0$) changes the result to $\Phi(\bes)/\Phi(\bes)_{SSM} < 0.1$ and
$\Phi(\ba)/\Phi(\ba)_{SSM} = 0.38 \pm 0.05$ using the solar model of Bahcall 
and Pinnsoneault \cite{bah95}.
\subsubsection{Astrophysical and nuclear solutions}
\begin{figure}
\begin{center}   
\epsfig{file=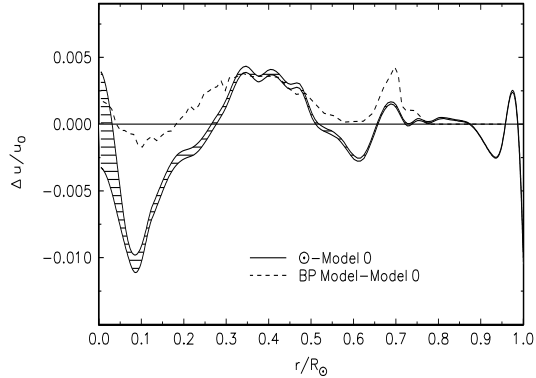,width=7cm,height=5cm}
\caption{Isothermal sound speed ($u=p/\rho$) profile compared with the
reference model of 
\protect \cite{dzi96} and the standard solar model of \protect \cite{bah95}. 
Shown is the deviation of sound
speed 
$\Delta u$ from the calculated value (from \protect \cite{dzi96}).}
\label{pic:soundsp}
\end{center}
\end{figure}

Typical astrophysical solutions try in some way to reduce the central
temperature in the sun to account for less high energetic \neus . For an overview
see \cite{bah89}.
Because of the strong
temperature dependence of the \ba -flux ($\propto T^{18}$), a reduction to 96 \% of
the SSM of BP value could explain the \sk data. However the ratio
$\Phi (\bes)/ \Phi (\ba) \propto T^{-10}$ increases in contrast to
experiments which basically imply no \bes \neus at all. This is the main
reason why \neu solutions are preferred. A way to circumvent this naive T
dependence is given by Cumming et al. \cite{cum96} assuming a slow mixing of the solar core on
timescales
characteristic 
of \hed equilibrium. The result is a remarkably different out-of equilibrium \hed -
profile in the
solar core leading to
two consequences: First of all, more helium is produced via 
the \hed(\hed,2p)\hev chain reducing the
\bes and \ba \neu flux and secondly, the short-living \bes is produced at higher temperature
favouring the \bes(p,$\gamma$)\ba reaction with respect 
to \bes(e,$\nu_e$)\lis. The combined effect
is a somewhat reduced \ba flux and a significantly reduced \bes flux.\\
All the
models experience
significant constraints from helioseismological data. The agreement between the measured and
calculated sound speed profiles are in good agreement even in the solar interior
(Fig. \ref{pic:soundsp}) \cite{cas97a}. In the region 
0.2 $< r/R_{\odot}< $ 0.65 the deviation from
expectation is less than 0.5 \% and even in the solar core it seems to be less than 4 \%. 

\subsubsection{Neutrino oscillations in matter}
\label{cha632}
\begin{figure}
\begin{center}  
\epsfig{file=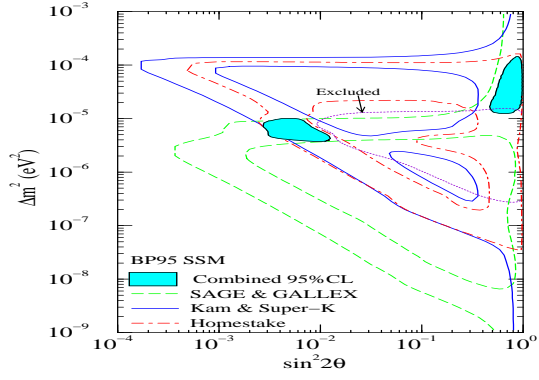,width=7cm,height=5cm}
\caption{MSW-solutions. The excluded region is given by the non-observation of any
day-night effect. The two islands (grey) are the allowed parameters in
agreement with all experimental data (from \protect \cite{hat97}). Also shown are
the allowed contours for
the single experiments.} 
\label{pic:mswsol}
\end{center}
\end{figure}

\begin{figure}
\begin{center}  
\epsfig{file=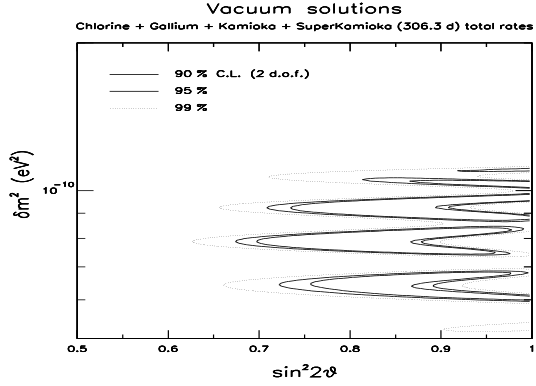,width=7cm,height=5cm}
\caption{\delm versus \sint plot if vacuum oscillations
are the explanation for the solar neutrino deficit (from \protect \cite{fog98}).}
\label{pic:vacsol}
\end{center}
\end{figure}

\begin{figure}
\begin{center}  
\epsfig{file=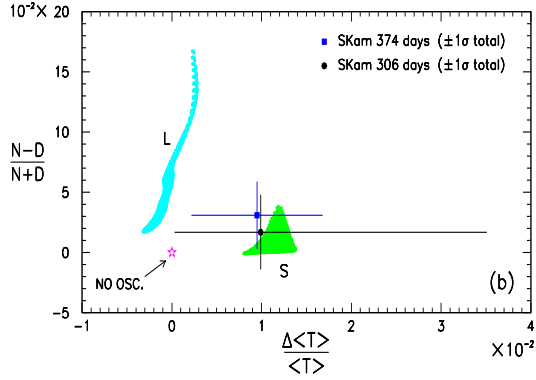,width=7cm,height=5cm}
\caption{Day-night effect versus the deviation of the electron recoil spectrum. By
an accurate measurement of these two flux-independent quantities it seems 
possible to distinguish the
large (L) and small (S) MSW solution. The no \osz scenario is marked as a star
(from \protect \cite{fog98}).}
\label{pic:fogli}
\end{center}
\end{figure}

An elegant solution to the \snp are \neu \oszs either in vacuum or in
matter (MSW-effect). The latter offers the chance to suppress the
\neus of intermediate energies completely, but leaving the low-energy \neus 
untouched and the high energy \neus only partly suppressed.
While in solar matter \nel can interact with electrons
via charged and neutral currents, other \neu flavours only have
neutral current interactions. This leads to different forward scattering amplitudes
making the mass eigenstates within the sun a
function of the electron density. For the \osz amplitude, a resonance
behaviour occurs, allowing maximal mixing even if the vacuum mixing angle 
is small \cite{wol78,mik86}. The resonance occurs at an electron
density $N_e$ of
\be
N_e = \frac{\delm cos 2 \theta}{2 \sqrt{2} G_F E}
\ee
In the adiabatic limit, where the electron density along the 
trajectory of the \neu changes slowly
enough, a full conversion is achieved, while in the non-adiabatic limit
there is a certain
\tran probability. In case of a linear change in electron density the probability is given by 
the Landau-Zener probability
\be
P (E) = exp (- \frac{\pi \delm \sint R_S}{E})
\ee
with $R_S = 6.6 \cdot 10^9$ m.  Because the running experiments have different
energy thresholds,
different
contours in the \delm - \sint plane arise, which overlap only in small regions.
Careful statistical analyses have been done by several authors
\cite{hat97,bah97a,lis97,bah98}.
The preferred solutions are a non-adiabatic or small
angle solution (SMA) at \delm $\approx 5 \cdot 10^{-6} eV^2$ and \sint
$\approx$ 0.008 and a large angle solution (LMA) at \delm $\approx
1.6 \cdot 10^{-5} eV^2$ and \sint $\approx$ 0.6 (Fig. \ref{pic:mswsol}). 
Also vacuum \oszs
(VAC) are not ruled out \cite{pet97}, giving parameters of \delm $\approx 6 \cdot
10^{-11} eV^2$ and \sint $\approx$ 0.9 (Fig. \ref{pic:vacsol}). A 
hybrid solution of vacuum \oszs and
MSW-effect seems possible in a three flavour scenario \cite{liu97}. 
The upcoming SNO detector will be able to distinguish between all
the different solutions. By looking for the electron recoil spectrum as
well as the ratio of charged currrent/ neutral current reactions,
they should in principle be able to distinguish the solutions with a
high confidence level. Also the upcoming radiochemical detectors
with their different energy thresholds provide further rejection
power between the different solutions.
The earth can regenerate some of the \neus for certain
parameter values. By looking for the resulting day(D)-night(N) effect, 
\sk gives a value of
\be
\frac{N-D}{N+D} \times 100  = -2.3 \pm 2.0 (stat.) \pm 1.4 (sys.)
\ee
showing no hints for such an effect and excluding certain parameter regions in
Fig. \ref{pic:mswsol}. Because
of the long
exposure time of the radiochemical 
experiments, they are not able to search for this effect. 
Including a second flux independent quantity 
$\Delta \langle T \rangle /\langle T \rangle$, the deviation
of the average measured kinetic energy of the electrons from the standard value, Fogli et al.
\cite{fog98}
obtain from the binned spectrum of \sk 
\be
\Delta \langle T \rangle /\langle T \rangle \times 100 = 0.95 \pm 0.73 \quad (1 \sigma \quad total)
\ee
resulting in a small preference for the SMA-solution (Fig.\ref{pic:fogli}).

\subsubsection{Neutrino magnetic moments in matter}
\begin{figure}
\begin{center}  
\epsfig{file=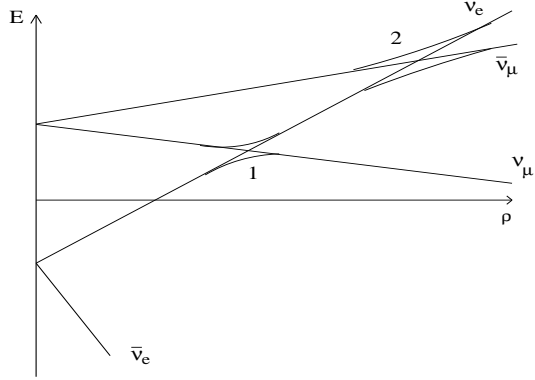,width=7cm,height=5cm}
\caption{Two resonances can occur for \neus coming
from the solar interior. First they experience the RSFP resonance and
afterwards, if the conversion is not complete, they go through the MSW
resonance.}
\label{pic:resonances}
\end{center}
\end{figure}

A similar resonance behaviour can also occur if the solar \neu problem is 
solved by a \neu magnetic moment. 
The pure spin-flavour precession $\nu_{eL} \ra \nu_{eR}$ in the solar magnetic field
cannot explain
the data because it results
in an energy independent suppression. By allowing spin-flavour \trans like $\nu_{eL} \ra
\bar{\nu}_{\mu R}$
it has been shown that a resonance behaviour can occur \cite{mar88}. The
transition probability can be written
as \cite{akh97}
\be
P(\nu_{eL} \ra {\bar \nu}_{\mu R}; r) = \frac{(2\mu B_\perp)^2}{(\delm /2E - 
\sqrt{2}G_F N_{eff})^2 +
(2\mu B_\perp)^2} sin^2(\frac{1}{2} \sqrt{D} r)
\ee
where $D$ is the denominator of the pre-sine factor and $N_{eff}$ is given 
by $N_e - N_n/2$ (Dirac)
and $N_e - N_n$ (Majorana) respectively. The most general case is 
the occurence of two resonances
(Fig. \ref{pic:resonances}).
By transversing the sun, \neus first undergo the spin-flavour precession and afterwards the
MSW-resonance. Depending on the involved \delm and E, the predicted 
conversion probability can be
quite complicated and detailed predictions for the \exps depend on the chosen parameters. It is
interesting to note, that in case of adiabacity in the resonant
spin-flavour precession scenario a MSW resonance never will occur. Assuming a maximal magnetic
field within the sun between 25 -50 kG and a \mamo of \munu = $10^{-11}$ \mub the observed data
can be explained if \delm is within a region of $4 \cdot 10^{-9} - 2 \cdot 
10^{-8} eV^2$. In case
both mechanisms are at work the \delm - region is shifted to $10^{-7} -10^{-8} 
eV^2$ and \sint $<$
0.25 by allowing a maximal magnetic field between 15-30 kG. Support for this scenario could come
from the detection of solar \bnel which can be produced via $\nu_{eL} \ra \bar{\nu}_{\mu R} \ra
\bar{\nu}_{eR}$. A detailed discussion can be found in \cite{akh97}.

\subsection{Future solar neutrino experiments}
From the discussion in the last section it seems obvious that the investigation 
of the \bes region needs
special attention. At present several \exps are under construction of which
some should be able to
investigate this region as well as finally solve the problem. 
\subsubsection{\SNO}
\begin{figure}
\begin{center}   
\epsfig{file=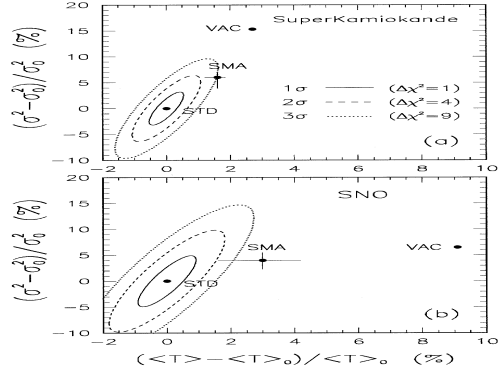,width=7cm,height=5cm}
\caption{Statistical power of \sk and SNO to distinguish between the
different solutions of the solar \neu problem. Shown are the moments of the
electron sprectrum (for details see  \protect \cite{bah97a}). The SMA can be
checked at the 3$\sigma$ level, the error bars correspond to the experimentally 
allowed MSW-region (with kind permission of J.N. Bahcall).}
\label{pic:snosigma}
\end{center}
\end{figure}

The next real-time solar \neu \expe which will be online is the \SNO (SNO). 
This detector will use
1000 t of D$_2$O and is installed in the Creighton mine near Sudbury,
Ontario. The big advantage of this
\expe is a model independent test of the \osz hypothesis by using weak neutral and
charged currents. The detection reactions are
\bea
\nu_e + d \rightarrow e^- + p + p \\
\label{eq:snonc} \nu + d \rightarrow \nu + p + n \\
\nu + e \ra \nu + e
\eea
While the first reaction is flavour sensitive, the second is not. 
To detect the neutron in the second reaction, two strategies are envisaged: Cl 
will be added to the
heavy water, to use
the $^{35}Cl(n,\gamma)^{36}Cl$ process and/or a set of He-filled proportional counters 
will be deployed. The
threshold will be
around 5-6 MeV and
start of measuring is expected in 1998. Beside the NC/CC ratio, the measured electron recoil
spectrum of 
the \ba \neus will provide strong discrimination power among the different scenarios
(Fig.\ref{pic:snosigma}).
\subsubsection{BOREXINO}
An experiment especially designed to explore the intermediate region
containing the \bes line at 862 keV is BOREXINO. It will use 300 t of
scintillator. The detection relies on
neutrino-electron scattering and the detector will be sensitive 
to \neu energies larger than 450 keV.
Of special importance is the produced ''Compton-Plateau'' from the 862 keV line with an expected
event rate of 50 per day
according to the SSM or about 10 per day in case of the MSW-effect. The detector is extremely
sensitive to impurities requiring a background level
of $10^{-16}$ g(U,Th)/g. The ability to achieve such low background levels 
could be demonstrated in a smaller pilot experiment (CTF).
BOREXINO is currently installed in the Gran Sasso Laboratory. Data taking 
should start around 2001. 
\subsubsection{ICARUS}
The \ica \expe plans to use in full scale about 3000t LAr in form of a TPC for solar \neu detection.
The
technique
offers precise tracking with high resolution dE/dx measurement, full event 
reconstruction in a bubble chamber-like quality and full sampling electromagnetic and
hadronic calorimetry. 
Besides neutrino-electron scattering
with a possible threshold of about 5 MeV, also the capture to the isotopic analogue state
in $^{40}K$  
\be
\nel + \arv \ra e^- + ^{40}K^{\ast}
\ee
will be used. The threshold for this reaction will be 5.9 MeV and allows 
the detection of photons
with a sum energy of 4.38 MeV from the de-excitation of $^{40}K$ in coincidence 
with the electron. By
comparing both
reaction rates, it is possible to get direct informations on the \osz hypothesis. A first 600t
module will be installed in Gran Sasso Laboratory 1999.
\subsubsection{HELLAZ}
A large real time \expe even for the detection of pp-\neus is the
proposed HELLAZ experiment. It will consist of a 12 t high pressure helium TPC
at liquid nitrogen temperature. A smaller prototype for demonstration of 
the basic principle exists.
\subsubsection{$^{127}I$}
A new detector relying on \ihsz is at present installed in the Homestake mine near the
chlorine \expe \cite{lan97}. The detection reaction is
\be
\nu_e + ^{127}I \rightarrow e^- +^{127}Xe
\ee
with a threshold of 789 keV. All the experimental mechanisms like extraction and detection
are similar to the chlorine \expe . The extraction will happen in two cycles (day/night) to two
different charcoal traps with an accumulation time of about 1 month. $^{127}Xe$ will decay to
excited states of $^{127}I$ making a coincidence detection of the gamma with the Auger electrons
possible. The detector will use 235 t of NaI solution with a total of 100 t iodine and
will be in
operation soon. The expected event rate is about 36.4 SNU where 14 SNU results from
\bes \neus . In contrast to the chlorine \expe , there is no bound isotopic
analogue state in the
I-Xe
system, increasing the relative contribution from \bes with respect to \ba.
\subsubsection{$^7Li$}
In comparison with chlorine it might be interesting to envisage a \lis experiment. Both
have nearly identical thresholds but the contribution of \bes and \ba \neus to the
signal is quite different. Such a detector is using the reaction
\be
\lis + \nel \ra \bes + e^-
\ee
and is currently under construction \cite{gal97}. The energy threshold is 860 keV making
this detector
sensitive to \bes and \ba \neusp The plan is to use 10 tons of metallic
Li. The solubility of \bes in \lis decreases with falling temperature, making an
extraction with cooled traps possible. Only 10.4 \% of EC from \bes produce a 478 keV photon
which could be used for detection by conventional techniques. Therefore cryogenic detectors
are necessary to measure the Auger electrons and nuclear recoil adding to 
an energy of 112 eV.
A prototype of 300 kg Li is constructed and presently under investigation. The predicted rate is
60.8$^{+7}_{-6}$ SNU \cite{gal97}.
\subsubsection{$^{176}Yb$}
A different approach to use low threshold detectors allowing solar \neu spectroscopy is
given by Raghavan \cite{rag97}. The idea is to use an isotope I and populate excited
states in the
neighbour isotope F which
will decay after a short time (10 - 100 ns) with $\gamma$-emission 
($\nu_e + I \ra e^- + F^{\ast}
\ra
F + \gamma$). This makes the use of
delayed coincidence techniques possible. To prevent the mother isotope from
decaying by single beta 
decay, the idea is to use double beta decay candidates like \seza , \gdhs or \yhss . By
using different excited states, it is even possible to compare different contributions of
the solar
\neu flux.
A 15 \% Yb loaded scintillator could be successfully created,
still fulfilling
all experimental requirements. A 100t scintillation detector containing 10t Yb 
is taken into consideration.
\subsubsection{$^{131}Xe$}
Another idea is to use the reaction
\be
\xehed + \nel \ra \cshed + e^-
\ee
The threshold would be 352 keV and the expected rate is about 45 SNU \cite{geo97}. For the
different
contributions see tab. \ref{tab:snuradio}. A 1 kton detector would result
in a detection rate of 1500
events/year
according to the SSM, where \bes \neus would contribute 37 \% of the signal. 
A liquid xenon detector
like those proposed for dark matter searches would be the appropriate choice.
\subsubsection{\gno}
Because GALLEX is finished and the wish to continue with measurements of pp-\neus over
the next decade, it was decided to continue as Gallium Neutrino Observatory (GNO). For the
first 2 years it will continue with the same setup as
GALLEX but then an increase of the gallium mass to 60 t and even 100 t as well as technical
improvements are foreseen.

\section{Astrophysical aspects of \neus}
\label{cha7}
\subsection{Neutrinos from \sne}
\begin{figure}  
\begin{center}  
\epsfig{file=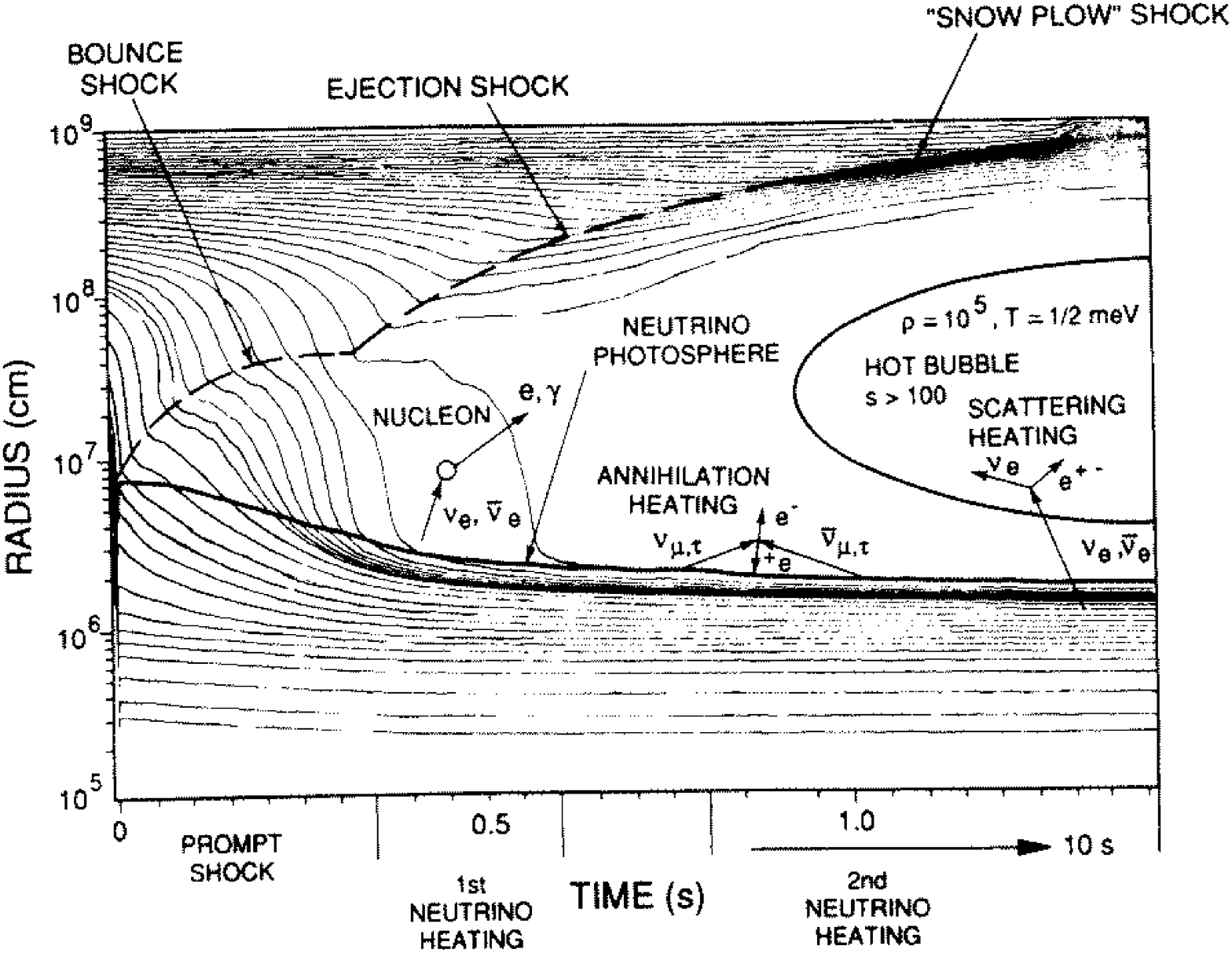,width=8cm,height=6cm}    
\caption{Time development of a type II
supernova explosion after core bounce. As the nuclear matter is over-compressed in the collapse, a
rebounce occurs and produces a shock wave (from \protect \cite{col90}).}  
\label{pic:snexpl}
\end{center}  
\end{figure}

Among the most violent stellar events are \sn explosions. Supernovae are products of the late
phase of stellar
evolution and can be divided in two classes. Supernova Type Ia are C,O white dwarfs in
binary systems, which 
exceed the Chandrasekhar-mass 
\be
M_{Ch} = 5.72 Y_e^2 M_{\odot}
\ee
by accreting matter from a companion, leading to a thermal deflagration of
the white dwarf. With this type of \sn no \neu emission and hydrogen-lines are connected.
Stars with masses
of $M \stackrel{>}{\sim} 8 \msun$ burn nuclear fuel up to iron group elements. If 
the iron core passes the
Chandrasekhar-mass,
it will collapse to a neutron star or even black hole, creating a \sn explosion. Because of
the ejected
outer hydrogen shell, hydrogen-lines can be observed and the \sn is called Type II. For a more
detailed
classification scheme see \cite{whe90}. Detailed discussions on the mechanism of \sn
explosions can be found in 
\cite{woo86,woo90}, only the principal scheme relevant for \neu
physics is outlined here.
Typical
values at the beginning of the
collapse are a central density of $\rho \approx 4 \cdot 10^9 gcm^{-3}$, 
a temperature of $8 \cdot 10^9$ K
and an electron per nucleon fraction of $Y_e \approx 0.42$. The gravitational force is
basically balanced by the pressure
of degenerated electrons. Photo-disintegration of iron via 
\be
^{56}Fe \rightarrow 13 ^4He + 4n - 124.4 \quad MeV 
\ee
and electron capture on free protons and on heavy nuclei
\be
\label{eq:delepto}e^- + p \rightarrow n + \nu_e , \quad e^- + ^ZA \rightarrow ^{Z+1}A + \nu_e
\ee
reduce the electron density.
Therefore the star loses its pressure support and it collapses.
This collapse stops when the iron core reaches nuclear density because of the
now stiff \eos.
Because of an overshoot this part of the core bounces back with an energy depending 
on the unknown \eos for overdense nuclear matter. The outer part of the 
core still continues to fall in,
thus producing a
pressure discontinuity at the sonic point, which develops into an outgoing shock wave.
Depending on the
energy of the shock wave, it is able to escape the iron core and to create the explosion
(''direct explosion
mechanism'') or it stalls in the core and needs some additional energy
input to be successful ("delayed
explosion mechanism") (Fig.\ref{pic:snexpl}). The total binding energy 
released in such an event is of
the order $5 \cdot
10^{53}$ erg, where \neus carry away about 99 \%.
\subsubsection{General remarks}
\begin{figure}
\begin{center}
\epsfig{file=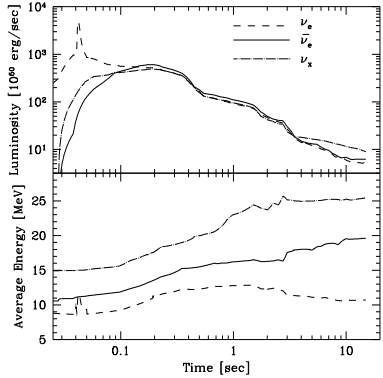,width=5cm,height=5cm}  
\caption{Time evolution of the \neu luminosity and
average energy of a 
\sn explosion model from \protect \cite{tot97a}. The core bounce is 3-4 ms before
the neutronisation burst of \nel (from \protect \cite{tot97a}).}
\label{pic:neutime}
\end{center}
\end{figure}

The emitted \neu spectrum consists basically of two parts. The first one is a 
peak of \nel resulting 
from
the deleptonisation period (eq. \ref{eq:delepto}) and lasting only a few ms. During the collapse
phase the core becomes opaque
even for \neus and they diffuse within the core. The outgoing shock wave dissociates the
infalling iron
nuclei, increasing the mean free path for the \neus which pile up behind the 
shock. When the shock
traverses the neutrinosphere (defined in a similar sense like the photosphere of the
sun) all this
\nel will be emitted at once. The mean energy is relatively high $\langle E_{\nu} \rangle
\approx 15 $MeV, but
the total energy released is only about $10^{51}$ erg. The second contribution comes from the
Kelvin-Helmholtz cooling phase of the proto-neutron star acting as blackbody 
source and producing \neus
dominantly by bremsstrahlung. 
All flavours are emitted in more or less equal
numbers (Fig.\ref{pic:neutime}) within a few seconds. Because \nmu and
\ntau
have no charged current interactions they have a lower opacity and decouple 
at higher temperature and
density. Also the opacity for \bnel is lower than for \nel because less protons are
available and
the opacity is dominated by $\nel + n \ra p + e^-$ and $\bnel + p \ra n + e^+$ respectively.
Therefore one typically finds $\langle \enu \rangle$ = 10 - 12 MeV (\nel), 
$\langle \enu \rangle$ = 14 -
17 MeV (\bnel) and
$\langle \enu \rangle$ = 24 - 27 MeV
(\nmu,\ntau). Because the energy is approximately equipartitioned 
between the flavours, the fluxes behave as $\Phi (\nel) >
\Phi (\bnel) > \Phi (\nmu,\ntau)$.
The \neu spectrum might well be described by a Fermi-Dirac shape
\be
\frac{dL_{\nu}}{d \enu} \propto \frac{\enu^3}{1+ exp (\enu/T_{\nu} - \eta_{\nu})}
\ee
where $\eta_{\nu}$ is a degeneracy parameter. A detailed study of the time evolution
of the \neu luminosity is rather
complex and depends on many parameters like the \eos , the mass of the 
collapsed core at bounce, the
amount of postbounce accretion and the temperature profile after collapse. 
So far the only event
which could be detected is \sna .\\
In addition past \sne could create a
relic \neu background which is discussed in \cite{mal97,har97}.
The total integrated flux predicted by Malaney\cite{mal97} using the
redshift evolution of HI-gas and the associated star formation rate is between 2-5.4 \cms
which is a factor of 10 less than previous estimates of
\cite{tot95,tot96}. The predicted
spectrum has a peak in the region 2-5 MeV which is significantly due to
\sne with redshifts larger than 1. Unfortunately the background of
terrestrial reactors, solar and atmospheric \neus will make a detection in
this region very unlikely. Experiments like \sk therefore have to search
in the region 15-40 MeV, where the theoretical predictions for the \sn background fluxes are
approximately
the
same.

\subsubsection{\sna}
\begin{figure}
\begin{center}  
\epsfig{file=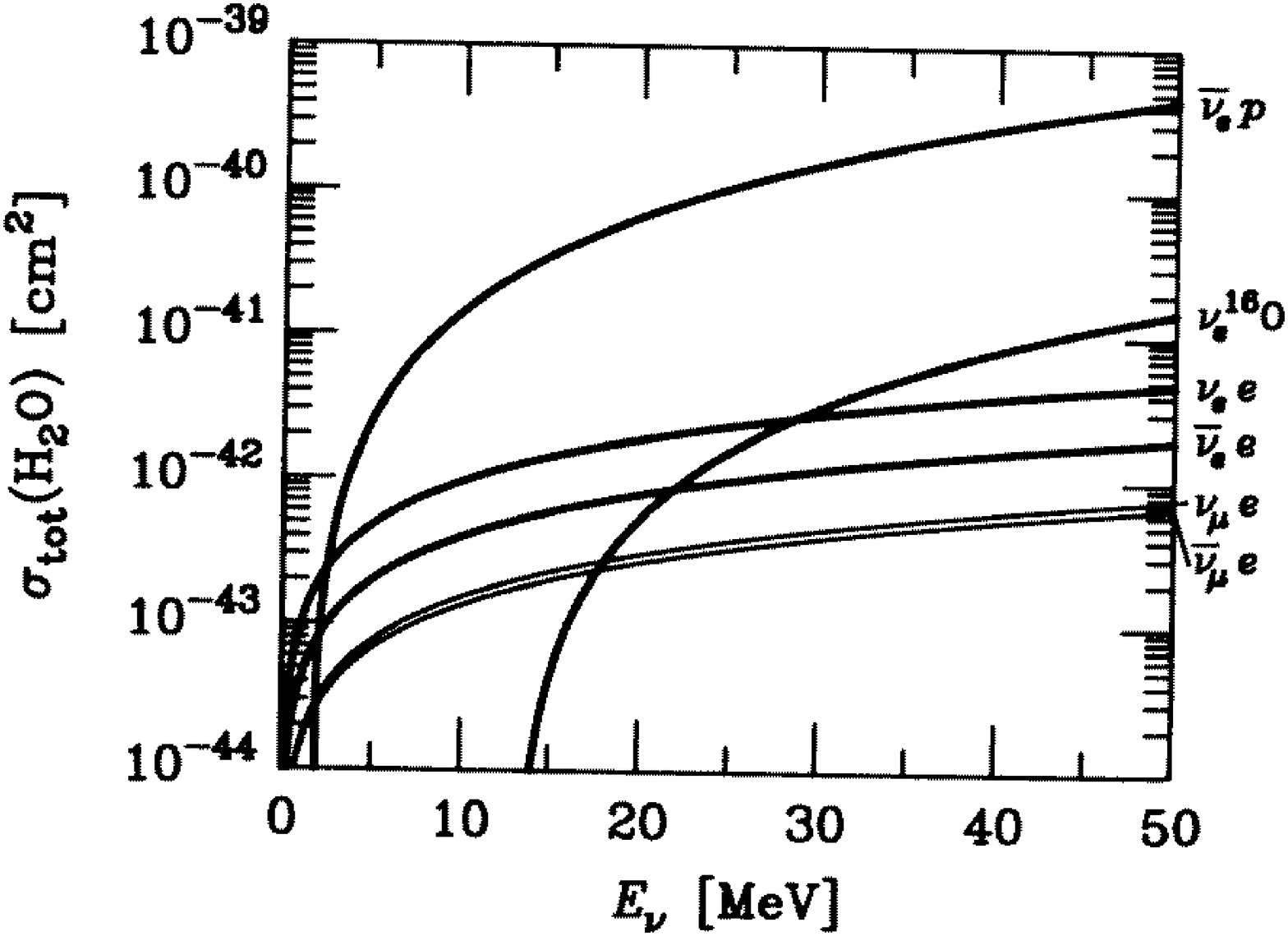,width=7cm,height=5cm}
\caption{Different cross sections involved in
the detection of \sn \neusp By far the largest cross section is $\bnel p \ra 
ne^+$ (from \protect \cite{raf96}).}
\label{pic:crosssections}
\end{center}
\end{figure}

\begin{figure}
\begin{center}  
\begin{tabular}{cc}
\epsfig{file=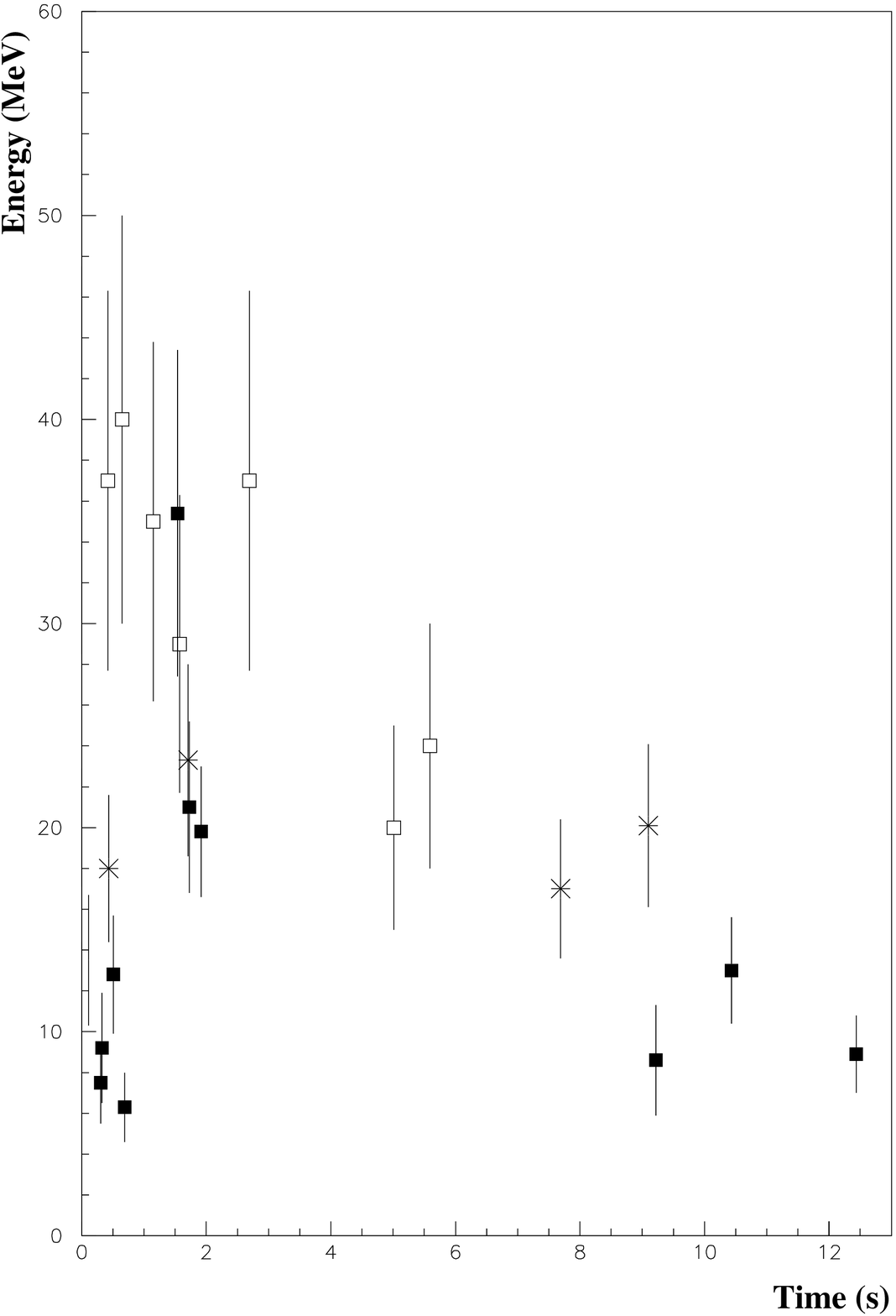,width=6cm,height=6cm} &
\epsfig{file=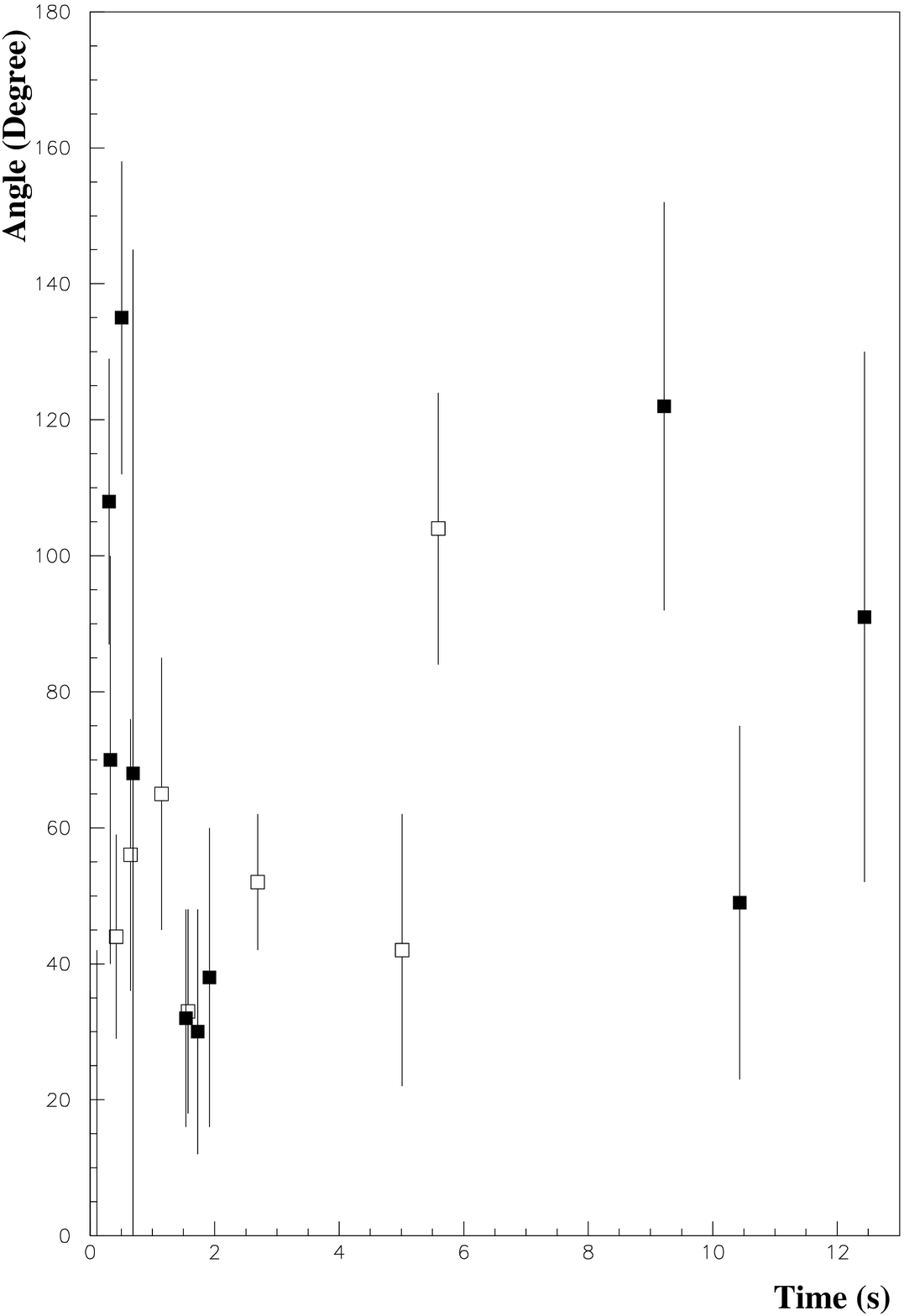,width=6cm,height=6cm} \\
\end{tabular}
\caption{Energy and observed angle as a function of
time for IMB (white), Kamiokande (black) and Baksan (stars). }
\label{pic:winkeldis}
\end{center}
\end{figure}

On February 23, 1987 the blue supergiant Sanduleak -69 202 exploded in the 
Large Magellanic Cloud 
at a distance of about 50
kpc \cite{arn89}.
For the first time
\neus could be detected, bringing theoretical \sn model calculations in the regime of
experimental verification. Four detectors claim observation of the \neu burst, namely
IMB \cite{bio87,bra88}, Kamiokande \cite{hir87,hir88}, the
Baksan
Scintillator Telescope \cite{ale88} and the Mont
Blanc Liquid Scintillator
detector \cite{agl88}. The Mont Blanc detection happened about 5 h earlier
than the detection
of the other \expsp Because of the relative low energy of the events, the non-observation
of any signal during this period in the much larger 
water-Cerenkov detectors and a missing astrophysical scenario for producing 
two \neu bursts, this
detection is normally considered as a background fluctuation.\\
The relevant interaction processes in the water detectors are $\bnel p \ra ne^+$, 
$\nu e - \nu e$
elastic scattering and $\nel \sfs \ra \ssz + e^-$ which becomes the dominant contribution
for
\nel for \enu
larger than about 30 MeV (Fig.\ref{pic:crosssections}).
For the
scintillator detectors the \sfs
reaction is absent but
at \enu larger than about 30 MeV the reaction $\nel \csz \ra \sszw + e^-$ contributes.
By far the largest cross section is $\bnel p \ra ne^+$
resulting in an isotropic event 
distribution and is suggesting that all observed events are due to \bnel interactions.\\
The observed numbers of \neus are 11 events within 12 s (Kamiokande), 
8 events in 5.5 s (IMB) and 5 events
in 14 s (Baksan). Some events were already attributed to background and are not included. 
A recent detailed maximum likelihood analysis was done by Loredo
and Lamb as described in \cite{raf96}. Their best-fit
values are 16.9 events plus 5.6 background for Kamiokande, 4 events 
at IMB and 1.8 plus 1 background at
Baksan. A two component cooling scheme consisting of the Kelvin Helmholtz 
cooling plus a low energy
component which mimics the neutrinos emitted during the stalled shock-phase 
results in a \neusph of 18
km and a total binding energy of $3.08 \cdot 10^{53}$ erg in good agreement with
theoretical expectations.\\
The observed signals contain some "anomalies". First of all there is a discrepancy in the \neu
energies between Kamiokande and IMB which imply a harder spectrum for IMB. 
Because of the rather high
threshold, IMB is sensitive to the high-energy tail of the assumed \neu spectrum, which
might have
substantial uncertainties. A second point is the large 7s gap between 
the first 8 and last 3 events of
Kamiokande. This might be a statistical fluctuation 
because IMB and Baksan do have events in this period.
The most striking is the deviation from isotropy which is expected 
if all events are due to \bnel
interactions (Fig.\ref{pic:winkeldis}). The observed distribution is only
at the per cent level in agreement with isotropy and an
explanation for this fact is still missing.
\\
Concerning \neu properties several things could be learned even by the
low statistics of the observed
event numbers. Because of the observed pulse length, mass effects on the propagation from
\sna to the earth
restrict the \neu mass to a conservative upper limit of about 25 eV, 
which could be improved under certain
model dependent assumptions down to 13 eV. Because the measured number 
of events corresponds to the
theoretical expectation, rather stringent bounds on \neu decay can be given, implying 
\be
\frac{\enu}{m_{\nu}} \tau_{\nu_e} \geq 5 \cdot 10^{12} s.
\ee
The radiative decay channel is additionally limited by the Solar Maximum Mission which
did not observe
any signal which could be related to the \neu burst of \sna \cite{chu89}. 
This was confirmed by a systematic
search for the radiative decay of \neus coming from Type II \sn using 
COMPTEL data \cite{mil96}. 
A stringent bound also exists on the electric charge of the \neusp 
An electric charge would cause
longer travel distances for lower energy \neus because effects
due to the galactic magnetic field are more important to them. Therefore the 
charge of the \neus can be
bounded by $Q_{\nu} < 10^{-17} e$.\\
Neutrino \oszs and the MSW-effect might also play a significant role in 
supernovae. Oscillations in the
channel
\nel - $\nu_X$ are able to reduce the prompt signal significantly. A detailed analysis,
assuming \delm
$\stackrel{<}{\sim} 3 \cdot 10^{-4} eV^2 (\enu/10 MeV)$ to allow for a
resonance outside the \neusph , reveals that a probability
for a conversion of more than 50 \% already occurs for 
$\delm sin^3 2 \theta \stackrel{>}{\sim} 4 \cdot
10^{-9} eV^2 (\enu/10 MeV)$ extending to large \delm values. Furthermore if the resonance
lies outside the
\neusph and within the shock wave (which is valid for \neu masses in the 
region of order 10 eV or
above and therefore a cosmological interesting region) the higher energy \nmu and \ntau , 
if converted to
\nel could help to revive the stalled shock \cite{ful92}. The minimum 
mixing angle necessary is
\sint \gsim $10^{-8}$
\enu /10 MeV. This would imply a reduction of the prompt \nel pulse. The \osz 
of \nel could furthermore 
influence the creation of r-process events. The hot bubble between the 
settled neutron star and the
escaping shock wave a few seconds after the core bounce seems to be a reliable 
place for r-process
nucleosynthesis. The p/n-ratio in the bubble is governed by the reactions $\nel n
\leftrightarrow pe^-$ and
$\bnel p \leftrightarrow n e^+$. Because \neus are much more common than electrons and
positrons, the p/n
ratio is ruled by the spectra and fluxes of the \neus. This normally results in a neutron rich
environment, because \bnel are more energetic than \nel. Oscillations of \nmu and \ntau to \nel
outside the \neusph could make the \nel flux more energetic than the \bnel . 
Even a 10 \% oscillation
effect would drive the medium to a proton-rich state. The parameters
obtained to allow the r-process also lie in
the cosmological region requiring \delm \gsim 2 eV and \sint \lsim
$10^{-5}$ for \nel - \nmu ,\ntau \oszs \cite{qia93}. At smaller \delm the \osz effect
has no impact on the r-process because it occurs at too large radii. 

\subsubsection{Experimental status}
The experimental observation of \sna launched several new searches
for \sn \neusp
Besides specially developed detectors, basically all new real-time solar 
\neu detectors like \sk,
ICARUS and SNO will be able to see such \neusp The predicted count rate for \sk of a
galactic \sn at
a distance of 10 kpc is 5000 - 10000 \bnel interactions! Such an event would open the
possibility to explore the mass of
\nmu and \ntau down to 50 eV by using the neutral current 
excitation of \sfs \cite{bea98}. SNO might be
especially
valuable because of the NC desintegration of D (eq. \ref{eq:snonc}) 
which will be dominated by \bnmu
and
\bntau . Detectors like LVD and MACRO also
offer additional discovery potential. The main component of all detectors will 
still be the \bnel
detection. A completely different scheme which is mainly sensitive to \nmu
and \ntau is the SNBO idea \cite{cli90} recently put into a realistic detector design in form of
OMNIS \cite{smi97}. The basic idea of \cite{cli90} is NC excitation of nuclei
\be
\nu A(Z,N) \ra A^{\ast}(Z,N) \ra A(Z,N-1) + \nu + n 
\ee
via the de-excitation by neutron emission. As target material large underground areas of
rock or
salt should be equipped with neutron detectors. The efficiency can be increased by using caverns
in
the rock for neutron detection. With about 200 tons of a Gd loaded scintillator, event rates of
more than 2000 \neu interactions for a galactic \sn in a distance 
of 8 kpc seem feasible. Because
of their higher energy, the signal would be dominated by \nmu and 
\ntau interactions. An extension
to
extragalactic \sne (an increase to about 4 Mpc in sensitivity would imply about 1 \sn per year)
unfortunately seems unrealistic at present times.

\subsection{Neutrinos from other astrophysical sources}
After describing experimentally observed astrophysical \neu
sources like the sun and \sne there might be other sources of \neus
of even higher energies ($E_{\nu} >100 MeV$). Their
existence may be closely related to the recently discovered TeV-
$\gamma$-sources and the sources of cosmic rays.
Neutrinos are produced via pp-collisions or photoproduction in cosmic beam-dump \exps 
due to the decay of the created charged pions and kaons. The associated production of
neutral pions allows a relation between expected photon and \neu fluxes
(eq.\ref{eq:flux}). 
The threshold for pion-photoproduction is 
rather high and the minimal proton energy required is given by
\be
E_p = \frac{(2m_p + m_{\pi})\cdot m_{\pi}}{4 E_{\gamma}} = 7 
\cdot 10^{16} (E_{\gamma} (eV))^{-1} (eV)
\ee
Using the cosmic microwave background (CMB) as target photons, a threshold
energy
for the proton of $E_p \approx 5 \cdot 10^{19}$ eV results
(GZK-cutoff) \cite{gre66,zat66}.
In contrast to photons and protons, \neu propagation is not influenced by
attenuation or deflection by magnetic fields. Neutrinos give direct information
on the source location and might reveal sources which have no $\gamma$-counterpart.
Even more, while high energy photons are influenced by $\gamma \gamma 
\ra e^+ e^-$ reactions on the
CMB and in the TeV-region by the same reaction with the presently unknown IR-background,
they have a limited range
for detection, whereas \neus can be observed to largest distances.\\
The main detection reaction on earth will be \ugm discussed in 
more detail in chapter \ref{cha722}. 
\subsubsection{Sources and predictions}
Several galactic and extragalactic sources are discussed for
highest energy \neus most of them are also investigated for creating and accelerating \crs. 
For a detailed discussion of sources see
\cite{gai95}.
The spectral shape of the primary \crs follows a power law according to
\be
\Phi(E) \propto E^{-(\gamma + 1)}
\ee
where the spectrum observed on earth is characterised by $\gamma
\approx 1.7$ (for E $< 10^{15}$ eV) then steepening to $\gamma 
\approx 2$ (''knee region'') and then
changing again at $10^{19}$ eV (''ankle'').
This spectrum is steeper than the accelerated one because of the energy
dependence of cosmic ray diffusion in the galaxy. Neutrinos produced in interactions 
with the interstellar gas should follow the shape of the primary cosmic ray spectrum up
to highest energies. On the other hand, if the production occurs at the acceleration
site,
there is no diffusion effect and the spectral shape follows the hard source spectrum
($\gamma \approx 2 - 2.2$). From the above mentioned interaction mechanisms it
is clear that
there are point sources in the sky and a diffuse component.
Two examples of possible galactic point sources are X-ray binaries (a
compact object like a neutron star or black hole is accreting
matter from a non-compact companion) like Cygnus X-3 or young
\sn remnants. To produce a detectable signal of a few upward
going muons per year in a $10^5 m^2$ detector, X-ray
binaries have to convert almost all energy in the acceleration of protons. The \sn
remnants on the other hand are rare events and typically produce a
signal only during a period on the scale of years after the explosion. A guaranteed source for
\neu
production is the galactic disc, where a diffuse
photon background due to interactions of cosmic ray
protons
with interstellar matter could be observed by EGRET on CGRO \cite{hun97}.
If \neus are coming from $\pi$-decay like the observed
photons the \neu flux can be determined by \cite{gai95}
\be
\label{eq:flux}
\Phi_{\nu} = C \cdot \left( 1-(\frac{m_{\mu}}{m_{\tau}})
\right)^{\alpha-1}
\frac{1}{1- A_{\gamma}}
\ee
where $A_{\gamma}$ is the energy dependent photon attenuation dominated by $\gamma\gamma \ra
e^+e^-$.
A second source is our sun, because of cosmic ray interactions
within the solar atmosphere. Moreover the sun could trap neutralinos $\chi$ as
dark matter in its interior. Their $\chi{\bar\chi}$-annihilation
can be a source of high energy \neusp
Predictions for the \neu -flux on earth for a 500 GeV
neutralino $\chi$
are of the order $\Phi_{\nu} \approx 2 \cdot 10^{-8} \cms$.\\
The most prominent extragalactic source candidates are active galactic
nuclei (AGN) and gamma ray bursters (GRBs). The present -here simplified- 
picture of \agn consists of a
central
supermassive black
hole ($\approx 10^8 -10^{10} \msun$) accreting matter near its Eddington limit. The
accelerated matter will
form a hot, dense plasma and will be partly sucked into the black hole and partly
redirected by magnetic
fields forming two opposite directed jets perpendicular to the 
accretion disc. In this scenario there are
basically two ways to accelerate particles by shock acceleration and to produce a \neu
flux. The first possibility is near the central engine as described in
\cite{gai95}. Energy losses take place due to processes like $p\gamma \ra \Delta^+ \ra n \pi^+$ and
$p \gamma \ra p+ e^+ + e^-$ in the radiation field as well as pp-collisions in the gas.
Both processes give rise to photons and \neus, but the produced photons from
$\pi^0$-decay
cascade down and are released as X-rays, because the central region is optically thick for
energies larger than $\sim$ 5 MeV. 
Crucial for this mechanism to work are the assumptions on proton
propagation and confinement in the core region. 
Note however, that the majority of X-ray emitting \agn does not show a nonthermal X-ray spectrum
(cascade origin) but a thermal spectrum peaking at $\sim$ 100 keV.
The second source which might explain the observation of
TeV-photons from seve\-ral
extragalactic sources like Markarian 421 are the highly relativistic jets. 
It is of outstanding interest to know whether the observed photons are
created from inverse Compton scattering or synchrotron emission of accelerated electrons or from
photoproduction of pions from accelerated protons \cite{man98}.
The observation of \neus would help to clarify the situation.\\
By integrating over all cosmological AGN, one should also see a diffuse 
background of \neus in the
same way as the diffuse $\gamma$-ray background is obtained. 
According to Stecker \cite{ste91} it remains
flat up to about
$10^7$ GeV and then starts to fall steeply. For energies larger than about $3 \cdot 10^4$ GeV it becomes dominant with respect to atmospheric \neusp 
While different models agree more or less
in their prediction at the high energy end of $\approx 10^{10}$ GeV, orders of magnitude
differences exist
in the lower energy region around $10^5$ GeV. Predicted fluxes of several 
models can be found in
\cite{nel93,sza94,man95,ber95,hil97}.
\begin{figure}
\begin{center}  
\epsfig{file=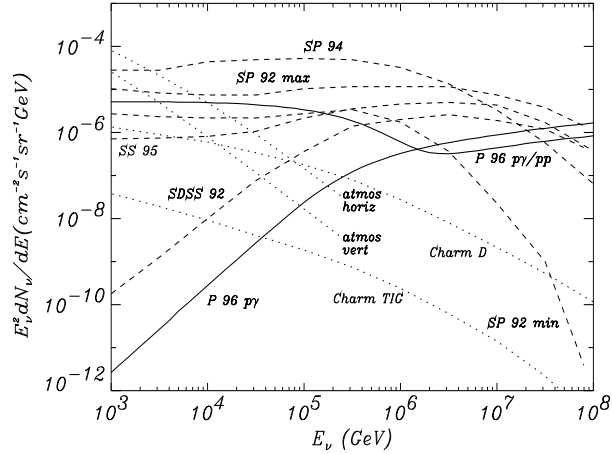,width=8cm,height=6cm}
\caption{Predicted diffuse isotropic \neu flux from the sum over all
active galactic nuclei predicted by various models. Also shown is the horizontal and vertical
atmospheric \neu flux, dominating in the region below 1 TeV. Two different 
models for prompt \neus
from charm decay of TIG \protect \cite{thu95} and model D of Zas et al.
\protect \cite{zas93} are also shown (from \protect \cite{hil97}).}
\label{pic:galneuspec}
\end{center}
\end{figure}
A combined high energy \neu spectrum for point and diffuse sources can be seen in Fig.
\ref{pic:galneuspec}. The typical estimates for their detection via \ugm
are in the order of 0.1-10 events per year for a 0.1
km$^2$ detector with rather large uncertainties. A flux limit of $d\Phi/dE_{\nu} < 7
\cdot 10^{-13} GeV^{-1} \cms sr^{-1} (90 \% CL)$ for \nmu in the energy region of
$E_{\nu} \approx $ 2.6 TeV exist from the Frejus-\expe \cite{rho96}, already ruling 
out the model of Szabo et al. \cite{sza94} 
and the maximal flux predictions of Bhattacharjee et al. \cite{bha92}.\\
Other partially more exotic sources of high energy \neus might exist. Neutrinos
associated with GRBs are discussed in Waxmann et al. \cite{wax97}. Further scenarios are
annihilation or decay of
superheavy particles like the neutralino $\chi$ \cite{ber96}. Also
evaporating black holes and radiation from topological defects 
like cosmic strings might be sources for
\neus \cite{bha92,sig97}. 

\subsubsection{Experimental search}
\label{cha722}
\begin{figure}
\begin{center}  
\begin{tabular}{cc}
\epsfig{file=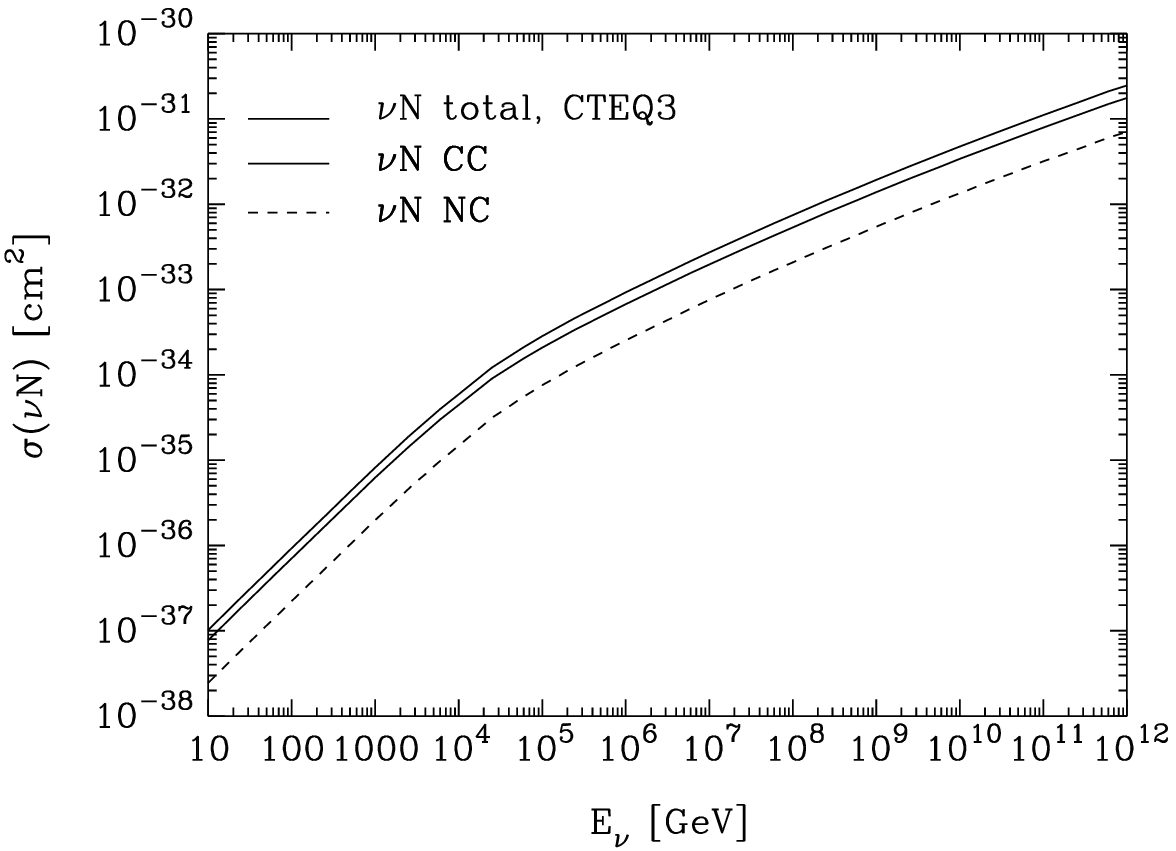,width=6cm,height=6cm} &
\epsfig{file=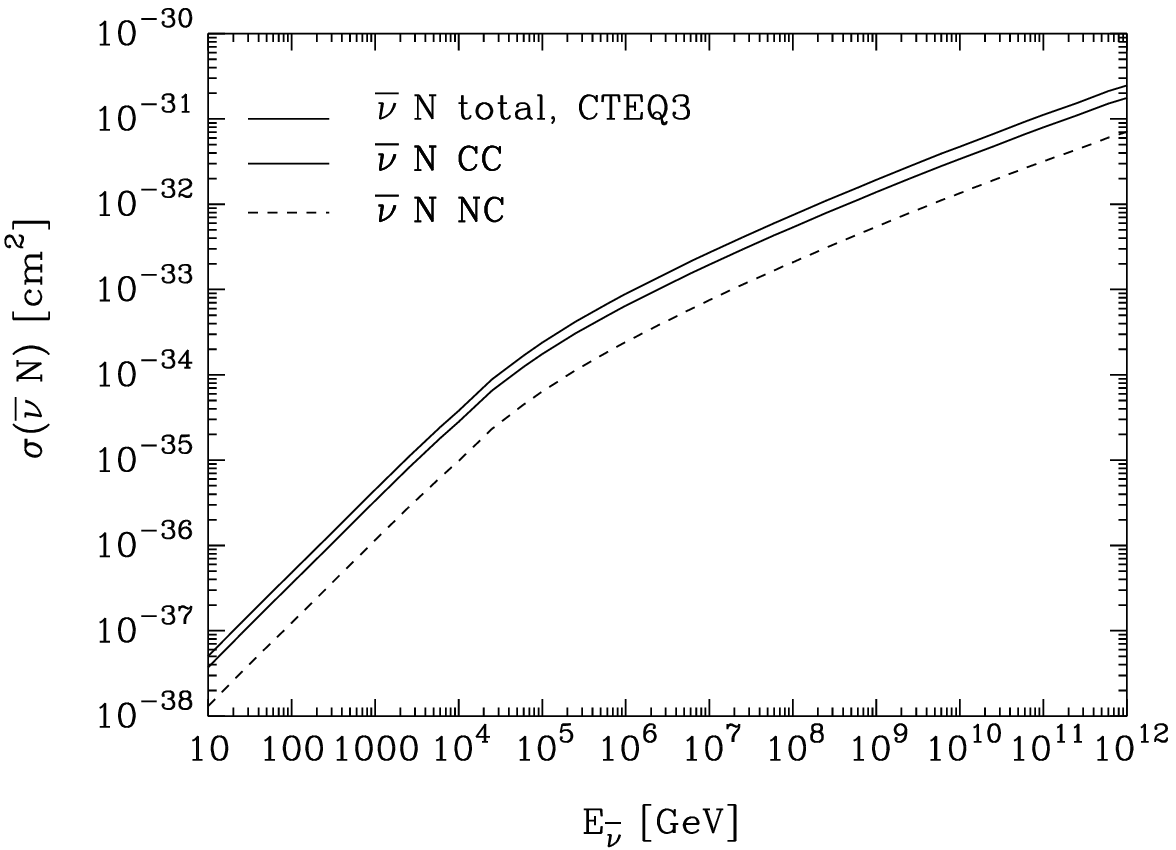,width=6cm,height=6cm} 
\end{tabular}
\end{center}
\caption{Total as well as NC and CC cross sections for
high enery \neus (left)
and antineutrinos (right) (from \protect \cite{gan96}).}
\label{pic:xsections}
\end{figure}

\begin{figure}
\begin{center}  
\epsfig{file=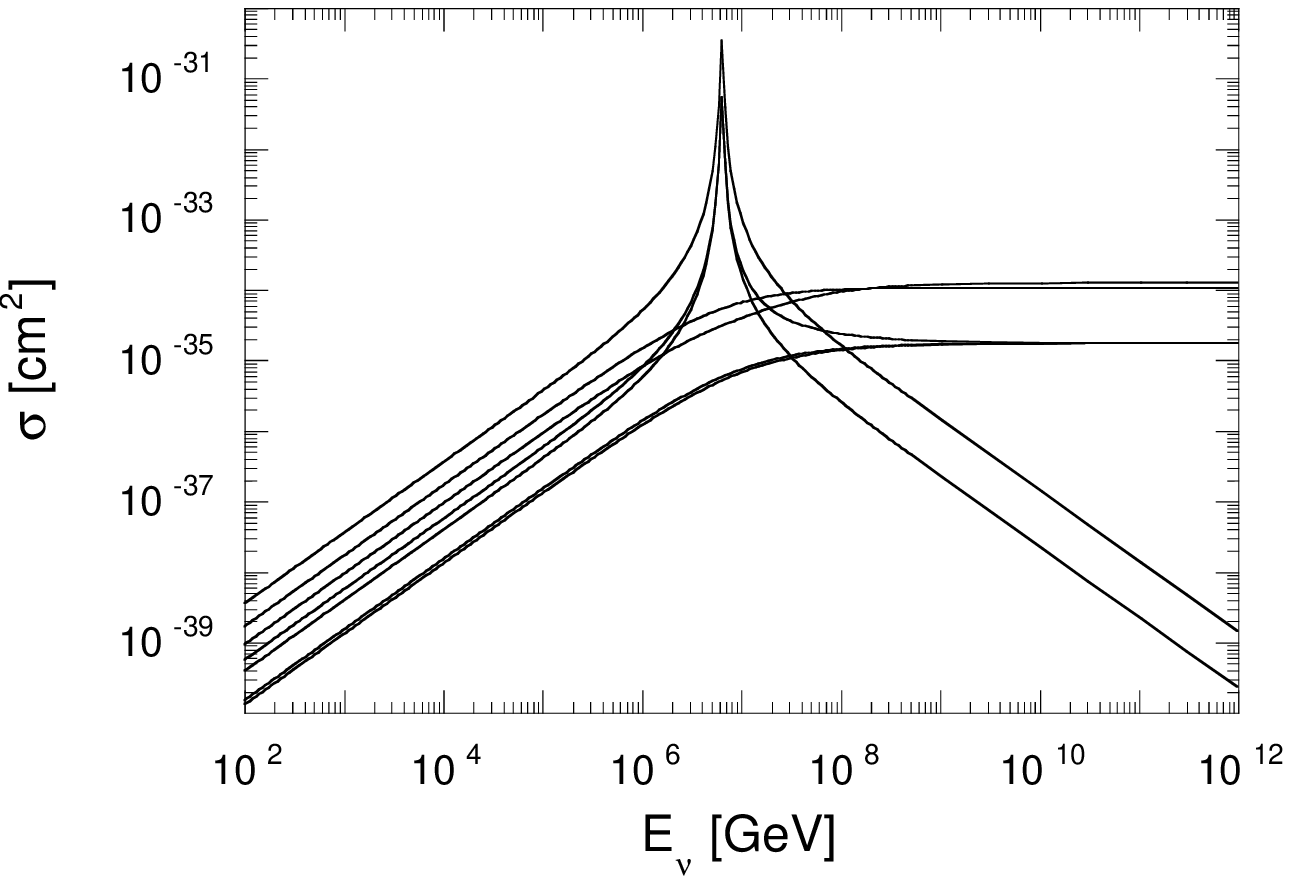,width=6cm,height=6cm} 
\end{center}
\caption{Glashow resonance in the $\bar{\nel} e$ cross section. The 
curves correspond, in the low energy region from
highest to lowest, to $(i) \bar{\nel} e \ra$ hadrons $(ii) \nmu e \ra \mu
\nel{} 
(iii) \nel e \ra \nel e{} (iv) \bnel e \ra \bnmu \mu{} (v)
\bnel
e \ra \bnel e{} (vi) \nmu e \ra \nmu e{} (vii) \bnmu e \ra \bnmu e$ (from
\protect \cite{gan96}).}
\label{pic:glashow}
\end{figure}

From the flux estimates of the last chapter, it is
rather clear that very large detectors are required.  The proposed detection reaction for
the VHE-\neus are $\nmu$ CC reactions producing \ugm . 
The corresponding cross-sections are shown in Fig. \ref{pic:xsections}. The 
\nmu CC cross section is given by
\be
\frac{d^2 \sigma}{dx dy} = \frac{2 G_F^2 m \enu}{\pi} (\frac{m_W^2}{Q^2 +
m_W^2}) (x q(x,Q^2) + x \bar{q}(x,Q^2) (1-y)^2) 
\ee
with $q(x,Q^2), \bar{q}(x,Q^2)$ as the quark distribution 
functions, the Bjorken variable $x=Q^2/2m \nu$, $m$ 
the nucleon mass and
$y= (\enu - E_{\mu}) / \enu$. Towards higher energies the contribution from the
 presently unknown small x-region is becoming more important and
theoretical extrapolations have to be used \cite{gan96}. 
To face the background from
the atmospheric \neu flux, which is at lower energies typically a factor of $10^6$ 
larger than cosmic sources (but falls off
steeper towards higher energies), only muons transversing the detector from below can be used
(upward going muons). Therefore cosmic sources of interest have to stay 
below the horizon for a significant time.\\
The effective detector size for detecting \ugm can be
enlarged because the surrounding material can be used as an additional target. The energy
loss rate of muons due to ionisation and catastrophic losses like bremsstrahlung,
pair-production and hadroproduction is given by 
\be 
\frac{dE_{\mu}}{dX} = - \alpha - \frac{E_{\mu}}{\xi} 
\ee 
with the critical energy $\epsilon = \alpha \xi \approx 500$ GeV. This
results in a range R of ($\xi \approx$ 2 km.w.e.)
\be R \approx \xi ln ( 1 + \frac{E_0}{\epsilon} )
\ee
For high energy muons the radiation losses are dominant, resulting in a change 
of energy dependence of the range from linear to logarithmic.
For a muon with initial energy 
larger than 0.5 TeV the
range exceeds 1 km.
For \neu ener\-gies larger than about 40 TeV the interaction
length becomes smaller than the diameter of the earth resulting in a
shadowing effect.\\
The existing underground detectors like MACRO, \sk and LVD have too
small sizes to measure statistically significant signals. Therefore natural resources of
water and ice are used to detect the Cerenkov-light of the passing muon. After the
termination of DUMAND, the Lake Baikal \expe (water) and AMANDA (ice) are the most
advanced projects.\\
The Baikal \neu telescope (NT) is located in a depth of 1.1 km at a
distance of 3.6 km from the shore \cite{bel97}. The final setup of NT-200
is finished recently
and consists of 8 strings forming a heptagonal array of seven strings 
around a central string. It
consists of 192 pairs of 8'' \pmts arranged in a distance of 5 and 7.5 m along the
strings. This rather small spacing allows a relatively low energy threshold of about 10
GeV. Besides the low threshold the main advantages of the \expe are the relatively cheap
deployment of tubes, because the frozen lake offers a good platform for deployment and
Lake Baikal is a sweet-water sea containing no \kav whose decay would produce
background. The effective area is between $1000 - 5000 m^2$ depending
on energy and an increase of the effective area by a factor 20-50 
is under consideration. Clear \ugm have been observed.\\
AMANDA \cite{bar97d,bir97} is at present operating 300 8'' \pmts
in the antarctic ice in a depth of 1500 - 2000 m. The basically bubble-free ice offers
extraordinary optical properties, e.g. the absorption length for $\lambda \approx 500$ nm
is around 100 m and the scattering length about 25 m. The detector offers 
an effective area of $10^4 m^2$ with a mean
angular resolution of 2.5$^o$. An upgrade to a 1 km$^3$ detector (ICECUBE) which would
consist of about 5000 \pmts mounted on 80 strings with a spacing of about 100 m is under
consideration. Good candidates for \ugm have been observed by AMANDA.\\
Two further water
\exps in the Mediterranean are in a kind of preparation phase, namely NESTOR
\cite{nes95} near Greece,
in a depth of 3800 m and ANTARES \cite{bla97} near Toulon (France), in a depth
of 2000m.\\
Associated with two of the
projects mentioned above are
alternative \exps using different detection techniques. They are called RICE
\cite{all97} (together
with AMANDA) and SADCO \cite{ded97} (together with NESTOR).  They rely on
detection of radio and
acoustic signals produced if $E_{\nu} \stackrel{>}{\sim}$ 1 PeV. In the
acoustic case, the produced shower particles in the \nel - interaction lose
energy through
ionisation leading to a local heating and density change. The density change propagates as
sound waves through the medium and can be detected with an array of detectors like
hydrophones allowing also a reconstruction of the event by measuring the arrival times and
amplitudes. The second method uses the effect of coherent radio Cerenkov radiation which is
produced as long as the
wavelength is large with respect to the spatial extension of the shower \cite{zas92}.
The neutral pions
create an electromagnetic shower of size 1m in ice via their $\pi^0
\ra \gamma \gamma$
decay, therefore producing frequencies in the
region 100 Mhz - 1 GHz. \\ 
For detection of very high energetic \neus the reaction 
\be \bar{\nel} + e^- \ra W^- \ra hadrons
\ee
can be used which shows a resonance
behaviour (Glashow - resonance) at $s=m_W^2$, meaning $E_{\nu} = 6.3 \cdot 10^6$ GeV. The
$\bar{\nu}_e e$ cross section at the resonance is about a factor 30 larger than the
corresponding $\nu$N cross section (Fig. \ref{pic:glashow}).
The field of high energy \neu astrophysics is still in its initial phase
but will provide important information in the future.

\subsection{Relic \neus}
\begin{figure}
\begin{center}  
\epsfig{file=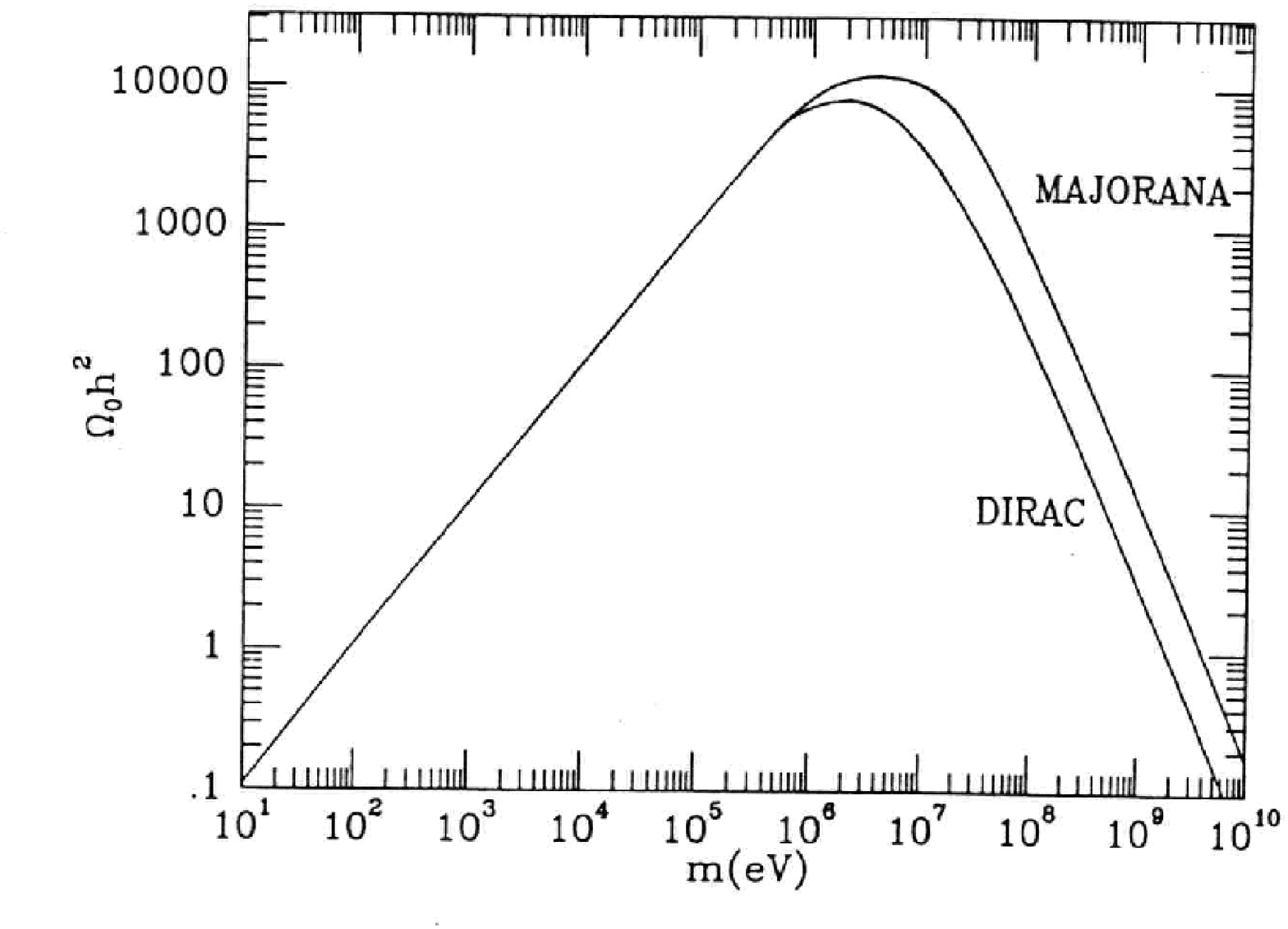,width=7cm,height=5cm}
\caption{Neutrino contribution to $\Omega$ as a
function of the \neu mass. Two regions remain for stable \neus to be
cosmological of interest. Either below 100 eV (acting as hot dark matter) or 
above 5 GeV (acting as cold dark matter). Otherwise they have to be unstable
because they would overclose the universe.} 
\label{pic:omeganeu}
\end{center}
\end{figure}

The thermal history of the universe according to the big bang model predicts not
only
a photon background but also a \neu background. While the photons are observed as the cosmic
microwave background, the \neu background is still undetected. The 
temperatures of both are related
by the relation 
\be
\tnu = (\frac{4}{11})^{\frac{1}{3}} \tg
\ee 
where \tg is measured quite accurately by COBE to be \tg = 2.728 $\pm$ 0.004
\cite{fix96}, thus predicting a \neu
background temperature of 1.95 K. The total number density and matter density 
(the flavour densities
are one third) are then given by
\bea
n_{\nu} = \frac{3 g_{\nu}}{22} n_{\gamma} = 337 cm^{-3}\\ 
\label{eq:rhonu} \rho_{\nu} = \frac{7 g_{\nu}}{8 g_{\gamma}} (\frac{4}{11})^{\frac{4}{3}}
\rho_{\gamma} = 0.178
\frac{eV}{cm^3}
\eea
with the statistical weights $g_{\gamma} =2, g_{\nu}$ = 2 for \majo \neus and 
light ($m_{\nu}$ \lsim 300
keV) Dirac \neus, otherwise $g_{\nu}$ = 4. The mean energy of the \neus 
today is $5.28 \cdot 10^{-4}$ eV
making a detection extremely difficult. The \neu contribution to the matter density is
given as
\be
\label{eq:omeganu}\Omega_{\nu} = \frac{\rho_{\nu}}{\rho_c} = 5.32 \cdot 10^{-3}
\frac{g_{\nu}}{h^2} \frac{m_{\nu}}{eV}
\ee
where $\rho_c$ is the critical density and $h=H_0 / 100 kms^{-1}Mpc^{-1}$ 
the normalised Hubble-constant.
For $m_{\nu}$ \gsim 1 MeV \neus become non-relativistic and their density is
suppressed by a 
Boltzmann-factor. The behaviour of $\Omega_{\nu}$ as a function of
$m_{\nu}$ is shown in Fig.\ref{pic:omeganeu}.
As can be seen, stable \neus only exist for $m_{\nu}$ \lsim 100 eV and for $m_{\nu}$ \gsim
2(5) GeV
in the Dirac (\majo) case (Lee-Weinberg-bound). Both allowed mass regions 
offer massive \neus as good dark
matter candidates, either as hot dark matter (\neus in the eV-range) 
or as cold dark matter (\neus in the
GeV range) \cite{kla97b}. Pure cold dark matter models predict to many galaxy
clusters,
which can be avoided by including a hot component. Because of their free
streaming in the early universe, light \neus wash out perturbations on
small
scales, reducing the power there. A mixed hot and cold dark matter model
with $\Omega=1$ and 5 eV \neus contributing $\Omega_{\nu} \approx 0.2$ for
$h=0.5$ seems to be a good description of the COBE normalised power
spectrum and the observed large scale structure. This can be
improved if the hot component consists of two \neu flavours having a mass
of about 2.5 eV \cite{pri95}. GeV \neus as cold dark matter are bounded by
\bb \exps excluding Dirac - \neus with standard interactions between 26
GeV and 4.7 TeV as the dominant component \cite{reu91,bec94}.
The linear contribution for light \neus to $\Omega_{\nu}$ (eq.\ref{eq:omeganu}) can be
converted in a \neu mass bound.
The condition not to overclose the universe requires for stable \neus
\be
\sum_i m_i (\frac{g_{\nu}}{2}) = 94 eV \Omega_{\nu} h^2
\ee
which is orders of magnitude more stringent for \nmu and \ntau than laboratory limits. 
The same condition also allows only certain ranges of lifetimes for unstable \neus\cite{moh91}.
Radiative decay channels are additionally
restricted because the created photons would
otherwise influence the thermodynamic evolution too strongly. Other decay channels might be
possible. The most common ones discussed for heavy \neu decays are
\bea
\nu_H \ra \nu_L + \gamma \\
\nu_H \ra \nu_L + l^+ l^- \quad l= e,\mu\\
\nu_H \ra \nu_L + \bar{\nu}_L + \nu_L \\
\nu_H \ra \nu_L + \chi
\eea
where $\chi$ corresponds to a light scalar like the majoron. Bounds on the radiative decay mode
exist from reactor \exps \cite{obe87}, LAMPF \cite{kra91} and from \sna \cite{blu92a}. The
decay mode
involving $e^+e^-$-pairs is restricted in the region 1-8 MeV by reactor data \cite{hag95} and at
higher energies by accelerator searches (see \cite{fei88}).\\
Experimental detection of the \neu background will be extremly difficult. One 
of the suggestions is to
take advantage of coherence in $\nu N$-scattering and by using cryogenic detectors, but
also this proposal is
far from practical realisation. A chance might be that eq.\ref{eq:rhonu} is modified during
decoupling in the early universe by incomplete
annihilation and finite temperature QED-effects, which might increase 
$\rho_{\nu}$ by about 1 \% \cite{dol97}. 
This effect might be detectable in the future satellite missions MAP and PLANCK by its influence
on the anisotropies in the cosmic microwave background \cite{lop98}.

\section{Conclusions and outlook}
The question whether \neus have a non-vanishing rest mass influences research
areas from particle physics up to cosmology, but is still an open question.
At present all hints for \neu masses are connected with \neu \osz effects namely
the solar \neu deficit, the atmospheric \neu anomaly and the evidence from \lsnd . 
The involved \delm scales are in the $10^{-5} (10^{-11})eV^2$
MSW-(vacuum)solution, $10^{-3} eV^2$ and
eV-region respectively. 
Several theoretical models have been developed to describe these evidences in a
consistent way \cite{car96,ack97}.
With the appearance of the new CHOOZ and \sk results most of these models
seem to be ruled out. Moreover to explain all data in addition to
the three standard \neus a new sterile \neu seems necessary
\cite{cal93,bil97,bar97,moh98}. The scheme of Barger et al. \cite{bar97}
proposes a nearly degenerate \nmu and \ntau in the eV-range and
much lighter \nel
and $\nu_S$. The splitting between the eV-states and the light ones is
determined by \lsnd data. The atmospheric
anomaly can be solved by \nmu - \ntau \oszs and both can act as hot dark
matter. The solar \neu problem is solved by \nel - $\nu_S$ \oszs which
require the $\nu_S$ to be slightly heavier than \nel.\\
The right answer will hopefully be given in the future, because of an increasing
number of new experiments.

\ack
I want to thank all my collegues for providing me with informations and help,
especially C.P. Burgess, T.K. Gaisser, C. G\"o$\beta$ling, W. Hampel, T.
Kirsten, H.V. Klapdor-Kleingrothaus, J.G. Learned, K. Mannheim, R.N.
Mohapatra, G. Raffelt, C. Spiering, C. Weinheimer and T. Ypsilantis 
for useful discussions and comments.


\begin{thebibliography}{99}
\bibitem{pau77} W. Pauli, F\"unf Arbeiten zum Ausschlie$\beta$ungsprinzip und zum Neutrino,
Texte zur Forschung Vol. 27, Wissenschaftliche Buchgesellschaft Darmstadt 1977, W. Pauli,
On the earlier and more recent history of the neutrino (1957) in: Neutrino physics, ed.
K. Winter, Cambridge University Press, 1991
\bibitem{rei56} F. Reines, C.L. Cowan, \Journal{\NAT}{178}{446}{1956}
\bibitem{dan62} G. Danby et al., \Journal{\PRL}{9}{36}{1962}
\bibitem{cer97} D. Abbaneo et al., CERN-PPE/97-154


\bibitem{wey29}H. Weyl, \Journal{\ZP}{56}{330}{1929}
\bibitem{maj37}E. Majorana, \Journal{\NCA}{14}{171}{1937}
\bibitem{bil87} S.M.Bilenky, S.T. Petcov, \Journal{\RMP}{59}{671}{1987},
\Journal{\RMP}{60}{575}{1988} (err),
\Journal{\RMP}{61}{169}{1989} (err)
\bibitem{gel78} M. Gell-Mann, P. Ramond, R. Slansky, in Proc. {\it Supergravity},
ed. P. van Nieuwenhuizen, D. Freedman, North Holland, 1979, p.315
\bibitem{yan80} T. Yanagida, \Journal{\PTP}{64}{1103}{1980}
\bibitem{lan88} P. Langacker in {\it Neutrinos} ed. H.V. Klapdor, Springer 1988
\bibitem{blu92} S. Bludman et al., \Journal{\PRD}{45}{1810}{1992}
\bibitem{moh91}R.N. Mohapatra, P. Pal, {\it Massive \neus in physics and astrophysics}, 
World Scientific, 1991
\bibitem{zee80} A. Zee, \Journal{\PLB}{93}{389}{1980}, \Journal{\PLB}{161}{141}{1985}
\bibitem{wol80}L. Wolfenstein, \Journal{\NPB}{175}{93}{1980}
\bibitem{bab88}K.S. Babu, \Journal{\PLB}{203}{132}{1988}
\bibitem{pat74}J.C. Pati, A. Salam, \Journal{\PRD}{10}{275}{1974}
\bibitem{moh95}R.N. Mohapatra, \ECT p.44
\bibitem{moh81}R.N. Mohapatra, G. Senjanovic, \Journal{\PRD}{23}{165}{1981}

\bibitem{fra95} A. Franklin, \Journal{\RMP}{67}{457}{1995}
\bibitem{wie96} F.E. Wietfeldt, E.B. Norman, \Journal{\PRP}{273}{149}{1996}
\bibitem{hol92} E. Holzschuh, \Journal{\RPP}{55}{851}{1992}
\bibitem{ott95} E.W. Otten, \Journal{\NPBP}{38}{26}{1995}
\bibitem{lob85} V.M. Lobashev et al., \Journal{\NIMA}{240}{305}{1985}
\bibitem{pic92} A. Picard et al., \Journal{\NIMB}{63}{345}{1992}
\bibitem{swi97} A.M. Swift et al., Proc. Neutrino'96, World Scientific, 1997, p.278
\bibitem{meu98} P. Meunier in astro-ph/9801320
\bibitem{deR81} A. deRujula, \Journal{\NPB}{188}{414}{1981}
\bibitem{spr87} P.T. Springer, C.L.Bennett, R.A. Baisden, \Journal{\PRA}{35}{679}{1987}
\bibitem{ass96} K. Assamagan et al., \Journal{\PRD}{53}{6065}{1996}
\bibitem{jec94} B. Jeckelmann et al., \Journal{\PLB}{335}{326}{1995}
\bibitem{len98} S. Lenz et al., \Journal{\PLB}{416}{50}{1998}
\bibitem{alb92} H. Albrecht et al., \Journal{\PLB}{292}{221}{1992}
\bibitem{cin93} R. Ammar et al., \Journal{\PLB}{431}{209}{1998}
\bibitem{ale96} G. Alexander et al., \Journal{\ZPC}{72}{231}{1996} hep-ex/9806035 
\bibitem{pas97} L. Passalacqua, \Journal{\NPBP}{55C}{435}{1997}
\bibitem{bar98} R. Barate et al., \Journal{\EPC}{2}{395}{1998}
\bibitem{gyu95} G. Gyuk, M.S. Turner, \Journal{\NPBP}{38}{13}{1995}
\bibitem{fie97} B.D. Fields, K. Kainulainen, K.A. Olive,
\Journal{\ASP}{6}{169}{1997}


\bibitem{doi83} M. Doi, T. Kotani, E. Takasugi, \Journal{\PTP}{69}{602}{1983}
\bibitem{doi85} M. Doi, T. Kotani, E. Takasugi, \Journal{\PTPS}{83}{1}{1985}
\bibitem{mut88}K.Muto, H.V. Klapdor in {\it Neutrinos}, ed. H.V. Klapdor, Springer 1988
\bibitem{goe35}M. Goeppert-Mayer, \Journal{\PR}{48}{512}{1935}
\bibitem{fur39}W.H. Furry, \Journal{\PR}{56}{1184}{1939}
\bibitem{sch82} J. Schechter, J.W.F. Valle, \Journal{\PRD}{25}{2591}{1982}
\bibitem{tak84} E. Takasugi, \Journal{\PLB}{149}{372}{1984}
\bibitem{kla95} H.V. Klapdor-Kleingrothaus, A. Staudt, {\it  Non accelerator 
particle physics}, IOP
Publ., Bristol 1995
\bibitem{boe92} F. Boehm, P. Vogel, {\it Physics of massive neutrinos},
Cambridge Univ. Press 1992
\bibitem{tre95}V.I. Tretyak, Y.G. Zdesenko, At. Dat. and Nucl. Dat. Tab. 61,43 (95)
\bibitem{heu95} G. Heusser, \Journal{\ARNP}{45}{543}{1995}
\bibitem{fio67}E. Fiorini et al. , \Journal{\PLB}{45}{602}{1967}
\bibitem{gue97} M. Guenther et al., \Journal{\PRD}{55}{54}{1997}
\bibitem{avi97}F.T. Avignone, Proc. Neutrino'96, World Scientific, 1997 p.361 
\bibitem{you91}K. You et al. , \Journal{\PLB}{265}{53}{1991}
\bibitem{geo95} A.S. Georgadze, \Journal{\PAN}{58}{1093}{1995}
\bibitem{ale97} E. Fiorini, Proc. Neutrino'96, World Scientific, 1997 p.352
\bibitem{far97} J. Farine, Proc. Neutrino'96, World Scientific, 1997 p.347 
\bibitem{eji97} H. Ejiri, Proc. Neutrino'96, World Scientific, 1997 p.342
\bibitem{arn95} R. Arnold et al., \Journal{\NPA}{636}{209}{1998}
\bibitem{kir86} T. Kirsten, Proc. {\it Int. Symp. on Nuclear beta decays 
and neutrinos}, eds. T. Kotani,
H. Ejiri, E. Takasugi, Osaka 1986 , World Scientific p.81
\bibitem{vog86} P. Vogel, M.R. Zirnbauer, \Journal{\PRL}{57}{3148}{1986}
\bibitem{gro89} K. Grotz, H.V. Klapdor, {\it Weak interactions in nuclear-,particle- and
astrophysics}, Adam Hilger, Bristol 1990
\bibitem{suh98} J. Suhonen, O. Civitarese, \Journal{\PRP}{300}{123}{1998}
\bibitem{sta90} A. Staudt, K. Muto, H.V. Klapdor-Kleingrothaus, \Journal{\EPL}{13}{31}{1990}
\bibitem{kir67} T. Kirsten, W. Gentner, O.A. Schaeffer, \Journal{\ZP}{202}{273}{1967}
\bibitem{kir68} T. Kirsten, W. Gentner, O.A. Schaeffer, \Journal{\PRL}{20}{1300}{1968}
\bibitem{lin88} W.J. Lin et al., \Journal{\NPA}{481}{477 and 484}{1988}
\bibitem{ber92} T. Bernatowicz et al., \Journal{\PRC}{47}{806}{1993}
\bibitem{tur92} A.I. Turkevich, T.E. Economou, G.A. Cowen, \Journal{\PRL}{67}{3211}{1992}
\bibitem{ell87} S.R. Elliott, A.A.Hahn, M.Moe, \Journal{\PRL}{59}{1649}{1987}
\bibitem{tom91} T. Tomoda, \Journal{\RPP}{54}{53}{1991}
\bibitem{vog95} P. Vogel in \ECT p.323
\bibitem{fae95} A. Faessler in \ECT p.339
\bibitem{mut91} K. Muto, E. Bender, H.V. Klapdor, 
\Journal{\ZPA}{39}{435}{1991}
\bibitem{bau97}L. Baudis et al., \Journal{\PLB}{407}{219}{1997}
\bibitem{hir96} M. Hirsch et al., \Journal{\PLB}{374}{7}{1996}
\bibitem{hir96a} M. Hirsch, H.V. Klapdor-Kleingrothaus, S. Kovalenko, 
\Journal{\PLB}{352}{1}{1995}, \Journal{\PRD}{53}{1329}{1996}
\bibitem{hir96b} M. Hirsch, H.V. Klapdor-Kleingrothaus, S. Kovalenko, 
\Journal{\PLB}{378}{17}{1996}, \Journal{\PRD}{54}{4207}{1996}
\bibitem{bar96a} A.S. Barabash in \ECT p.502 
\bibitem{hal83} A.Halprin, S.T.Petcov, S.P.Rosen, \Journal{\PLB}{125}{335}{1983}
\bibitem{zub97} K. Zuber, \Journal{\PRD}{56}{1816}{1997}
\bibitem{doi88} M. Doi, T. Kotani, E. Takasugi, \Journal{\PRD}{37}{2575}{1988}
\bibitem{chi80} Y. Chikashige, R.N. Mohapatra, R.D. Peccei, \Journal{\PRL}{45}{1926}{1980}
\bibitem{san88}R. Santamaria, J.W.F. Valle, \Journal{\PRL}{60}{397}{1988}
\bibitem{gel81}G.B. Gelmini, M. Roncadelli, \Journal{\PLB}{99}{411}{1981}
\bibitem{bur94}C.P. Burgess, J.M. Cline, \Journal{\PRD}{49}{5925}{1994}
\bibitem{hir96c} M. Hirsch et al., \Journal{\PLB}{372}{8}{1996}
\bibitem{car93} C.D. Carone, \Journal{\PLB}{308}{85}{1993}
\bibitem{moh88} R.N. Mohapatra, E. Takasugi, \Journal{\PLB}{211}{192}{1988}
\bibitem{zub92} K. Zuber, Talk presented at PASCOS'92, \Journal{\ANYAS}{688}{509}{1993}
\bibitem{tan93} J. Tanaka, H. Ejiri, \Journal{\PRD}{48}{5412}{1993}
\bibitem{gue96}M. Guenther et al. , \Journal{\PRD}{54}{3641}{1996}
\bibitem{nor84} E.B. Norman, M.A. DeFaccio, \Journal{\PLB}{148}{31}{1984}
\bibitem{hir94} M. Hirsch et al., \Journal{\ZPA}{347}{151}{1994} 
\bibitem{ale98} A. Alessandrello et al., \Journal{\PLB}{420}{109}{1998}
\bibitem{bar97b} A.S. Barabash, Proc. Neutrino'96, World Scientific, 1997 p.374
\bibitem{rag94} R.S. Raghavan, \Journal{\PRL}{72}{1411}{1994}
\bibitem{fio98} E. Fiorini, private communication
\bibitem{kla98} H.V. Klapdor-Kleingrothaus, J. Hellmig, M. Hirsch, 
\Journal{\JPG}{24}{483}{1998}
\bibitem{moe91} M. Moe, \Journal{\PRC}{44}{R931}{1991}
\bibitem{mit96} L. Mitchell, Talk presented at Topical Workshop on Neutrino 
Physics, Adelaide 1996
\bibitem{lee77} B.W. Lee, R.E. Shrock, \Journal{\PRD}{16}{1444}{1977}
\bibitem{mar77} W.J. Marciano, A.I. Sanda, \Journal{\PLB}{67}{303}{1977}
\bibitem{pal92}P.B. Pal, \Journal{\IJMP}{A7}{5387}{1992}
\bibitem{kra90} D. Krakauer et al., \Journal{\PLB}{252}{171}{1990}
\bibitem{abe87} K. Abe et al., \Journal{\PRL}{58}{636}{1987}
\bibitem{kim88} C.S. Kim, W.J. Marciano, \Journal{\PRD}{37}{1368}{1988}
\bibitem{vol86} M.B. Voloshin, M.I. Vysotsky, L.B. Okun, \Journal{\SJNP}{44}{440}{1986}
\bibitem{bar88} R. Barbieri, R.N. Mohapatra, \Journal{\PRL}{61}{27}{1988}
\bibitem{lat88} J.M. Lattimore, J. Cooperstein, \Journal{\PRL}{61}{23}{1988}
\bibitem{raf90} G. Raffelt, \Journal{\PRP}{198}{1}{1990}
\bibitem{ams97}C. Amsler et al., \Journal{\NIMA}{396}{115}{1997}
\bibitem{bed97} A.G. Beda et al., \Journal{\PAN}{61}{66}{1998}
\bibitem{bar96} I. Barabanov et al., \Journal{\ASP}{5}{159}{1996}
\bibitem{jar90} C. Jarlskog, \Journal{\PLB}{241}{579}{1990}
\bibitem{tom94} D. Tommasini et al., \Journal{\NPB}{444}{451}{1994}
\bibitem{hof96} A. Hoefer, L.M. Seghal, \Journal{\PRD}{54}{1944}{1996}
\bibitem{buc91} W. Buchm\"uller, C. Greub, \Journal{\NPB}{363}{345}{1991}
\bibitem{glu97} J. Gluza, M. Zralek, \Journal{\PRD}{55}{7030}{1997}
\bibitem{ma89} E. Ma, J. Pantaleone, \Journal{\PRD}{40}{2172}{1989}
\bibitem{dic91} D.A. Dicus, P. Roy, \Journal{\PRD}{44}{1593}{1991}
\bibitem{alm97} F.M.L. Almeida et al., \Journal{\PLB}{400}{331}{1997}

\bibitem{pon57} B. Pontecorvo, \Journal{\ZETZ}{33}{549}{1957}, \Journal{\ZETZ}{34}{247}{1958}
\bibitem{kay89} B. Kayser, F. Gibrat-Debu, F. Perrier, {\it Physics of massive \neus}, 
World Scientific 1989
\bibitem{kim93} J.C.W. Kim, A. Pevzner, {\it Neutrinos in physics and astrophysics}, 
Harwood Academic, New York 1993
\bibitem{zub97a} K. Zuber, Proc. 4th Int. Solar Neutrino Conference,
ed. W. Hampel, MPIK Heidelberg, 1997 hep-ph/9807468
\bibitem{sch85} K. Schreckenbach et al., \Journal{\PLB}{160}{325}{1985}
\bibitem{kla82} H.V. Klapdor, J. Metzinger, \Journal{\PRL}{48}{127}{1982}
\bibitem{dec94} Y. Declais et al., \Journal{\PLB}{338}{383}{1994}
\bibitem{chooz} H. de Kerret et al.,The CHOOZ Experiment, Proposal, LAPP Report (1993)
\bibitem{apo98} M. Apollonio et al., \Journal{\PLB}{420}{397}{1998}
\bibitem{pave}G. Gratta, Proc. Neu\-tri\-no'96, World Scientific 1997, p. 248
\bibitem{kamland}M. Nakahata, Talk presented at Conf. EPS'97, Jerusalem, Aug. 1997
\bibitem{karmen}G. Drexlin et al., \Journal{\PNPP}{32}{351}{1994}
\bibitem{arm98} B. Armbruster et al., \Journal{\PRC}{57}{3414}{1998}
\bibitem{lsnddet}C. Athanassopoulos et al., \Journal{\NIMA}{388}{149}{1997}
\bibitem{lsnd}C. Athanassopoulos et al., \Journal{\PRL}{77}{3082}{1996}
\bibitem{lsnd1}C. Athanassopoulos et al., LA-UR-97-1998 (1997), subm. to Phys. Rev. C
\bibitem{chu97} E. Church et al., nucl-ex/9706011
\bibitem{dyd84} F. Dydak et al., \Journal{\PLB}{134}{281}{1984}
\bibitem{all88} J.W. Allaby et al., \Journal{\ZPC}{40}{171}{1988}
\bibitem{xxx97} N. Armenise et al., CERN-SPSC/97-21
\bibitem{amb98} G. Ambrosini et al., \Journal{\PLB}{420}{225}{1998}
\bibitem{van96} B. VandeVyver, P. Zucchelli, CERN-PPE 96-113
\bibitem{gon96} M.C. Gonzales-Garcia, J.J. Gomez-Cadenas, CERN-PPE 96-114
\bibitem{chorusdet}E. Eskut et al., \Journal{\NIMA}{401}{7}{1997}
\bibitem{chores} E. Eskut et al., \Journal{\PLB}{424}{202}{1998} hep-ex/9807024
\bibitem{nomaddet} J. Altegoer et al., \Journal{\NIMA}{404}{96}{1998}
\bibitem{nomadres} J. Altegoer et al., \Journal{\PLB}{431}{291}{1998}, D. Autiero, Talk
presented at ICHEP98, Vancouver
\bibitem{ccfr}A. Romosan et al., \Journal{\PRL}{78}{2912}{1997}
\bibitem{tosca} A.S. Ayan et al., CERN-SPSC-97-5
\bibitem{rev97} J.P. Revol e al., ICARUS-TM-97/01
\bibitem{aut97} D. Autiero et al., CERN-SPSC/97-23 
\bibitem{keksk} Y. Oyama, hep-ex/9803014
\bibitem{minos}E. Ables et al., Fermilab-Proposal P875 (1995)
\bibitem{icarus}C. Rubbia, \Journal{\NPBP}{48}{172}{1996}
\bibitem{noe}M. Ambrosio et al., \Journal{\NIMA}{363}{604}{1995}, G. de Cataldo et al., 
http://www1.na.infn.it/wsubnucl/accel/noe/noe.html
\bibitem{tom}T. Ypsilantis et al., Preprint LPC/96-01, CERN-LAA/96-13
\bibitem{niwa} H. Shibuya, et al., CERN-SPSC-97-24, LNGS-LOI 8/97
\bibitem{gee97} S. Geer, \Journal{\PRD}{57}{6989}{1998}
\bibitem{thu95} M. Thunman, G. Ingelman, P. Gondolo, \Journal{\ASP}{5}{309}{1996}
\bibitem{bar89} G. Barr, T.K. Gaisser, T. Stanev, \Journal{\PRD}{39}{3532}{1989}
\bibitem{bug89} E.V. Bugaev, V.A. Naumow, \Journal{\PLB}{232}{391}{1989}
\bibitem{hon90} M. Honda et al., \Journal{\PLB}{248}{193}{1990}
\bibitem{per94} D.H. Perkins, \Journal{\ASP}{2}{249}{1994}
\bibitem{agr96} V. Agrawal et al. , \Journal{\PRD}{53}{1314}{1996}
\bibitem{gai96} T.K. Gaisser et al., \Journal{\PRD}{54}{5578}{1996}
\bibitem{dau95} K. Daum et al., \Journal{\ZPC}{66}{417}{1995} 
\bibitem{bel96} R. Bellotti et al., \Journal{\PRD}{53}{35}{1996}
\bibitem{cir97} M. Circella et al., 25th ICRC Vol.7 p. 117, Durban 1997
\bibitem{bug98} E. V. Bugaev et al., \Journal{\PRD}{58}{054001}{1998} 
\bibitem{lip95} P. Lipari et al., \Journal{\PRL}{74}{4384}{1995}
\bibitem{fuk98} Y. Fukuda et al., \Journal{\PLB}{433}{9}{1998}
\bibitem{kas96} S. Kasuga et al., \Journal{\PLB}{374}{238}{1996}
\bibitem{fuk98a} Y. Fukuda et al., Preprint hep-ex/9805006 hep-ex/9807003 
\bibitem{kea98} E. Kearns, hep-ex/9803007, to appear in Proc. TAUP'97
\bibitem{man98a} G. Mannocchi et al., Preprint hep-ph/9801339 
\bibitem{yas98} O. Yasuda, Preprint hep-ph/9804400
\bibitem{gon98} M.C. Gonzales-Garcia et al., \Journal{\PRD}{58}{033004}{1998} 
\bibitem{foo97} R. Foot, R.R. Volkas, O. Yasuda, \Journal{\PLB}{421}{245}{1998}
\bibitem{vis97} F. Vissani, A.Y. Smirnov, Preprint hep-ph/9710565
\bibitem{lip97} P. Lipari, M. Lusignoli, \Journal{\PRD}{57}{3842}{1998}
\bibitem{fog97} G.L. Fogli, E. Lisi, A. Marrone, \Journal{\PRD}{57}{5893}{1998}
\bibitem{hon95} M. Honda et al., \Journal{\PRD}{52}{4985}{1995}
\bibitem{glu94} M. Gl\"uck, E. Reya, A. Vogt, \Journal{\ZPC}{67}{433}{1995}




\bibitem{cla68} D.D. Clayton, {\it Principles of stellar evolution}, New York,McGraw-Hill, 1968
\bibitem{bah89} J.N. Bahcall, {\it Neutrino Astrophysics}, Cambridge Univ. Press 1989 
\bibitem{bah95} J.N. Bahcall, R.M. Pinnsoneault, \Journal{\RMP}{67}{781}{1995}
\bibitem{iau90} IAU Proc. {\it Inside the sun}, eds. G. Berthomieu, M. Cribier, 
Berlin, Kluwer, 1990 
\bibitem{tur93a} S. Turck-Chieze et al., \Journal{\PRP}{230}{57}{1993} 
\bibitem{tur93b} S. Turck-Chieze, I. Lopes, \Journal{\APJ}{389}{478}{1993}
\bibitem{ric96} O. Richards et al., \Journal{\AA}{312}{1000}{1996}
\bibitem{dar96} A. Dar, G. Shaviv, \Journal{\APJ}{486}{933}{1996} 
\bibitem{cas97} V. Castellani et al., \Journal{\PRP}{281}{309}{1997} 
\bibitem{gre93} N. Grevesse, in {\it Origin and Evolution of elements}, ed. N. 
Prantzos,E. Vangioni Flam
and M. Casse, Cambridge Univ. Press 1993, p.15-25
\bibitem{gre93a}N. Grevesse, A. Noels, \Journal{\PS}{T47}{133}{1993}
\bibitem{ale94}D.R. Alexander, J.W. Ferguson, \Journal{\APJ}{437}{879}{1994}
\bibitem{igl96}C.A. Iglesias, F.J. Rogers, \Journal{\APJ}{464}{943}{1996}
\bibitem{tri97}S.C. Tripathy, J. Christensen-Dalsgaard, astro-ph/9709206
\bibitem{rog96}F.J. Rogers, F.J. Swenson, C.A. Iglesias, \Journal{\APJ}{456}{902}{1996}
\bibitem{rol88} C.E. Rolfs, W.S. Rodney, {\it Cauldrons in the cosmos}, The university 
of chicago press, 1988
\bibitem{adl98} E.G. Adelberger et al., astro-ph/9805121
\bibitem{arp97} M. Junker et al., \Journal{\PRC}{57}{2700}{1998}
\bibitem{ham98} F. Hammache et al., \Journal{\PRL}{80}{928}{1998}
\bibitem{kav69} R.W. Kavanagh et al., \Journal{\BAPS}{14}{1209}{1969}
\bibitem{fil83} B.W. Filippone et al., \Journal{\PRL}{50}{412}{1983} 
and \Journal{\PRC}{28}{2222}{1983}
\bibitem{bro97} C. Broude et al., CERN/ISC 97-17
\bibitem{dav97} R.Davis Jr. et al., Proc. 4th Int. Solar Neutrino Conference,
ed. W. Hampel, MPIK Heidelberg, 1997
\bibitem{cle98} B.T. Cleveland et al., \Journal{\APJ}{496}{505}{1998}
\bibitem{auf94}M. Aufderheide et al., \Journal{\PRC}{49}{678}{1994}
\bibitem{bah96}J.N. Bahcall et al., \Journal{\PRC}{54}{411}{1996}
\bibitem{lan97} K. Lande et al., Proc. 4th Int. Solar Neutrino Conference,
ed. W. Hampel, MPIK Heidelberg, 1997
\bibitem{tot97} Y. Totsuka, Talk presented at 35th Int. School 
on Subnuclear Physics, Erice 1997, Y. Suzuki, Talk presented at Neutrino'98
\bibitem{fuk98b} Y. Fukuda in astro-ph/9801320
\bibitem{abd97} J.N. Abdurashitov et al, Proc. 4th Int. Solar Neutrino
Conference, ed. W. Hampel, MPIK Heidelberg, 1997
\bibitem{kir97} T. Kirsten, Talk presented at TAUP97, Gran Sasso 1997
\bibitem{bah97}J.N. Bahcall, M.H. Pinsonneault, Proc. Neutrino'96, 
World Scientific, 1997, p.56
\bibitem{bah91}J.N. Bahcall, \Journal{\PRD}{44}{1644}{1991}
\bibitem{hat97} N. Hata, P. Langacker, \Journal{\PRD}{56}{6107}{1997}
\bibitem{hat95} N. Hata, P. Langacker, \Journal{\PRD}{52}{420}{1995}
\bibitem{cum96} A. Cumming, W.C. Haxton, \Journal{\PRL}{77}{4286}{1996}
\bibitem{cas97a}V. Castellani et al., astro-ph/9712174, to appear in Proc. TAUP'97  
\bibitem{dzi96} W. A. Dziembinski, \Journal{\BASI}{24}{133}{1996}
\bibitem{wol78} L. Wolfenstein, \Journal{\PRD}{17}{2369}{1978}
\bibitem{mik86} S.P. Mikheyev, A.Y. Smirnow, \Journal{Nuovo Cimento C}{9}{17}{1986}
\bibitem{bah97a} J.N. Bahcall, P.I. Krastev, \Journal{\PRD}{56}{2839}{1997}
\bibitem{lis97} E. Lisi, D. Montanino, \Journal{\PRD}{56}{1792}{1997}
\bibitem{bah98} J.N. Bahcall, P.I. Krastev, A.Y. Smirnov, hep-ph/9807216
\bibitem{pet97} S.T.Petcov, Proc. 4th Int. Solar Neutrino Conference, ed. W. Hampel, 
MPIK Heidelberg, 1997
\bibitem{liu97} Q.Y.Liu, Proc. 4th Int. Solar Neutrino Conference, ed. W.
Hampel, MPIK Heidelberg, 1997 
\bibitem{fog98} G.L. Fogli, E. Lisi, D. Montanino, hep-ph/9803309
\bibitem{mar88} C.S. Lim, W.J. Marciano, \Journal{\PRD}{37}{1368}{1988}
\bibitem{akh97}E. Akhmedov, Proc. 4th Int. Solar Neutrino Conference, ed. W.
Hampel, MPIK Heidelberg, 1997
\bibitem{gal97}M. Galeazzi et al., \Journal{\PLB}{398}{187}{1997}
\bibitem{rag97}R. S. Raghavan, \Journal{\PRL}{78}{3618}{1997}
\bibitem{geo97}A.S. Georgadze et al., \Journal{\ASP}{7}{173}{1997}



\bibitem{whe90} J.C.Wheeler in {\it Supernovae}, Jerusalem School, World
Scientific 1990
\bibitem{woo86} S.E. Woosley, T.A. Weaver, \Journal{\ARAA}{24}{205}{1986}
\bibitem{woo90} S.E. Woosley (ed.), {\it Supernovae}, Springer, 1990
\bibitem{col90} S.A. Colgate in {\it Supernovae}, Jerusalem School, World
Scientific 1990
\bibitem{tot97a} T. Totani et al., astro-ph/9710203
\bibitem{mal97} R.A. Malaney, \Journal{\ASP}{7}{127}{1997}
\bibitem{har97} D.H. Hartmann, S.E. Woosley, \Journal{\ASP}{7}{137}{1997}
\bibitem{tot95} T. Totani, K. Sato, \Journal{\ASP}{3}{367}{1995}
\bibitem{tot96} T. Totani, K. Sato, Y. Yoshii, \Journal{\APJ}{460}{303}{1996}
\bibitem{arn89} W.D. Arnett et al., \Journal{\ARAA}{27}{629}{1989}
\bibitem{bio87} R.M. Bionta et al., \Journal{\PRL}{58}{1494}{1987}
\bibitem{bra88} C.B. Bratton et al., \Journal{\PRD}{37}{3361}{1988}
\bibitem{hir87} K.S. Hirata et al., \Journal{\PRL}{58}{1490}{1987}
\bibitem{hir88} K.S. Hirata et al., \Journal{\PRD}{38}{448}{1988}
\bibitem{ale88} E.N. Alekseev et al., \Journal{\PLB}{205}{209}{1988}
\bibitem{agl88} M. Aglietta et al., \Journal{\EPL}{3}{1315}{1988}
\bibitem{raf96} G. Raffelt, {\it Stars as laboratories for
fundamental physics}, University of Chicago Press, 1996
\bibitem{chu89} E.L. Chupp, W.T.Vestrand, C. Reppin, \Journal{\PRL}{62}{505}{1989}
\bibitem{mil96} R.S. Miller, J.M. Ryan, R.C. Svoboda, \Journal{\AAS}{120}{635}{1996}
\bibitem{ful92} G.M. Fuller et al., \Journal{\APJ}{389}{517}{1992}
\bibitem{qia93} Y.Z. Qian et al., \Journal{\PRL}{71}{1965}{1993}
\bibitem{bea98} J.F. Beacom, P. Vogel, \Journal{\PRD}{58}{05130}{1998}
\bibitem{cli90} D.B. Cline, \Journal{\NPB}{14A}{348}{1990}
\bibitem{smi97} P.F. Smith, \Journal{\ASP}{8}{27}{1997}
\bibitem{gre66} K. Greisen, \Journal{\PRL}{16}{748}{1966}
\bibitem{zat66} G.T. Zatsepin, V.A. Kuzmin, \Journal{\JETP}{4}{53}{1966}
\bibitem{gai95} T.K. Gaisser, F.Halzen, T. Stanev, \Journal{\PRP}{258}{173}{1995}
\bibitem{hun97} S.D. Hunter et al., \Journal{\APJ}{481}{205}{1997}
\bibitem{man98} K. Mannheim, \Journal{\SCI}{279}{684}{1998}
\bibitem{ste91} F.W. Stecker et al., \Journal{\PRL}{66}{2697}{1991} and
\Journal{\PRL}{69}{2738(E)}{1992}
\bibitem{nel93} L. Nellen, K. Mannheim, P. Biermann, \Journal{\PRD}{47}{5270}{1993}
\bibitem{sza94} A.P. Szabo, R.J. Protheroe, \Journal{\ASP}{2}{375}{1994}
\bibitem{man95} K. Mannheim, \Journal{\ASP}{3}{295}{1995}
\bibitem{ber95} V. Berezinsky, \Journal{\NPBP}{38}{363}{1995} and references therein
\bibitem{hil97} G.C. Hill, \Journal{\ASP}{6}{215}{1997}
\bibitem{zas93} E. Zas, F. Halzen, R.A. Vasquez, \Journal{\ASP}{1}{297}{1993}
\bibitem{rho96} W. Rhode et al., \Journal{\ASP}{4}{217}{1996}
\bibitem{bha92} P. Bhattacharjee et al., \Journal{\PRL}{69}{567}{1992}
\bibitem{wax97} E. Waxmann, J.N. Bahcall, \Journal{\PRL}{78}{2292}{1997}
\bibitem{ber96} V. Berezinsky et al., \Journal{\ASP}{5}{333}{1996}
\bibitem{sig97} G. Sigl, S. Lee, D.N. Schramm, \Journal{\PLB}{392}{129}{1997}
\bibitem{gan96} R. Gandhi et al., \Journal{\ASP}{5}{81}{1996} hep-ph/9807264
\bibitem{bel97} I.A. Belolaptikov et al., \Journal{\ASP}{7}{263}{1997}
\bibitem{bar97d} S. Barwick et al., 25th ICRC Vol.7 p. 1, Durban 1997 
\bibitem{bir97} A. Biron et al., DESY-PRC 97-05
\bibitem{nes95} NESTOR-Proposal, May 1995
\bibitem{bla97} F. Blanc et al., ANTARES-Proposal, astro-ph/9707136
\bibitem{all97} C. Allen et al., astro-ph/9709223 
\bibitem{ded97} L.G. Dedenko et al., astro-ph/9705189 
\bibitem{zas92} E. Zas, F. Halzen, T. Stanev, \Journal{\PRD}{45}{362}{1992}
\bibitem{fix96} D.J. Fixsen et al., \Journal{\APJ}{473}{576}{1996}
\bibitem{kla97b} H.V. Klapdor-Kleingrothaus, K. Zuber, {\it Particle
astrophysics}, IOP Publ., Bristol 1997
\bibitem{pri95} J.R. Primack et al., \Journal{\PRL}{74}{2160}{1995}
\bibitem{reu91} D. Reusser et al., \Journal{\PLB}{255}{143}{1991}
\bibitem{bec94} M. Beck et al., \Journal{\PLB}{336}{141}{1994}
\bibitem{obe87} L. Oberauer et al., \Journal{\PLB}{198}{113}{1987}
\bibitem{kra91} D.A. Krakauer et al., \Journal{\PRD}{44}{R6}{1991}
\bibitem{blu92a} S. Bludman, \Journal{\PRD}{45}{4720}{1992}
\bibitem{hag95} C. Hagner et al., \Journal{\PRD}{52}{1343}{1995}
\bibitem{fei88} F. von Feilitzsch, in {\it Neutrinos} ed. H.V.Klapdor, Springer 1988
\bibitem{dol97} A.D. Dolgov, S.H. Hansen, D.V. Semikoz, \Journal{\NPB}{503}{426}{1997}
\bibitem{lop98} R. E. Lopez et al., astro-ph/9803205

\bibitem{car96} C.Y. Cardall, G.M. Fuller, \Journal{\PRD}{53}{4421}{1996}
\bibitem{ack97} A. Acker, S. Pakvasa, \Journal{\PLB}{397}{209}{1997}
\bibitem{cal93}D.O. Caldwell, R.N. Mohapatra,\Journal{\PRD}{48}{3259}{1993}
\bibitem{bil97} S.M. Bilenky, C. Giunti, W. Grimus, hep-ph/9711311
\bibitem{bar97}V. Barger, T.J. Weiler, K. Whisnant, \Journal{\PLB}{427}{97}{1998}
\bibitem{moh98} S.C. Gibbons et al., \Journal{\PLB}{430}{296}{1998} 

\end{thebibliography}
\end{document}